\documentclass[12pt]{article}
\usepackage[margin=1in]{geometry} 
\usepackage{natbib}

\usepackage{appendix,soul}
\usepackage{graphicx,placeins}
\usepackage{tabularx,booktabs,array,dcolumn}
    \newcolumntype{L}[1]{>{\raggedright\let\newline\\\arraybackslash\hspace{0pt}}m{#1}}
    \newcolumntype{C}[1]{>{\centering\let\newline\\\arraybackslash\hspace{0pt}}m{#1}}
    \newcolumntype{R}[1]{>{\raggedleft\let\newline\\\arraybackslash\hspace{0pt}}m{#1}}
    \newcolumntype{d}[1]{D{.}{.}{0}}
\usepackage{caption,subcaption}
\usepackage{chngpage}
\usepackage{float,rotating}
\usepackage{color,xcolor}
\usepackage{amsmath,amsthm,amssymb,bbold,amsfonts,mathtools}
    \allowdisplaybreaks
\usepackage{setspace}
\usepackage[bottom]{footmisc}   
\usepackage[pdftex,
            pdfauthor={Alexandros Theloudis, Jorge Velilla, Pierre-Andre Chiappori, Jose Ignacio Gimenez-Nadal, Jose Alberto Molina},
            pdftitle={Commitment and the Dynamics of Household Labor Supply},
            pdfsubject={D12; D13; D15; J22; J31},
            pdfkeywords={Household behavior; Intertemporal choice; Commitment; Collective model; Family labor supply; Dynamics; Wages; PSID},
            hyperfootnotes=false,bookmarksopen=true]{hyperref}
\hypersetup{colorlinks=true,allcolors=black}    

\newcommand{\fsz}{\footnotesize}

\newcommand{\mcl}{\multicolumn}

\title{\textbf{Commitment and the\\Dynamics of Household Labor Supply}\thanks{This paper develops and supersedes our previous work in \citet{ChiapporiEtAl2020_CommitmentIZA}. We thank the editor, Francesco Lippi, three anonymous referees, and seminar and conference participants in the Structural Econometrics Group at Tilburg, Essex, KU Leuven, the ISR/PSID, Porto, Wuppertal, Barcelona Summer Forum (Income Dynamics and the Family \& Structural Microeconometrics workshops), SOLE, SED, the Labor \& Marriage Markets workshop at Aarhus, the Gender \& Family Webinar by CY Cergy, and the Econometric Society North American Summer Meeting at UCLA for helpful comments. We benefited from many discussions with Jaap Abbring, Richard Blundell, Jeff Campbell, Sam Cosaert, George-Levi Gayle, Krishna Pendakur, Luigi Pistaferri, and Mich\`ele Tertilt.}}

\author{
Alexandros Theloudis\thanks{Tilburg University, 5000 LE Tilburg, The Netherlands; email: \href{mailto:a.theloudis@gmail.com}{a.theloudis@gmail.com}.}
\and 
Jorge Velilla\thanks{University of Zaragoza, 50005 Zaragoza, Spain; email: \href{mailto:jvelilla@unizar.es}{jvelilla@unizar.es}.}
\and
Pierre-Andr\'e Chiappori\thanks{Columbia University, New York, NY 10027, USA; email: \href{mailto:pc2167@columbia.edu}{pc2167@columbia.edu}.} 
\and 
J. Ignacio Gim\'enez-Nadal\thanks{University of Zaragoza, 50005 Zaragoza, Spain; email: \href{mailto:ngimenez@unizar.es}{ngimenez@unizar.es}.}
\and
Jos\'e Alberto Molina\thanks{University of Zaragoza, 50005 Zaragoza, Spain; email: \href{mailto:jamolina@unizar.es}{jamolina@unizar.es}.}
}

\date{Revised February 7, 2024}

\begin{document}

\maketitle

\begin{abstract}
The extent to which individuals commit to their partner for life has important implications. This paper develops a lifecycle collective model of the household, through which it characterizes behavior in three prominent alternative types of commitment: full, limited, and no commitment. We propose a test that distinguishes between all three types based on how contemporaneous and historical news affect household behavior. Our test permits heterogeneity in the degree of commitment across households. Using recent data from the Panel Study of Income Dynamics, we reject full and no commitment, while we find strong evidence for limited commitment.

\bigskip\bigskip

\noindent \textbf{Keywords}: Household behavior; Intertemporal choice; Commitment; Collective model; Family labor supply; Dynamics; Wages; PSID.

\bigskip\smallskip

\noindent \textbf{JEL classification}: D12; D13; D15; J22; J31.
\end{abstract}
\pagebreak


\section{Introduction}

Large amounts of commitment are necessary for investing in common assets, for producing goods at home efficiently, and for pooling risk across family members. By contrast, limits to commitment typically induce investments in private assets, prevent the partners from economically abusing each other, and offer a way out from a bad marriage. In this paper, we develop a lifecycle collective model of the household, through which we characterize behavior in three alternative regimes: full, limited, and no commitment. We show that current and past news affect behavior differently in each case and we propose a test that distinguishes between all three. Using recent PSID data, we reject full and no commitment, while we find strong evidence for limited commitment with large heterogeneity across households. 

Consider two or more parties who interact repeatedly sharing risk, such as spouses who offer each other intra-household insurance \citep{Mazzocco2007,LiseYamada2019}, village households who transfer goods or income among them \citep[][]{Townsend1994RiskInsurance,Ligon2002}, workers who supply labor and firms that offer employment \citep{ThomasWorrall1988SelfEnforcingWageContracts,BeaudryDiNardo1991EffectImplicitContracts}, or agents who trade assets \citep{KehoeLevine1993DebtConstrainedAssetMarkets,AlvarezJermann2000Efficiency}. The extent to which the parties commit to some future behavior is clearly crucial as it shapes the degree of risk sharing between them. 

This paper investigates this issue from a household perspective. We follow the by now dominant approach to household behavior, the collective model \citep{Chiappori1988,Chiappori1992}, which relies on \emph{cooperative} game theory. While in its static version this approach assumes that individuals reach Pareto-efficient agreements, its dynamic aspects are more complex. Although \emph{ex post} efficiency is generally maintained, \emph{ex ante} efficiency strongly depends on assumptions about the spouses' ability to commit. Following \citet{Mazzocco2007}, the literature has considered three alternative modes of commitment. At one extreme, full commitment is assumed: the spouses commit to a plan that disciplines the sharing of resources regardless of shocks that may affect them differently \citep[e.g.][]{ChiapporiostaDiasMeghir2018MarriageMarket}. Full commitment has nice normative properties but its realism is disputable; in particular, there exists few (if any) legal ways to enforce this type of agreement. At the other extreme, no form of commitment is possible, and the relationship amounts to a sequence of short-term bargaining games (see \citealp{LundbergPollak2003EfficiencyMarriage}, for a theoretical analysis, or \citealp{GousseEtAl2017Marriage}, for an empirical application). Clearly, such models result in a high level of dynamic inefficiency; in particular, they imply very strong restrictions on the ability to share risk within the group.

In the middle are limited commitment models, in which the spouses commit to a plan up to the point that some shock reduces one's individual welfare below their outside option \citep[][]{Mazzocco2007}. They then renegotiate the plan or unilaterally switch to their outside option as in the case of unilateral divorce \citep[e.g.][]{Voena2015}. Such models have mostly been developed in a non-cooperative context \citep{ThomasWorrall1988SelfEnforcingWageContracts,Kocherlakota1996}; as such, they typically require an infinite horizon for the existence of non-trivial subgame perfect equilibria. Yet, they admit a straightforward, \emph{cooperative} interpretation; namely, they are ex ante efficient in a second best sense, i.e. taking into account the limitations that exist from the spouse's ability to commit. Nevertheless, limited commitment models still assume quite a lot of commitment. For instance, if an agent's outside option increases (although not enough to become binding), she will not try to renegotiate a better outcome, even though the static bargaining solution will have evolved in her favor.\footnote{In the non-cooperative setting of \citet{ThomasWorrall1988SelfEnforcingWageContracts} or \citet{Kocherlakota1996}, partial commitment is enforced via the infinite horizon and the possibility of future retaliations in \emph{any} period. In our finite horizon setting, only trivial subgame perfect equilibria exist, and partial commitment has to be assumed, in line with the mostly axiomatic approach of cooperative game theory. This point becomes clearer below.}

Most studies of the dynamics of household behavior make specific assumptions about commitment; their predictions are thus dependent on such assumptions. Yet, a few papers develop comparisons between models \citep[][]{Mazzocco2007,LiseYamada2019}. Relying on a parametric framework, these works try to identify the Pareto weight in each period and contrast its dynamics between \emph{full} and \emph{non}-full commitment. The goal of our paper is to propose a more direct, reduced form approach by focusing on which variables affect behavior in \emph{each} (full, limited, no) commitment mode. Specifically, we do three things.

First, we characterize household behavior in each commitment mode. We establish that shocks to the economic environment of the family affect behavior differently in each case. The differences manifest via the Pareto weight on each person's preferences that disciplines the sharing of marital surplus. In full commitment, shocks, whether current or past, do not affect the Pareto weight, which remains constant over time or across states of the world. In limited commitment, current shocks may shift the Pareto weight if they trigger a renegotiation; this depends on the history of the couple, therefore also on past shocks, because individual welfare from marriage is a function of the past sharing of resources in the couple. In no commitment, by contrast, current shocks shift the Pareto weight continuously regardless of past shocks or circumstances. These restrictions on the Pareto weight translate into analogous restrictions on family labor supply, as we establish subsequently.

Second, we show that the set of variables that enter the full commitment Pareto weight (for which neither current nor historical information matters) is nested within the set that enters the no commitment weight (for which current information matters but historical information does not), which in turn is nested within the set that enters the limited commitment weight (for which all information matters). Thus the direction of nesting is not what one would expect by the names of the commitment regimes alone. Nesting, a common recursive form across modes (which we derive), and natural exclusion restrictions from current and past news allows us to devise a test that, for the first time, separates the three regimes. 

Our test is about the presence of effects from current and past shocks, as well as the sign of such effects. In limited commitment, current shocks to distribution factors (variables that enter the Pareto weight) affect behavior in a way that is tightly determined by the assignability of the shock. Consider a cash transfer to women \citep[e.g.][]{ArmandEtAl2020}. If the transfer triggers a renegotiation, the female Pareto weight should increase, raising her leisure and reducing her labor supply. By contrast, the male weight would decrease, reducing his leisure and increasing his labor supply. These \emph{asymmetric} effects reflect a power shift in the couple, where favorable news empower its recipient and simultaneously weaken their partner. The renegotiation, however, depends on the individual welfare from marriage today, which depends positively on the Pareto weight \emph{until} today, which is itself determined by previous renegotiations. Therefore, the Pareto weight has memory and \emph{past} shocks to distribution factors matter for current behavior in the same asymmetric way between spouses as current shocks do. By contrast, history does not matter in no commitment and past shocks are bygones. Neither current nor past shocks matter in full commitment.

Third, we confirm that wages enter the Pareto weight naturally even though they are not conventional distribution factors (they also affect the budget set). We show that their bargaining effects (effects \emph{through} the Pareto weight) are distinguishable from conventional income and substitution effects and can thus be used to test for commitment. This is appealing because wages are more easily available in household data than conventional distribution factors, they are assignable, and they typically vary considerably over time.

We implement our test empirically in a sample of married couples from the Panel Study of Income Dynamics in the US over the last two decades. Our main outcome of interest is individual labor supply in the household while our main source of shocks is male and female wage shocks. Our primary empirical exercise can thus be seen as one that investigates the dynamic effects of (current and past) wages on family labor supply.  

We consistently reject full and no commitment. By contrast, we find strong evidence for limited commitment. Favorable shocks reduce one's own labor supply \emph{and} increase the partner's; this is \emph{simultaneously} true for current \emph{and} historical shocks from multiple periods, precisely as limited commitment postulates. These effects, which cannot be explained on the basis of substitution, income, wealth, or tax adjustments, are consistent with power shifts in which favorable shocks improve the bargaining power of the recipient spouse. History matters under limited commitment, so shocks that shifted past bargaining power have lasting effects on behavior in a very specific way. This is exactly what we find in the data.

The simplest form of our test can be implemented fairly easily in \emph{reduced form}, without parameterizing or estimating individual preferences. This is appealing because, on one hand, the test does not rely on a specific functional form for utility and, on the other hand, it can help quickly inform about commitment without simulating a rather involved dynamic model. However, estimation of parts of the \emph{underlying structure} permits and reveals large heterogeneity in the degree of commitment across households. The overall evidence for limited commitment masks in fact that many couples exhibit full commitment (null bargaining effects), while others strongly exhibit limited commitment. 

This paper contributes to the literature on household behavior and, in particular, to its intertemporal aspects. Bargaining and in particular collective models have recently become the norm in this literature. \citet{Voena2015} develops a limited commitment labor supply model to study the impact of unilateral divorce. \citet{FernandezWong2017FreetoLeave} have a similar goal, though their choice of model is one of no commitment. \citet{ChiapporiostaDiasMeghir2018MarriageMarket} develop a full commitment labor supply model to study the labor and marriage market implications of education choice. \citet{LiseYamada2019} use a time use model with no commitment to study resource sharing. \citet{Foerster} builds a household model with limited commitment to study how alimony affects parents' welfare.\footnote{\citet{Bronson2014DegreesAreForEver} and \citet{MazzoccoRuizYamaguchi2014LaborSupplyWealthMarriageDecisions} use limited commitment (LC) lifecycle models to study education choices and household specialization; \citet{GousseEtAl2017Marriage} study home production under no commitment; \citet{LowMeghirPistaferriVoena2018} study welfare reforms with LC; \citet{BlasuttoKozlov2021ChangingMarriage} and \citet{Reynoso2022} study how unilateral divorce affects cohabitation and the marriage market under LC; \citet{DeRockKovalevaPotoms} study demand for housing in a LC setting. \citet{Chiappori2017} review the literature.} While these excellent works select a priori the commitment technology available to agents (so their conclusions are conditional on that choice), we take a step back and \emph{test} for the extent of commitment in married couples. As such, the closest paper to ours is the seminal work of \citet{Mazzocco2007}, who tests for full vs. non-full commitment based on whether current news affects consumption sharing. \citet{LiseYamada2019} do similarly in a model with home production. They all find evidence for this and reject full commitment. While this is often seen as evidence for limited commitment, in reality these tests cannot separate no from limited commitment. By contrast, the test we propose distinguishes between all three alternatives based on the additional role of historical information. Moreover, our test is not only about the presence of effects from current and past news but also about the sign of such effects that is strictly disciplined by theory.\footnote{Other tests include \citet{Townsend1994RiskInsurance} in the context of risk sharing in village economies, \citet{Blau2016} in a household context, and \citet{Walther2018NoncooperativeDecisionMakingMalawi} in a setting with production inefficiencies. \citet{LafortuneLow2020} show empirically that home ownership affects the degree of commitment.}

The test is motivated by our characterization of behavior across commitment modes. The paper thus also relates to the macro and development literatures that study transfers without commitment \citep[e.g.][]{CoateRavallion1993ReciprocityWithoutCommitment,Kocherlakota1996,Ligon2002,Dubois2008}.\footnote{Our paper is also related to an expanding family literature in macroeconomics; see \citet{DoepkeTertilt2016FamiliesInMacroeconomics} and \citet{GreenwoodGunerVandenbroucke2017FamilyEconWritLarge} for excellent reviews.} \citet{Mazzocco2007} and \citet{AdamsEtAl2014ConsumeNowLater} do similarly in a full information, collective household context without, however, considering all three regimes we analyze here.\footnote{\citet{Chiappori2006} and \citet{CherchyeDeRockVermeulen2007characterization} characterize static collective behavior.} However, there are alternative modes of behavior, e.g. cooperative models with asymmetric information \citep{AshrafFieldVoenaZiparo2022}, moral hazard \citep[][]{Kinnan44}, or general infinite horizon non-cooperative models, to which our test is not directly applicable.

The paper is finally related to the literature in labor economics that concerns the labor supply response to wages, particularly to the partner's wages, as in \citet{Lundberg1985AddedWorker} and \citet{Hyslop2001IntrafamilyEarnings}; \citet{BellouKaymak2012WagesImplicitContracts}, \citet{BlundellPistaSaporta2016Family_Labour} and \citet{WuKrueger2021ConsumptionInsurance} are recent contributions. Our distinctive feature is the focus not only on responses to current wages but also on dynamic responses to spousal wages multiple periods in the past.

The paper develops as follows. Section \ref{Section::Illustration_Simple_Setting} illustrates the main ideas in a simple setting. Section \ref{Section::Household_Lifecycle_Behavior} presents the full lifecycle model and the complete characterization of behavior. Section \ref{Section::Test_Commitment} presents the formal test for commitment, section \ref{Section::Implementation} discusses its implementation, and section \ref{Section::Results} shows the results. Section \ref{Section::Discussion} discusses extensions and section \ref{Section::Conclusion} concludes.

\section{Illustration of main ideas}\label{Section::Illustration_Simple_Setting}

This section illustrates the main ideas in a simple parametric framework with two periods $t\in\{t_1,t_2\}$, no discounting, and no savings. The next section relaxes these restrictions. 

A household consists of two individuals, a male and a female, respectively subscripted by $j\in\{1,2\}$. Each partner has logarithmic indirect utility $v_j (x_{jt}) =\ln x_{jt}$, where $x_{jt}$ may be any assignable expenditure on goods or leisure; we will be more specific about this in section \ref{Section::Household_Lifecycle_Behavior}. Individual expenditure must satisfy the budget constraint $x_{1t} + x_{2t} = z_{1t} + z_{2t} = y_{t}$ in each period, where $z_{jt}$ denotes the realization of $j$'s income at $t$. Each spouse has in each period an outside option $\bar{U}_{jt} = \phi (z_{jt})$, which may depend on his/her current income realization. We will now derive individual consumption in each commitment case.

Under \emph{full commitment}, the couple decides its full (contingent) allocation plan ex ante. Specifically, it solves a program of the form $\max_{x} v_{1}(x_{1t_1}) + v_{1}(x_{1t_2})$, subject to the budget constraint in each period $t$ and an ex ante participation constraint for individual $2$, given by $v_{2}(x_{2t_1}) + v_{2}(x_{2t_2}) \geq \bar{U}_{2}$. Here $\bar{U}_{2}$ is defined ex ante and may reflect the situation on the marriage market \citep[as in][]{ChiapporiostaDiasMeghir2018MarriageMarket} or stem from some initial, cooperative bargaining game, e.g., over the partners' expectations of future income.

If $\mu_2$ denotes the Lagrange multiplier on the last constraint, the household program can be written as $\max_{x} \mu_1 \big(v_{1}(x_{1t_1}) + v_{1}(x_{1t_2})\big) + \mu_2 \big(v_{2}(x_{2t_1}) + v_{2}(x_{2t_2})\big)$, subject to the budget constraint in each period. $\mu_1$ is the multiplier on spouse 1's utility. Equivalently, the household solves, in each period $t$, the program $\max_{x} \mu_1 v_{1}(x_{1t}) + \mu_2 v_{2}(x_{2t})$ subject to the budget constraint at $t$, with solutions given by 
\begin{equation}\label{Eq::Illustration.FC}
    \frac{x_{jt_1}}{y_{t_1}} ~ \text{(period 1)} =  \frac{x_{jt_2}}{y_{t_2}} ~ \text{(period 2)} =   \frac{\mu_j}{\mu_1+\mu_2}, \quad \text{for }j\in\{1,2\}.  
\end{equation}
The crucial point is that the Pareto weights, $\mu_1$ and $\mu_2$, remain constant; in particular, they do \emph{not} depend on the realization of individual income in any period.\footnote{Individual expenditure thus depends only on total income, not its individual realizations. This property, called the \emph{mutuality principle}, characterizes efficient risk sharing, enabled by the full commitment assumption.} Under log utility, this implies a \emph{constant} share of total income accruing to each individual.

\emph{Limited commitment} requires deriving the \emph{ex ante}, \emph{second best} optimal contract. Different cases must be distinguished, depending on whether some participation constraint in marriage is violated (and, if so, in which period) by the full commitment solution. For instance, if the latter satisfies all participation constraints, then it coincides with the limited commitment solution. Alternatively, if income realizations result in a second period constraint being violated by the first best solution \eqref{Eq::Illustration.FC}, then the second period Pareto weight will be modified such that the constraint becomes exactly binding; as a result, the second period allocation will be affected by the second period income shocks.

A complete resolution is left to section \ref{Section::Household_Lifecycle_Behavior}; here, we simply focus on a particularly interesting case when a first period participation constraint is violated. For brevity, we let the outside options at $t_2$ be very low (so that second period constraints can be omitted). The ex ante, second best program can be written as $\max_{x} \mu_1 \big(v_{1}(x_{1t_1}) + v_{1}(x_{1t_2})\big) + \mu_2 \big(v_{2}(x_{2t_1}) + v_{2}(x_{2t_2})\big)$, subject to the budget constraint in each period and the first period participation constraints $v_{j}(x_{jt_1}) + v_{j}(x_{jt_2}) \geq \phi (z_{jt_1})$, for $j\in\{1,2\}$. Suppose, now, that the full commitment solution \eqref{Eq::Illustration.FC} violates one participation constraint -- say, of $j=1$, namely $\ln \frac{\mu_1}{\mu_1+\mu_2}y_{t_1} + \ln \frac{\mu_1}{\mu_1+\mu_2}y_{t_2} < \phi (z_{1t_1})$, while the constraint of $j=2$ is satisfied. At the \emph{second best} solution, spouse 1's constraint must be binding; its Lagrange multiplier $\nu (z_{1t_1})$ will be positive (while that on spouse 2's constraint is nil) and depend on the income realization $z_{1t_1}$. The household program then becomes $\max_{x} (\mu_1 + \nu (z_{1t_1})) \big(v_{1}(x_{1t_1}) + v_{1}(x_{1t_2})\big) + \mu_2 \big(v_{2}(x_{2t_1}) + v_{2}(x_{2t_2})\big) - \nu (z_{1t_1})\phi (z_{1t_1})$ subject to the budget constraint in each period (details in appendix \ref{Appendix::RecursiveFormulation}). Its solutions are 
\begin{align*}
\frac{x_{1t_1}}{y_{t_1}} ~ \text{(period 1)}   &= \frac{x_{1t_2}}{y_{t_2}}  ~ \text{(period 2)}   = \frac{\mu_1 + \nu (z_{1t_1})}{\mu_1 + \nu (z_{1t_1}) + \mu_2},\\
\frac{x_{2t_1}}{y_{t_1}} ~ \text{(period 1)}   &= \frac{x_{2t_2}}{y_{t_2}}  ~ \text{(period 2)}   = \frac{\mu_2}{\mu_1 + \nu (z_{1t_1}) + \mu_2},
\end{align*}
and the expenditure shares at $t_2$ depend on the realization of \emph{period 1} income $z_{1t_1}$ through $\nu (z_{1t_1})$. Technically, spouse $1$'s weight increases from $\mu_1$ to $\mu_1+\nu (z_{1t_1})$ in period 1 following the binding constraint, and this change persists in the absence of new violations. Note that the model assumes quite a lot of commitment; the period 2 outside option may have increased, but not enough to bind, so the spouses do not try to further renegotiate. It is precisely this level of commitment that generates the memory property we describe.

Under \emph{no commitment}, in the definition of `bargaining in marriage' of \citet{LundbergPollak2003EfficiencyMarriage} and \citet{Pollak2019BargainingInMarriage}, the spouses bargain in each period. Cooperative bargaining theory requires that, in each period, the allocation be Pareto efficient in the \emph{ex post} sense; therefore the household in period $t$ maximizes a weighted sum of utilities $\mu_{1t} v_{1} + \mu_{2t} v_{2}$, giving the solution
\begin{equation*}
    \frac{x_{jt}}{y_{t}} = \frac{\mu_{jt}}{\mu_{1t}+\mu_{2t}}, \quad \text{for }j\in\{1,2\} \text{ and } t\in\{t_1,t_2\}.  
\end{equation*}
In order to determine the Pareto weights $\mu_{jt}$ and the allocation of resources in each period, the model is solved backwards. In period 2, the allocation, as summarized by the weights $\mu_{jt_2}$, depends only on the period 2 bargaining framework, i.e. total income $y_{t_2}$ and the second period reservation utilities $\phi (z_{1t_2})$ and $\phi (z_{2t_2})$.\footnote{The exact form of $\mu_{jt_2}\big(y_{t_2}, \phi (z_{1t_2}), \phi (z_{2t_2})\big)$ depends on the specific bargaining concept at play (Nash, Kalai–Smorodinsky, etc.), on which no assumptions are made beyond cooperation.} In period 1, $\mu_{jt_1}$ and the allocation depend on $y_{t_1},$ $\phi (z_{1t_1})$ and $\phi (z_{2t_1})$, as well as on the expectations of individual future incomes $\tilde{z}_{1t_2}$, $\tilde{z}_{2t_2}$ given $z_{1t_1}$, $z_{2t_1}$, but, obviously, not on the \emph{realizations} of $z_{1t_2}$ and $z_{2t_2}$. 

Two remarks are due here. First, from a non-cooperative viewpoint, in any \emph{finite horizon} model of this type, the \emph{only} subgame perfect equilibrium leads to the above no commitment outcome. Second, the period 2 allocation does \emph{not} vary with period 1 incomes. In period 2 the spouses play a static, cooperative bargaining game, the outcome of which depends only on \emph{current} total income $y_{t_2}$ and outside options, as summarized by $\mu_{jt_2}\big(y_{t_2}, \phi (z_{1t_2}), \phi (z_{2t_2})\big)$. None of the past period's incomes matter anymore -- bygones are bygones.

In summary, under full commitment, idiosyncratic income realizations cannot affect expenditure shares in any period. Under limited commitment, expenditure shares may reflect current realizations (if a current participation constraint is violated) and past ones (if those have modified the past Pareto weight but current realizations have left it intact). Under no commitment, idiosyncratic realizations in each period influence the expenditure shares for that period, but past ones influence neither current nor future expenditure shares.

\section{Household lifecycle behavior}\label{Section::Household_Lifecycle_Behavior}

We now switch to our full setting, namely the lifecycle collective model in which forward-looking spouses make consumption/hours choices subject to idiosyncratic wage risk.

A household consists of a male ($j=1$) and a female ($j=2$) spouse, who get married at time $t=0$ and live for $\bar{t}$ periods. In each period $t\in\{0,\dots,\bar{t}\}$, each person enjoys utility from joint consumption $q_{t}$ and disutility from labor hours $h_{jt}$, as per individual preferences $u_{j}(q_{t},h_{jt})$. We assume that $u_{j}$ has continuous first/second partial derivatives with $u_{j[q]}>0$, $u_{j[h]}<0$ (disutility of work), and $u_{j[qq]}<0$, $u_{j[hh]}<0$ (concavity); utility is \emph{not} quasi-linear, a class of preferences for which we find no empirical support anyway. 

The couple's budget constraint, common across commitment alternatives, is given by
\begin{equation}\label{Eq::Budget_Constraint}
(1+r)a_{t} + \tau(y_{t}) = q_{t} + a_{t+1}, \quad \forall t, 
\end{equation}
where $a_{t}$ is common financial assets at the start of the period and $r$ is the deterministic interest rate.\footnote{The extension to risky assets is straightforward and does not affect the subsequent discussion. Assets in marriage are jointly held reflecting the community property regime in the US \citep[][]{Mazzocco2007}.} $\tau$ maps gross household earnings $y_{t} = w_{1t} h_{1t} + w_{2t} h_{2t}$ into disposable income $y_{t}^{D}$, accounting for joint taxation and benefits (e.g. EITC). $w_{jt}$ is the (individual, stochastic) price of an hour of market labor and $W_{t}=\{w_{1t},w_{2t}\}$ is the set of wages in the couple. We assume the spouses hold identical beliefs about future stochastic elements.

Both $\tau$ and individual utility $u_j$ depend on demographics, such as fertility. We suppress those here for brevity (we thus fix a household type), but we do account for them in the empirical application. Appendix \ref{Appendix::RecursiveFormulation} shows the role of demographics in the model explicitly.

\subsection{Commitment modes}

The next three subsections present the collective model in the three alternative commitment regimes. For now, we disregard divorce and thus model behavior conditional on continued marriage. Divorce is discussed in appendix \ref{Appendix::RecursiveFormulation}.

\subsubsection{Full commitment}

Upon marriage, the individuals commit fully to all future but state-contingent allocations of resources between them. In other words, they commit at $t=0$ to a plan that disciplines their actions in the future. Choices made under full commitment are therefore ex ante efficient and can be represented by the solution to the following problem at $t=0$:
\begin{align}\label{Eq::Definition.FC}
    &V_{0}^\text{FC}(\Omega_{0}) = 
    \max_{\{C_{t}\}_{0\leq t \leq \bar{t}}} 
    \mu_{1} (\Theta_{0}) \left(\mathbb{E}_{0} \sum_{t=0}^{\bar{t}} \beta^{t} u_{1}(q_{t}, h_{1t})\right) + 
    \mu_{2} (\Theta_{0}) \left(\mathbb{E}_{0} \sum_{t=0}^{\bar{t}} \beta^{t} u_{2}(q_{t}, h_{2t})\right)\\
    \notag
    &\text{subject to the budget constraint \eqref{Eq::Budget_Constraint} $\forall t$,}
\end{align}
where $C_{t} = \{q_{t}, h_{1t}, h_{2t}, a_{t+1}\}$ is the set of household choice variables in period $t$. For brevity, we do not show their dependence on contingent states -- we relegate this to appendix \ref{Appendix::RecursiveFormulation}.\footnote{We maintain a common discount factor between spouses; see \citet{AdamsEtAl2014ConsumeNowLater} for a generalization.}

The Pareto weights $\mu_{1}$ and $\mu_{2}$ are the utility weights the household places on each person's preferences at marriage. They determine ex ante the relative allocation of resources between spouses. $\Theta_0$ is the set of variables that affect the Pareto weights; because the weights are determined at marriage, it follows that $\Theta_0$ must only include information known or predicted at the time the household is formed, that is, at time $t=0$.\footnote{The discussion that follows only requires that some variables known at $t=0$ affect the Pareto weights at marriage; it does not require to specify why exactly this happens. One may think that the weights arise from a bargaining game played at marriage or from equilibrium conditions in the marriage market \citep[e.g.][]{ChiapporiostaDiasMeghir2018MarriageMarket}. In either case, information at $t=0$ (e.g. variables summarizing the marriage market) affects the Pareto weights; we thus allow $\mu_{1}$ and $\mu_{2}$ to depend on $\Theta_0$ to reflect, in reduced form, such a link.} 

The expectations, common between spouses as we assume throughout, are taken over the stochastic elements of future states in $\Omega_{t}$, $t>0$, such as future wages. Generally, the state space differs across commitment alternatives. We have purposefully not defined what explicitly goes into $\Omega$ but we will return to this point in the following sections.

\subsubsection{Limited commitment}

When an individual can unilaterally walk away from their partner (e.g. unilaterally divorce), not all future allocations are feasible in all states of the world. Certain plans may make one person better off outside the household. Assuming there is always a positive marital surplus to be shared, the household must make sure to not implement those plans.

Upon marriage, the spouses commit to future and state-contingent allocations of resources up to the point that one's marital participation constraint is violated. Choices under limited commitment are ex ante second-best efficient (ex ante efficient subject to participation constraints) and can be represented by the solution to the following problem at $t=0$: 
\begin{align}\label{Eq::Definition.LC}
    &V_{0}^\text{LC}(\Omega_{0}) = 
    \max_{\{C_{t}\}_{0\leq t \leq \bar{t}}} 
    \mu_{1} (\Theta_{0}) \left(\mathbb{E}_{0} \sum_{t=0}^{\bar{t}} \beta^{t} u_{1}(q_{t}, h_{1t})\right) + 
    \mu_{2} (\Theta_{0}) \left(\mathbb{E}_{0} \sum_{t=0}^{\bar{t}} \beta^{t} u_{2}(q_{t}, h_{2t})\right)\\
    \notag
    &\text{subject to the budget constraint \eqref{Eq::Budget_Constraint} $\forall t$, and the participation constraints:}
\end{align}
\vspace{-8ex}
\begin{alignat*}{3}
    \nu_{jt}~:~ 
    \underbrace{\mathbb{E}_{t} \sum_{\tau=t}^{\bar{t}} \beta^{\tau-t} u_{j}(q_{\tau}, h_{j\tau})}_{\mathclap{\substack{\text{individual inside value}\\\text{from marriage at $t$}}}}
    &~~\geq~~ 
    &&\underbrace{\widetilde{V}_{jt}(\Omega_{jt})}_{\mathclap{\substack{\text{individual outside value}\\\text{at $t$ (e.g. single/divorced)}}}} 
    \qquad &&\forall t>0,j\in \{1,2\},
\end{alignat*}
where $C_{t}=\{q_{t},h_{1t},h_{2t},a_{t+1}\}$ is the set of household choice variables in period $t$ and $\widetilde{V}_{jt}(\Omega_{jt})$ is the reservation utility of person $j$ at $t$, defined over  $\Omega_{jt}\subseteq \Omega_{t}$ -- more on this below. 

The participation constraints, one per individual and time period, ensure that each person enjoys at least as much value inside their joint household as they can possibly get from their outside option, i.e. by walking away from the relationship. In other words, the participation constraints ensure individual rationality in the relationship.
The constraints consist of two parts, the \emph{inside} (left hand side) and \emph{outside} (right hand side) values, defined on the basis of the forward-looking continuation value of each option.

The \emph{inside} value of person $j$ reflects the share of marital surplus that accrues to him/her given the choices made by the household in the period. Naturally, this value varies with the applicable Pareto weight in the period: a larger Pareto weight on person $j$ implies household choices more tailored to $j$'s tastes, thus accruing a larger share of marital surplus to him/her, and vice versa. This link between the Pareto weight and the individual inside value from marriage disciplines the dynamics of the Pareto weight in limited commitment, a point to which we return in the next sections.

The \emph{outside} value of person $j$ reflects how he/she may fare in life outside the current relationship; its precise form depends on specific assumptions about that situation. For instance, we may, as in \citet{Voena2015}, assume that the outside option is divorce, and that the corresponding value is the present value of future expected utility of a single/divorced person -- in which case we have:
\vspace{-2ex}
\begin{align*}
    &\widetilde{V}_{jt}(\Omega_{jt}) = 
    \max_{\{q_{j\tau},h_{j\tau},a_{j\tau+1}\}_{\tau=t,\dots,\bar{t}}}
    \mathbb{E}_{t} \sum_{\tau=t}^{\bar{t}} \beta^{\tau-t} \widetilde{u}_{j}(q_{j\tau}, h_{j\tau})\\
    &\text{subject to }~(1+r)a_{j\tau} + \tau(w_{j\tau}h_{j\tau}) = q_{j\tau} + a_{j\tau+1}, ~\forall \tau, \text{ and } a_{1t}+a_{2t}=a_{t}.
\end{align*}
Here, we let preferences depend on marital status (i.e. $\tilde{u}_{j}$ may differ from $u_{j}$) to capture marital preference shifts. $\widetilde{u}_{j}$ may include stochastic elements that reflect, in reduced form, the utility flow from possible future remarriage or grieving following the breakup. 

Other interpretations are however possible and the precise form of $\widetilde{V}_{jt}$ does not matter for the subsequent discussion. The key aspect is that $\widetilde{V}_{jt}$ is defined over the (single's) state space $\Omega_{jt}\subseteq \Omega_{t}$. Three distinct sets of variables enter $\Omega_{jt}$, whose role we describe below: the individual wage rate $w_{jt}$ (wages matter for $j$'s budget as single and for his/her labor market and possibly remarriage prospects), distribution factors $Z_{jt}$, and marital assets $a_{t}$. The expectations are taken over the stochastic elements of future states in $\Omega _{j\tau }$, $\tau>t$.

Distribution factors $Z_{jt}$ are exogenous stochastic variables that affect the singles' lifecycle prospects but not preferences or the couple's budget set conditional on household income \citep[][]{Bourguignon2009}.\footnote{Examples include relative non-labor incomes (e.g. \citealp{Thomas1990, Attanasio2014}), the sex ratio in the marriage market \citep[][]{Chiappori2002}, or divorce and property division laws \citep{Voena2015}.} These variables thus affect the outside but not the inside options in the household. The difference between $Z_{jt}$ and $\Theta_{0}$ is that the former variables vary during the course of the relationship while the latter do not. As the variables in $Z_{jt}$ vary stochastically over time, the single's outside value varies in response, which may make a participation constraint occasionally bind. To satisfy the constraint, the inside value of the constrained spouse must adjust, thus creating a link from the time-varying distribution factors to the choices made in the couple as we illustrate subsequently.

Upon household break-up, financial wealth accumulated during marriage is split between partners according to some fixed rule. Marital assets thus determine the wealth that a newly single possesses upon break-up, so wealth $a_{t}$ enters $j$'s outside value at $t$. This renders the outside options endogenous to choices made during marriage. A couple in our setting makes savings choices accounting for the implications of those choices for the outside options, in addition to the standard lifecycle/precautionary motives present also in full commitment.\footnote{The dependence of the outside options on wealth may move the household away from second-best efficiency; for example, the couple may overinvest in financial assets to improve their outside options. \citet{Chiappori2017} have a lengthy discussion of this as well as of an interesting opposite case.}

\subsubsection{No commitment}

While full and limited commitment feature some form of commitment at marriage to a future plan (the plan contingent on the Pareto weights at marriage and, in the case of limited commitment, the participation constraints), no commitment features no such marital `contract'. Upon marriage, the spouses do not commit to a future plan, that is, they do not guarantee each other a certain or minimum allocation of resources.

Without commitment, new information that arises over time changes the
division of marital surplus between spouses according to the bargaining game
they play. Choices under no commitment can be represented by the solution to
the following problem at $t=0$: 
\begin{align}\label{Eq::Definition.NC}
    &V_{0}^\text{NC}(\Omega_{0}) = \\
    \notag
    &\max_{\{C_{t}\}_{0\leq t \leq \bar{t}}} 
    \left(\mathbb{E}_{0} \sum_{t=0}^{\bar{t}} \beta^{t} \mu_{1} (\Theta_{0},W_{t},Z_{t},a_{t}) u_{1}(q_{t}, h_{1t})\right) + 
    \left(\mathbb{E}_{0} \sum_{t=0}^{\bar{t}} \beta^{t} \mu_{2} (\Theta_{0},W_{t},Z_{t},a_{t}) u_{2}(q_{t}, h_{2t})\right)\\
    \notag
    &\text{subject to the budget constraint \eqref{Eq::Budget_Constraint} $\forall t$,}
\end{align}
where $C_{t}=\{q_{t},h_{1t},h_{2t},a_{t+1}\}$ is the set of household choice variables in period $t$.

The underlying premise of no commitment is that the spouses engage in some form of repeated bargaining over the marital surplus in a way that reflects the prevailing economic environment, captured by wages $W_{t}$, distribution factors $Z_{t}=\{Z_{1t},Z_{2t}\}$, and wealth $a_{t}$. As new information arises over time, a person's bargaining position shifts given the bargaining game played by the spouses. The variables in $\Theta_{0}$ are fixed after marriage but they enter the Pareto weights because they may influence the type of game the couple plays.

Choices under no commitment are ex ante inefficient. This is because ex ante efficiency, at least in the first-best sense, implies that there exist no time-contingent transfers that improve both spouses' expected utilities \citep{Browning2014}. This requires the Pareto weights be the same over time, which is clearly not the case here. Choices are dynamically inefficient also because the bargaining weights depend on wealth. The spouse whose future weight increases more with wealth has an incentive to overinvest in assets, thus creating inefficiencies over time. Whether such inefficiency appeals to couples is ultimately an empirical question that our test for commitment helps address. Nevertheless, choices in a given period are ex post efficient in the sense that they maximize a weighted sum of individual period utilities.

Our representation of no commitment (in particular \emph{how} new information impacts the Pareto weights) is arguably abstract and lacks the precise microfoundations of full or limited commitment. Nevertheless, this abstractness enables it to be consistent with several popular underlying structures, such as resource allocations with Nash bargaining over the marriage market \citep{GousseEtAl2017Marriage}, household sharing when labor market shocks shift the balance of power \citep[][]{LiseYamada2019}, equilibrium allocations with bargaining over a default arrangement \citep{KatoRiosRull2022MarkovEquilibria}, and other. We return to this point below.

\subsection{Common recursive formulation}

The household problem in each commitment mode is a dynamic planning problem over the allocation of resources between spouses and across periods of time. Following \citet{Marcet2019}, we can recast each problem in the common recursive form: 
\begin{align}\label{Eq::Definition.RecursiveForm}
    &V_{t} (\Omega_{t}) = 
    \max_{C_{t}}~
    \mu_{1t} u_{1}(q_{t}, h_{1t}) + \mu_{2t} u_{2}(q_{t}, h_{2t}) +
    g_{t} (a_{t}) +
    \beta \mathbb{E}_{t} V_{t+1}(\Omega_{t+1})\\
    \notag
    &\text{subject to the budget constraint \eqref{Eq::Budget_Constraint}, and}\\
    \notag
    &\text{restrictions on the Pareto weights $\mu_{jt}$, $j\in\{1,2\}$, defined subsequently,}
\end{align}
with details reported in appendix \ref{Appendix::RecursiveFormulation}. In each period, the household maximizes a weighted sum of period utilities, an appropriate continuation value, and an additional term described below. Expectations are over the stochastic elements in $\Omega_{t+1}$, given the realization of $\Omega_{t}$.

The additional term, $g_{t}(a_{t})$, aggregates the singles' endogenous outside options in limited commitment. It is given by $g_{t}(a_{t}) = -\nu_{1t}\widetilde{V}_{1t}(w_{1t},Z_{1t},a_{t}) - \nu_{2t}\widetilde{V}_{2t}(w_{2t},Z_{2t},a_{t})$ in limited commitment, where $\nu_{jt}$ is the Lagrange multiplier on spouse $j$'s participation constraint at $t$, and by $g_{t}(a_{t})=0$ otherwise. $g_{t}(a_{t})$ highlights that a couple makes savings choices in limited commitment taking into account the effect of those choices on the partners' outside options. \citet{Chiappori2017} provide a further discussion of this motive.

The solution to \eqref{Eq::Definition.RecursiveForm} is a set of time-consistent\footnote{The policy functions depend on time because the horizon is finite.} optimal policy functions $q_{t}^{*}(\Omega_{t})$, $h_{jt}^{*}(\Omega_{t})$, $\forall j$, and $a_{t+1}^{*}(\Omega_{t})$, which depend on the state space $\Omega_{t}$ in each commitment alternative. Our test for commitment relies on estimating equations derived directly from \eqref{Eq::Definition.RecursiveForm} and its corresponding policy functions, given restrictions that the Pareto weight imposes on the state space in each case. The crucial point in \eqref{Eq::Definition.RecursiveForm} is that there is a pair of applicable Pareto weights $\mu_{1t}$ and $\mu_{2t}$ in each period, the dynamics of which we will now characterize.

\subsection{Characterization of the Pareto weight and the state space}

Bargaining power is relative inside the household since the sum $\mu_{1t} + \mu_{2t}$ can be normalized to a constant. Therefore, we subsequently refer to the Pareto weight in singular. Moreover, any variable that affects one person's Pareto weight must simultaneously and mechanically enter and affect the partner's weight in the opposite direction.\footnote{One can show this formally by explicitly including the restriction $\mu_{1t} + \mu_{2t} = \text{\emph{constant}}$ in program \eqref{Eq::Definition.RecursiveForm}.}

\textbf{Full commitment.} The Pareto weight is determined at marriage as a function of the initial bargaining variables $\Theta_{0}$. The weight remains constant over time, namely 
\begin{equation*}
    \mu_{jt} = \mu_{j}(\Theta_{0}), \qquad j\in\{1,2\}, \forall t, 
\end{equation*}
as we show in appendix \ref{Appendix::RecursiveFormulation}. The variables in $\Theta_{0}$ vary in the cross-section reflecting the couple's characteristics at marriage, local
marriage market conditions, or other heterogeneity that the individuals base their initial bargaining on. So $\mu_{j}(\Theta_{0})$ also varies in the cross-section of households. $\Theta_{0}$ includes at least one variable $\theta_{0}\in \Theta_{0}$ that improves $j$'s initial bargaining power ($\partial \mu _{j}/\partial \theta _{0}>0$) and, consequently, worsens that of the partner ($\partial \mu_{-j}/\partial \theta_{0}<0$). These bargaining variables remain fixed for $t>0$ after marriage, so $\mu_{j}(\Theta_{0})$ also remains fixed within a given family over time. The initial weight at marriage thus serves as the spouses' intra-household bargaining power over their entire lifecycle.

The time and state invariance of the Pareto weight (conditional on $\Theta_{0}$) is a well-known implication of first-best efficiency \citep{Browning2014}. Intuitively, once the initial bargaining weight is set at marriage, future shocks (for example, shocks to wages) do not change the allocation of marital surplus between spouses, who fully share any
idiosyncratic risk between them. Of course efficiency requires that the spouses exploit the economic opportunities that arise from variation in wages or other shocks; but with full commitment, those effects remain compatible with ex ante efficiency and the Pareto weight does not change in response. This implies that policies that seek to empower, say, women, e.g.
cash transfers targeted to women, cannot affect the division of marital surplus if they are implemented after marriage. The only policies that matter under full commitment are those that affect $\Theta_0$.

\textbf{Limited commitment.} The Pareto weight is given by:
\begin{equation*}
\begin{array}{lll}
            &\mu_{jt} = \mu_{jt-1} + \nu_{jt},    &\qquad j\in\{1,2\},\forall t>0,\\ 
    \text{with}
            &\mu_{j0} = \mu_{j}(\Theta_0),        &\qquad j\in\{1,2\},
\end{array}
\end{equation*}
where $\nu_{jt}$ is the Lagrange multiplier on spouse $j$'s participation constraint at $t$. We derive this in appendix \ref{Appendix::RecursiveFormulation}. The Pareto weight shifts when the continuation of the previous allocation of resources, summarized by the past weight $\mu_{jt-1}$, violates one's participation constraint. In such case, the constrained spouse's weight jumps by $\nu_{jt}>0$, i.e. the multiplier on the binding constraint. If no participation constraint binds, then $\nu_{jt}=0$, and the Pareto weight remains unchanged. Whether a participation constraint binds depends on the variables underlying the constraint, so we may write $\nu_{jt}\equiv \nu_{j}(W_{t},Z_{t},a_{t},\mu_{jt-1})$ as we explain below.

Consider the distribution factors. As $Z_{t}=\{Z_{1t},Z_{2t}\}$ shift the outside options, a person's participation constraint may bind for some realization of $Z_{t}$. To relax the constraint, the constrained person's bargaining power increases by $\nu_{jt}$, which shifts household decisions towards her preferences and improves her inside value. The increase in power is the smallest possible that makes $j$ indifferent between staying in the relationship and leaving. This follows from second-best efficiency as analyzed in \citet{Ligon2002}.\footnote{The present analysis is conditional on the spouses remaining married (marital surplus remains positive), so at most one participation constraint can bind in a given period \citep{Kocherlakota1996}. Therefore, following the increase in $j$'s power, there exists a feasible allocation at which her partner's constraint is also satisfied.} 

Wages also impact the participation constraints, though their workings are more nuanced. While $Z_{jt}\subseteq Z_{t}$ typically includes at least one variable $z_{jt}\in Z_{jt}$ that improves $j$'s outside option, increases her bargaining power ($\partial \mu_{jt}/\partial z_{jt}>0$) and symmetrically worsens the partner's ($\partial \mu_{-jt}/\partial z_{jt}<0$), wages simultaneously affect $j$'s outside \emph{and} inside values. An increase in $w_{jt}\in W_{t}$, however, should improve $j$'s outside value more, because any value from wages inside the relationship must be shared with her partner. As her participation constraint may thus bind, we expect $\partial \mu_{jt}/\partial w_{jt}>0$ and, in turn, $\partial \mu_{-jt}/\partial w_{jt}<0$. The wage $w_{jt}$ also affects the \emph{partner's} inside value through sharing; a wage rise loosens the partner's constraint while a wage cut tightens it (both income effects), so $\partial \mu_{-jt}/\partial w_{jt}<0$ like above.

Wealth $a_{t}$ also enters the participation constraints; but it does not clearly favor one party unless policies or explicit agreements dictate this (e.g. prenuptial contracts).

The extent to which a participation constraint binds in response to $Z_t$ or $W_{t}$ depends on the person's inside value, which, as we established earlier, varies with the applicable Pareto weight. Suppose $\mu_{jt-1}$ is the Pareto weight at the \emph{start} of period $t$, after shocks manifest but before decisions are made in the period. A relatively larger $\mu_{jt-1}$ implies a relatively larger share of marital surplus for $j$, thus making her outside option less desirable for a given realization of $Z_t$ or $W_{t}$; and vice versa. So whether $j$'s constraint binds at the \emph{start} of period $t$ depends on $\mu_{jt-1}$. Consequently, $\nu_{jt}$ and the updating of the Pareto weight upon decision making later on in the period also depend on it.

Where does $\mu_{jt-1}$ come from? The nature of decision making is such that the participation constraints are always satisfied at the updated $\mu_{jt}$ at the \emph{end} of the period. No further updating takes place before decision making in the \emph{following} period, therefore $\mu_{jt}$ at the \emph{end} of $t$ is also the applicable weight at the \emph{start} of $t+1$. By deduction, $\mu_{jt-1}$ is therefore the weight that materialized at the \emph{end} of $t-1$. By the time of decision making in period $t$, $\mu_{jt-1}$ summarizes the history of the household through binding past constraints and renegotiations, which is an artifact of the recursive nature of the participation constraints. Intuitively, a given shock that improves $j$'s outside option will not trigger a renegotiation if $j$ has been `happy' inside her relationship, that is, if she has historically earned a `good' share of the marital pie. So certain shocks trigger renegotiation in certain histories but not in other. History is summarized by a single variable, $\mu_{jt-1}$, and conditional on it, older information does not matter for decision making today \citep{Kocherlakota1996}.

This step-like movement of the Pareto weight in response to binding participation constraints is a well-known feature of limited commitment \citep[][]{Mazzocco2007}. The couple commits to the resource allocation enacted at marriage for as long as the participation constraints remain slack. Therefore, a given allocation and Pareto weight can be quite
persistent. The spouses thus fully share risk up to the point when a renegotiation takes place to satisfy a constrained spouse. After the renegotiation, the new weight may itself persist until another constraint binds. The Pareto weight thus varies cross-sectionally but also longitudinally, \emph{within} a given family. Policies that seek to empower women, e.g. targeted cash transfers, can thus be implemented during the relationship and may have lasting effects on the balance of power if they successfully improve women's outside options. Persistence and history are features of limited commitment that we subsequently exploit for testing.

\textbf{No commitment.} The Pareto weight is determined in each period given the prevailing information in the period; by construction, it is given by 
\begin{equation*}
    \mu_{jt} = \mu_{j}(\Theta_0,W_{t},Z_{t},a_{t}), \qquad j\in\{1,2\}, \forall t.
\end{equation*} 
The Pareto weight varies with the initial bargaining variables $\Theta_0$, which influence the type of game the spouses play. It also varies with new information that reveals over time, i.e. shocks to wages and distribution factors (the bargaining effects of which are typically assignable), and assets. For example, an increase in $w_{jt}\in W_{t}$ should empower spouse $j$ ($\partial \mu_{jt}/\partial w_{jt}>0$) and simultaneously weaken her partner ($\partial \mu_{-jt}/\partial w_{jt}<0$). How precisely this is done depends on the exact bargaining game on which no assumption is made beyond cooperation.

The continuous response of the Pareto weight to contemporaneous information is a well-known feature of no commitment and cooperative bargaining \citep[e.g.][]{LiseYamada2019}. The main implication is that the spouses cannot share risk efficiently as they cannot commit to transfer resources from one period to another. This lack of history, at least conditional on assets, makes the Pareto weight transitory in nature. It implies that policies that seek to empower women may be implemented during the relationship (e.g. by improving elements in $W_{t}$ or $Z_{t}$ that are assignable to women) but their effect is temporary. As soon as the policy disappears, any gains in the Pareto weight also disappear. Lack of history stems from cooperative bargaining and the finite horizon of our setting, as we illustrated in section \ref{Section::Illustration_Simple_Setting}. This absence of history is a feature of no commitment that we exploit for testing.

\textbf{State space.} The policy functions derived from \eqref{Eq::Definition.RecursiveForm} vary with wages $W_{t}$ and assets $a_{t}$, which affect the budget set in all commitment modes and, through it, the trade-off between consumption, savings, and work. The policy functions also vary with the applicable Pareto weight $\mu_{jt}$. In all alternatives, a relatively larger $\mu_{jt}$ implies period $t$ choices that favor spouse $j$, and vice versa. Consequently, the state space is given by $\Omega_{t} =\{W_{t},a_{t},\mu_{jt}\}$, common across modes.\footnote{In practice, only one person's weight enters the state space because of the restriction $\mu_{1t}+\mu_{2t}=\text{\emph{constant}}$.} However, each commitment alternative imposes different restrictions on $\mu_{jt}$. For example, $\mu_{jt}$ in full commitment is exclusively determined by the initial bargaining variables $\Theta_{0}$. We may thus replace $\mu_{jt}$ in the state space with $\Theta_{0}$, so $\Omega_{t}=\{W_{t},a_{t},\Theta_{0}\}$ in this case. By a similar argument, the limited commitment state space is given by $\Omega_{t}=\{W_{t},a_{t},Z_{t},\mu_{jt-1}\}$ while the no commitment state space by $\Omega_{t}=\{W_{t},a_{t},\Theta_{0},Z_{t}\}$. We show subsequently that these sets are nested, which enables us to test for the type of commitment.

\subsection{Nesting and restrictions on household behavior}\label{SubSection::Household_Lifecycle_Behavior_Nesting}

To understand how the different commitment modes are related, it is useful to pool together the corresponding Pareto weights, namely 
\begin{equation*}
\begin{array}{ll}
\text{full commitment:}     & \mu_{jt} = \mu_{j}(\Theta_{0})\\
\text{limited commitment:}  & \mu_{jt} = \mu_{j}(W_{t},Z_{t},a_{t},\mu_{jt-1}) \\
\text{no commitment:}       & \mu_{jt} = \mu_{j}(\Theta_0,W_{t},Z_{t},a_{t}),
\end{array}
\end{equation*}
where $\mu_{jt} = \mu_{jt-1} + \nu_{j}(W_{t},Z_{t},a_{t},\mu_{jt-1}) \equiv \mu_{j}(W_{t},Z_{t},a_{t},\mu_{jt-1})$ in limited commitment.

The limited commitment Pareto weight depends on its past value, which summarizes the history of the household from marriage until today. If observed and accounted for, $\mu_{jt-1}$ is a sufficient statistic for the past \citep[][]{Kocherlakota1996}. In practice, however, $\mu_{jt-1}$ is unobserved. From its law of motion, we can substitute the past weight recursively until $t=0$ (marriage) to obtain $\mu_{jt} = \mu_{j}(W_{t},Z_{t},a_{t},~\mu_{j}(W_{t-1},Z_{t-1},a_{t-1},~\mu_{j}(W_{t-2},Z_{t-2},a_{t-2},\dots,~\mu_{j}(\Theta_0))))$. Unconditional on its past value, $\mu_{jt}$ thus depends on the entire information set since marriage, which includes all historical wages $W_{t-\tau}$, distribution factors $Z_{t-\tau}$, and assets $a_{t-\tau}$, $\tau \in \{1,\dots,t-1\}$, that affected past participation constraints and, through them, the historical dynamics of bargaining power in the household. In other words, any variable that affects the unaccounted $\mu_{jt-1}$ must also affect the Pareto weight today. Consolidating the variables that enter the limited commitment weight and reordering the modes, we obtain 
\begin{equation*}
\begin{array}{ll}
\text{full commitment:}     & \mu_{jt} = \mu_{j}(\Theta_0)\\
\text{no commitment:}       & \mu_{jt} = \mu_{j}(\Theta_0,W_{t},Z_{t},a_{t})\\
\text{limited commitment:}  & \mu_{jt} = \mu_{j}(\Theta_0,W_{t},Z_{t},a_{t},\underbrace{W_{t-1},Z_{t-1},a_{t-1},\underbrace{W_{t-2},Z_{t-2},a_{t-2},\dots}_{\text{enters through $\mu_{jt-2}$}}}_{\text{enters through $\mu_{jt-1}$}}),
\end{array}
\end{equation*}
which reflects the nesting of the \emph{sets of variables} that matter for bargaining in each case.\footnote{Let $I_{t}=\{W_{t},Z_{t},a_{t}\}$. To be precise, we should write $\mu_{jt} = \widetilde{\mu}_{j}(\Theta_0,I_{t},I_{t-1},I_{t-2},\dots)$ in limited commitment, where $\widetilde{\mu}_{j}$ is the reduced form of $\mu_{jt} =
\mu_{j}(I_{t},\mu_{j}(I_{t-1},\mu_{j}(I_{t-2},\dots,~\mu_{j}(\Theta_0))))$. To avoid unnecessary notation (namely the tilde), we reinstate $\mu_{j}$ as the reduced form of its structural counterpart.}

\emph{Contemporaneous} information (e.g. information in $W_{t}$ or $Z_{t}$) does not matter for the Pareto weight in full commitment but it does matter in no and limited commitment. Full commitment is thus nested (in terms of the variables that matter for bargaining) within both non-full commitment alternatives, which is a well-known result since \citet{Mazzocco2007}. This implies that if contemporary variables affect the Pareto weight, this effect serves as evidence against full commitment. In other words, current information is a natural exclusion restriction that can help separate full from non-full commitment.

\emph{Historical} information (e.g. information in $W_{t-\tau}$ or $Z_{t-\tau}$, $\tau\geq1$) does not matter for the Pareto weight in no (or full) commitment but it does matter in limited commitment. No commitment is thus nested within limited commitment, which is a new result in the literature. This implies that if historical variables affect the contemporaneous Pareto weight, this effect serves as evidence against no commitment. In other words, history is a natural exclusion restriction that can help separate no from limited commitment.

The problem with this approach is that the Pareto weight is unobserved. From the theory of dynamic programming, however, the optimal labor supply policies that solve \eqref{Eq::Definition.RecursiveForm} are functions of the state space $\Omega_{t}=\{W_{t},a_{t},\mu_{jt}\}$. Therefore, the previous exclusion restrictions on the Pareto weight immediately become exclusion restrictions on the (typically observed) individual labor supplies $h_{1t}$ and $h_{2t}$ in the household, given by 
\begin{equation*}
\begin{array}{ll}
\text{full commitment:}     & h_{jt} = h_{jt}^{*}(W_{t},a_{t},\mu_{j}(\Theta_0))\\
\text{no commitment:}       & h_{jt} = h_{jt}^{*}(W_{t},a_{t},\mu_{j}(\Theta_0,W_{t},Z_{t},a_{t}))\\
\text{limited commitment:}  & h_{jt} = h_{jt}^{*}(\underbrace{W_{t},a_{t}}_{\mathclap{\text{non-bargaining state space}}},\mu_{j}(\underbrace{\Theta_0,W_{t},Z_{t},a_{t},W_{t-1},Z_{t-1},a_{t-1},\dots}_{\mathclap{\text{bargaining state space}}})),
\end{array}
\end{equation*}
for $j\in\{1,2\}$. This holds for all choices in the household; however, the assignability of labor supply allows us to exploit the properties of the so-called \emph{sharing rule}.

The sharing rule $(\rho_{1t},\rho_{2t})$ summarizes the share of total income each spouse can devote to own expenditure in a given period. If leisure is a normal good and utility not quasi-linear (as we assume throughout), an increase in $\rho_{jt}$ increases $j$'s leisure and decreases her supply of labor. Moreover, there is an one-to-one increasing relationship between $j$'s Pareto weight $\mu_{jt}$ and share $\rho_{jt}$, therefore $\partial h_{jt} /\partial \mu_{jt}<0$. This also implies $\partial h_{-jt}/\partial \mu_{jt}>0$ as bargaining power is relative. In words, an improvement in $j$'s bargaining position increases her leisure and reduces her labor supply; in parallel, her partner's bargaining position deteriorates, which reduces his leisure and increases his labor supply. This allows us to characterize how the various variables that enter the Pareto weight affect household labor supply.

Suppose we observe at least one initial bargaining variable $\theta_{0}\in\Theta_0$ and at least one time-varying distribution factor $z_{t}\in Z_{t}$; suppose that theory or intuition suggest that both $\theta_{0}$ and $z_{t}$ empower $j$. Under \emph{limited} commitment, we must jointly observe 
\begin{equation*}
\begin{array}{rllll}
    (\text{I}):     &\partial h_{jt}/\partial \theta_{0} < 0    &\text{ and } &\partial h_{-jt}/\partial \theta_{0} > 0    &\\
    (\text{II}):    &\partial h_{jt}/\partial z_{t} < 0         &\text{ and } &\partial h_{-jt}/\partial z_{t} > 0         &\\
    (\text{III}):   &\partial h_{jt}/\partial z_{t-\tau} < 0    &\text{ and } &\partial h_{-jt}/\partial z_{t-\tau} > 0    &\text{ for }\tau\geq1,
\end{array}
\end{equation*}
because, as $\theta_{0}$ and $z_{t}$ empower $j$, they must decrease her labor supply and increase her partner's. This must be true also for past distribution factors in (III) because, as changes in bargaining power are persistent in limited commitment, any variable that affected bargaining power in the past will have a lasting effect on behavior in the future. Moreover, the recursive nature of the weight (and the fact that more recent shifts in distribution factors may undo previous shifts) implies that the effects of historical variables should diminish in magnitude the further back in time we go. We return to this point subsequently.

Full commitment implies that effects (II) and (III) are absent, while no commitment implies that (III) is absent. This is a consequence of the type of information that matters for the Pareto weight in each case. Effect (I) can only be observed cross-sectionally while effects (II) and (III), if present, can be observed cross-sectionally \emph{and} longitudinally. Finally, effect (II) of current information is present in both no and limited commitment. Therefore, testing for current information alone, as in \citet{Mazzocco2007} and \citet{LiseYamada2019}, does not inform whether limited \emph{or} lack of commitment is the right framework through which household behavior should be analyzed.

Two final remarks are in order. First, additional distribution factors increase the number of restrictions on household labor supply. Second, wages $W_{t}$ are additional time-varying variables that enter the Pareto weight outside of full commitment. As is evident in the non-bargaining state variables in $h_{jt}^*$, of which $W_{t}$ is part, wages are not conventional distribution factors because they also affect the budget set. However, \emph{past} wages do not enter the state space outside of bargaining in limited commitment, so they satisfy the exclusion restriction of history. As they are also assignable, we should observe effect (III) in limited commitment also through past wages. Moreover, we will show that, conditional on household income, the \emph{partner}'s \emph{current} wage $w_{-jt}$ does not enter the state space outside of bargaining in no/limited commitment, so it satisfies the exclusion restriction of contemporaneous information. The \emph{partner}'s \emph{current} wage thus serves as an additional distribution factor, inducing analogous effects to (II) in no and limited commitment. This role of wages is appealing because wages are readily available in household data while conventional distribution factors are harder to find. With these points in mind, we now turn to our test for commitment.

\section{Test for commitment}\label{Section::Test_Commitment}

Limited commitment describes an environment in which current and historical distribution factors affect behavior. No commitment is a special case as history does not matter whereas current distribution factors do, while full commitment is a special case of no commitment in that current distribution factors do not matter. One may thus test for commitment by testing whether current and historical values of variables that enter the Pareto weight affect household behavior -- in this case labor supply -- and with what sign. 

There are two main ways to estimate the optimal policy functions $h_{jt}^*$, which are the objects over which we implement our test. The first approach involves the full or partial specification of preferences, expectations, and bargaining. The alternative approach leaves preferences and expectations unspecified while it only specifies the reduced form dependence of the Pareto weight on its arguments. This is inspired by \citet{BlundellPistaSaporta2016Family_Labour} who study how wage shocks transmit into labor supply and consumption in a unitary context.\footnote{\citet{Theloudis2017_PhDThesis} extends this approach to a collective setting.}

While the first approach enables the recovery of the specification and the assessment of counterfactuals, its main drawback is that it requires the estimation of preferences together with bargaining. Therefore, any test for commitment is ultimately a \emph{joint} test of commitment \emph{and} the specification used for preferences. The second approach avoids this as it does not require the specification of preferences -- consequently, it is unable to recover deep parameters or evaluate counterfactuals. As our goal here is to \emph{test} for commitment rather than recover preferences or bargaining primitives, it is natural to follow the second path. This entails the derivation of estimable labor supply equations from the model's first order conditions and the reduced form specification of the Pareto weight, both of which we describe below.

\subsection{Dynamics of household labor supply}

We use the general form of the household problem in \eqref{Eq::Definition.RecursiveForm} to derive the static optimality conditions for male and female hours. These conditions depend on the tax/benefits function $\tau (y_{t})$ through the budget constraint. It is hard to make progress without restricting $\tau$, so we follow \citet{HeathcoteStoresViolante2014WithPartialInsurance} and \citet{BlundellPistaSaporta2016Family_Labour} and approximate $\tau$ as $y_{t}^{D} \equiv \tau(y_{t}) \approx (1-\chi_{t})y_{t}^{1-\kappa_{t}}$. Tax/benefits parameters $\chi_{t}$ and $\kappa_{t}$ reflect the proportionality and progressivity of the tax and benefits system. A progressive system has a strictly positive progressivity parameter $\kappa_{t}$ while a proportional tax system has $\kappa_{t}=0$. In that case, the spouses are taxed separately at the proportionality rate $\chi_{t}$.

Except a few special cases of utility, the optimality conditions are \emph{implicit} functions of hours and cannot be directly estimated in the data. We follow \citet{BlundellPistaSaporta2016Family_Labour} and carry out a standard log-linearization of $-u_{j[h]}$, the marginal utility of hours, around the most recent values of consumption and hours.\footnote{Unlike \citet{BlundellPistaSaporta2016Family_Labour}, we do not log-linearize the intertemporal budget constraint.} We show in appendix \ref{Appendix::DerivationOptimalityConditions} that this operation yields a closed-form expression for the growth rate of male and female hours in terms of changes in the Pareto weight and other variables, given by
\begin{align}\label{Eq::EstimableEquation.BeforeReplacingMu}
\begin{split}
\Delta \log h_{jt}
    &= \underbrace{\delta_{jt} h_{jt-1}^{-1} \Delta \log (1-\chi_{t})}_{\mathclap{\text{1: tax effects}}}
    - \underbrace{\delta_{jt} \kappa_{t} s_{-jt-1}h_{jt-1}^{-1}\Delta \log y_{-jt}}_{\mathclap{\substack{\text{2: tax disincentives}\\\text{from partner earnings}}}}
    + \underbrace{\delta_{jt} h_{jt-1}^{-1} \Delta \log \lambda_{t}}_{\mathclap{\substack{\text{3: wealth and}\\\text{income effects}}}}\\[5pt]
    &- \underbrace{\delta_{jt} \zeta_{j} q_{t-1} h_{jt-1}^{-1} \Delta \log q_{t}}_{\mathclap{\substack{\text{4: consumption complementarities}}}}
    + \underbrace{\delta_{jt} (1-\kappa_{t} s_{jt-1}) h_{jt-1}^{-1} \Delta \log w_{jt}}_{\mathclap{\substack{\text{5: substitution effects}}}}
    - \underbrace{\delta_{jt} h_{jt-1}^{-1} \Delta \log \mu_{jt}}_{\mathclap{\substack{\text{6: bargaining effects}}}},
\end{split}
\end{align}
where $j\in\{1,2\}$, $-j$ indicates $j$'s partner, and $\Delta$ is the first difference between $t-1$ and $t$.\footnote{We assume the progressivity tax parameter does not change between proximate periods, so $\kappa_{t}=\kappa_{t-1}$. This assumption is innocuous (we show this in appendix \ref{Appendix::DerivationOptimalityConditions}) and makes the notation more compact.} The first two terms reflect the disincentives from shifts in, respectively, the proportionality of taxes and the partner's earnings due to progressive joint taxation. The third term reflects the wealth and income effects from shifts in the marginal utility of wealth $\lambda_{t}$ (the Lagrange multiplier on the sequential budget constraint). The fourth term captures consumption-hours complementarities. The fifth term reflects the substitution effects on labor supply from shifts in own wage $w_{jt}$, accounting for progressive taxation. Finally, the sixth term reflects the bargaining effects on labor supply from shifts in the Pareto weight $\mu_{jt}$.

Parameter $\delta_{jt}$ is given by one over $\alpha_{j}^{-1} + \kappa_{t} s_{jt-1} h_{jt-1}^{-1}$, where $\alpha_{j}>0$ is approximately equal to $j$'s Frisch elasticity of labor supply scaled by his/her hours of work, and $s_{jt}\geq0$ is $j$'s share of family earnings. It follows that $\delta_{jt}>0$, which helps sign most of the terms above. For example, an increase in the proportionality of taxes reduces labor supply due to tax disincentives (term 1) while an improvement in $j$'s Pareto weight reduces his/her hours reflecting the bargaining effects we described earlier (term 6). Finally, $\zeta_{j}$ reflects the nature of the consumption-hours complementarity and can thus be of any sign.\footnote{We show in appendix \ref{Appendix::DerivationOptimalityConditions} that $\alpha_{j} = u_{j[h]}/(u_{j[hh]}\exp(-\boldsymbol{\pi}_{j}^{h\prime}\boldsymbol{\xi}_{jt-1}))$ and $\zeta_{j} = (u_{j[hq]}\exp(-\boldsymbol{\pi}_{j}^{q\prime}\boldsymbol{\xi}_{jt-1}))/u_{j[h]}$.}

Terms 1 through 5 appear also in the dynamic unitary model of \citet{BlundellPistaSaporta2016Family_Labour} while term 6 is unique to the dynamic collective model. Expression \eqref{Eq::EstimableEquation.BeforeReplacingMu} is common across commitment modes in the latter case, with differences in behavior across modes arising mostly through the last term, i.e. through the way the Pareto weight changes in each mode.\footnote{Couples in limited commitment make savings choices taking into account the effects of those choices on the outside options. Therefore, the marginal utility of wealth $\lambda$ behaves differently in limited commitment compared to the other modes but this is a feature we do not exploit for testing.} Then specifying an expression for the (reduced form) dependence of $\mu$ on its arguments, which we do in the next section, allows us to use \eqref{Eq::EstimableEquation.BeforeReplacingMu} to test for commitment.

The \emph{partner}'s current wage does not explicitly appear in \eqref{Eq::EstimableEquation.BeforeReplacingMu} even though $w_{-jt}\in W_{t}$ was previously part of the non-bargaining state space of the problem. This is because, aside of bargaining, $w_{-jt}$ only induces tax disincentives and income effects that are fully accounted for by the partner's earnings and the marginal utility of wealth. Therefore, conditional on $y_{-jt}$ and $\lambda_{t}$, $w_{-jt}$ does not affect spouse $j$'s hours outside of bargaining. This is in contrast to $w_{jt}$ which is the price of time and thus affects hours irrespective of bargaining.

Two final remarks are due here. First, the nature of the log-linearization is such that the outcome equation is in terms of hours growth rather than of hours levels. While \eqref{Eq::EstimableEquation.BeforeReplacingMu} is fully consistent with the policy function $h_{jt}^*$, it is nonetheless not the policy function itself which generally disciplines the levels of hours. Second, the Euler equation in our model is very complicated because wealth enters the Pareto weight outside of full commitment.\footnote{\citet{Mazzocco2007} uses a simpler Euler equation by assuming wealth does not affect the outside options.} Our use of the static optimality condition avoids this complication at the cost of introducing a term for the marginal utility of wealth, $\Delta \log \lambda_{t}$, to which we return in section \ref{SubSection::Implementation_Estimation}.

\subsection{Dynamics of the Pareto weight}\label{SubSection::Test_Commitment_Dynamics_Pareto}

Our goal is to specify the reduced form dependence of the Pareto weight on its arguments, which will enable us to take \eqref{Eq::EstimableEquation.BeforeReplacingMu} to the data. Consider the structural version of the most general Pareto weight, i.e. of limited commitment, given by $\mu_{jt} = \mu_{j}(W_{t},Z_{t},a_{t},\mu_{jt-1})$ with $\mu_{j0} = \mu_{j}(\Theta_0)$, $j\in\{1,2\}$. To simplify the discussion, let $\mu_{jt}$ be a function of one stochastic distribution factor $z_{t}\in Z_{t}$ and the past Pareto weight only, i.e. $\mu_{jt} = \mu_{j}(z_{t},\mu_{jt-1})$, and let $\mu_{j0}$ be a function of one initial factor $\theta_{0}\in \Theta_{0}$. Assume without loss of generality that both factors empower spouse $j$, i.e. $\partial \mu_{jt} / \partial z_{t}>0$ and $\partial \mu_{j0} / \partial \theta_{0}>0$. We generalize the discussion to multiple distribution factors (as well as wages and assets) in appendix \ref{Appendix::ApproximationParetoWeight}.

Suppose momentarily that $\ddot{\mu}_{j}(z_{t},\mu_{jt-1})$ is the smooth approximation of $\mu_{j}(z_{t},\mu_{jt-1})$. If the steps in $\mu$ are sufficiently small, $\ddot{\mu}$ will be a reasonable approximation of the true dynamics of the Pareto weight. Appendix \ref{Appendix::ApproximationParetoWeight} shows that a log-linearization of $\mu_{jt}\approx\ddot{\mu}_{j}(z_{t},\mu_{jt-1})$ yields $\Delta \log \mu_{jt} \approx e_{\mu_{j},z} \Delta \log z_{t} + e_{\mu_{j},\mu_{jL}} \Delta \log \mu_{jt-1}$, where $e_{\mu_{j},z}$ is the elasticity of $\mu_{j}$ $\text{w.r.t.}$ $z$ and $e_{\mu_{j},\mu_{jL}}$ is its elasticity $\text{w.r.t.}$ the past weight (the subscript $L$ denotes the lag). Economic theory disciplines the signs of these elasticities; in this case $e_{\mu_{j},z}>0$ due to the assignability of $z$ while $e_{\mu_{j},\mu_{jL}}>0$ reflecting persistence in the Pareto weight in limited commitment. The elasticities depend on the \emph{past levels} of the distribution factor due to the nature of the log-linearization and they thus vary in the cross-section. This expression is useful because it relates the contemporaneous shifts in the Pareto weight to the contemporaneous shifts in the distribution factor and the most recent historical dynamics of the weight. 

Exploring the recursive nature of $\Delta \log \mu_{jt}$, we can substitute the past weight backwards until we reach $t=0$, i.e. marriage, and $\Delta \log \mu_{j0}$ on the right hand side. $\Delta \log \mu_{j0}$ describes the formation of the initial Pareto weight at marriage, i.e. the difference between the initial weight $\mu_{j0}$ and a generic one available to all individuals when they first meet and start dating at $t=-1$.\footnote{Let newly met individuals start dating with the same bargaining power (an innocuous normalization); the weight $\mu_{j0}$ of pairs that marry at $t=0$ differs from the generic value at $t=-1$, with the difference determined by the distribution factor $\theta_0$ realized at $t=0$.} 
The difference is driven by the initial distribution factor $\theta_{0}$ and, as different couples differ in $\theta_{0}$, $\Delta \log \mu_{j0}$ varies in the cross-section as a function of it. We adapt the simple log-linear formulation $\Delta \log \mu_{j0} = e_{\mu_{j},\theta} \theta_{0}$, where $e_{\mu_{j},\theta}$ reflects the loading factor of $\theta_0$ onto the initial weight. We expect $e_{\mu_{j},\theta}>0$ from the assignability of $\theta_0$. 

Combining these steps (exact derivation in online appendix \ref{Appendix::ApproximationParetoWeight}) yields
\begin{equation}\label{Eq::Definition.ApproxStructuralParetoWeight_Final}
    \Delta \log \mu_{jt} ~\approx~ 
    \underbrace{\sum_{\tau=0}^{t-1} (e_{\mu_{j},\mu_{jL}})^\tau e_{\mu_{j},z} \Delta \log z_{t-\tau}}_{\mathclap{\substack{\text{cumulative growth in Pareto weight}\\\text{due to time-varying distribution factors}}}}
    ~+~ 
    \underbrace{(e_{\mu_{j},\mu_{jL}})^t e_{\mu_{j},\theta} \theta_{0}}_{\mathclap{\substack{\text{formation of initial}\\\text{Pareto weight}\\\text{at marriage}}}},
\end{equation}
where $t$ reflects the number of periods since marriage. \eqref{Eq::Definition.ApproxStructuralParetoWeight_Final} shows that the limited commitment Pareto weight at $t$ is the accumulation of gradual shifts in the weight over time as a result of shifts in current ($\tau=0$) and historical ($\tau=1,\dots,t-1$) distribution factors $z_{\tau}$, as well as $\theta_{0}$ that reflects the formation of bargaining power at marriage. The effect of the distribution factor $z$ on the current weight is given by $\partial \Delta \log \mu_{jt}/\partial \Delta \log z_{t-\tau} = (e_{\mu_{j},\mu_{jL}})^\tau e_{\mu_{j},z}>0$. In appendix \ref{Appendix::ApproximationParetoWeight}, we show that $e_{\mu_{j},\mu_{jL}}\leq1$, so historical distribution factors have a gradually smaller effect as the length of time increases, with the rate of decay determined by $e_{\mu_{j},\mu_{jL}}$.

Expression \eqref{Eq::Definition.ApproxStructuralParetoWeight_Final} encapsulates the alternative commitment modes and our nesting argument. Limited commitment has $e_{\mu_{j},\mu_{jL}}>0$, $e_{\mu_{j},z}>0$, $e_{\mu_{j},\theta}>0$, with the latter two signed by the assignability of $z$ and $\theta_{0}$. No commitment has $e_{\mu_{j},\mu_{jL}}=0$, $e_{\mu_{j},z}>0$, $e_{\mu_{j},\theta}>0$; as such no commitment is nested within limited commitment in terms of the variables that enter the Pareto weight. Full commitment has $e_{\mu_{j},\mu_{jL}}=0$, $e_{\mu_{j},z}=0$, $e_{\mu_{j},\theta}>0$; as such full commitment is nested within no commitment. Finally, the unitary model has $e_{\mu_{j},\mu_{jL}}=0$, $e_{\mu_{j},z}=0$, $e_{\mu_{j},\theta}=0$; as such the unitary model is nested within the full commitment collective model. Conditional on identifying $e_{\mu_{j},\mu_{jL}}$, $e_{\mu_{j},z}$, $e_{\mu_{j},\theta}$, these points suggest the type of hypotheses one can formulate and assess in the data. 

Informed by \eqref{Eq::Definition.ApproxStructuralParetoWeight_Final}, our choice of specification for the reduced form dependence of the Pareto weight on its arguments is
\begin{equation}\label{Eq::Definition.ReducedFormParetoWeight}
    \Delta \log \mu_{jt} \approx
    \sum_{\tau=0}^{t-1} \eta_{j\tau}^{z}\Delta \log z_{t-\tau} + \eta_{jt}^\theta \theta_{0}.
\end{equation}
The $\eta_{j\tau}^{z}$'s and $\eta_{jt}^\theta$ are reduced form elasticities for the response of $j$'s Pareto weight to distribution factors: $\eta_{j\tau}^{z}$ captures the effect of the $z$ factor $\tau$ periods in the past; $\eta_{jt}^\theta$ captures the effect of the $\theta$ initial factor $t$ periods after marriage. With additional distribution factors (and wages and assets) affecting bargaining, the number of parameters increases considerably as we show in appendix \ref{Appendix::ApproximationParetoWeight}. Moreover, in our richest specification subsequently, we let the $\eta_{j\tau}^z$'s depend on the past levels of the distribution factors, that is $\eta_{j\tau}^z = \eta_{j\tau}^z(z_{t-\tau-1})$, to mimic the dependence of the elasticities $e_{\mu_{j},z}$ and $e_{\mu_{j},\mu_{jL}}$ on the past levels of the factors. A given shift in a distribution factor may thus shift the Pareto weight or not (that is, in spite of the smooth formulation in \eqref{Eq::Definition.ReducedFormParetoWeight}), depending on the factors' historical values.

\subsection{Formulation of test}\label{SubSec::Test_Commitment_Formulation}

Our final equation for hours combines the dynamics of household labor supply in \eqref{Eq::EstimableEquation.BeforeReplacingMu} with the Pareto weight in \eqref{Eq::Definition.ReducedFormParetoWeight}. After introducing additional distribution factors $z_{1t},z_{2t}\in Z_{t}$ and $\theta_{10},\theta_{20}\in \Theta_{0}$, reinstating wages $w_{1t},w_{2t}\in W_{t}$ and assets $a_{t}$ as arguments in $\mu_{jt}$, and pooling common terms, the combined hours equation for spouse $j\in\{1,2\}$ is given by
\begin{align}\label{Eq::EstimableEquation.Final}
\begin{split}
\Delta \log h_{jt}
    &~=~ \delta_{jt} h_{jt-1}^{-1} \Delta \log (1-\chi_{t})
    ~-~ \delta_{jt} \kappa_{t} s_{-jt-1}h_{jt-1}^{-1}\Delta \log y_{-jt}\\[5pt]
    &~+~ \delta_{jt} h_{jt-1}^{-1} \Delta \log \lambda_{t}
    ~-~ \delta_{jt} \zeta_{j} q_{t-1} h_{jt-1}^{-1} \Delta \log q_{t}\\[5pt]
    &~\underbrace{+~ \delta_{jt} (1-\kappa_{t} s_{jt-1}-\eta_{j0}^{w_{j}})}_{\mathclap{\substack{\beta_{j[w_{jt}]}:\text{ substitution and bargaining effects}\\\text{of own current wage}}}} h_{jt-1}^{-1} \Delta \log w_{jt} 
    ~\underbrace{-~ \delta_{jt} \eta_{j0}^{w_{-j}}}_{\mathclap{\substack{\beta_{j[w_{-jt}]}:\text{ bargaining effect}\\\text{of partner's current wage}}}} h_{jt-1}^{-1} \Delta \log w_{-jt}\\[5pt]
    &~\underbrace{-~ \sum_{\tau=1}^{t-1}\delta_{jt} \eta_{j\tau}^{w_{j}}}_{\mathclap{\substack{\beta_{j[w_{jt-\tau}]}:\text{ bargaining effects}\\\text{of own past wages}}}} h_{jt-1}^{-1} \Delta \log w_{jt-\tau} 
    ~\underbrace{-~ \sum_{\tau=1}^{t-1}\delta_{jt} \eta_{j\tau}^{w_{-j}}}_{\mathclap{\substack{\beta_{j[w_{-jt-\tau}]}:\text{ bargaining effects}\\\text{of partner's past wages}}}} h_{jt-1}^{-1} \Delta \log w_{-jt-\tau}\\[5pt]
    &~\underbrace{-~ \sum_{\tau=0}^{t-1}\delta_{jt} \eta_{j\tau}^{z_{j}}}_{\mathclap{\substack{\beta_{j[z_{jt-\tau}]}:\text{ bargaining effects}\\\text{of current and past }z_{j}}}} h_{jt-1}^{-1} \Delta \log z_{jt-\tau} 
    ~\underbrace{-~ \sum_{\tau=0}^{t-1}\delta_{jt} \eta_{j\tau}^{z_{-j}}}_{\mathclap{\substack{\beta_{j[z_{-jt-\tau}]}:\text{ bargaining effects}\\\text{of current and past }z_{-j}}}} h_{jt-1}^{-1} \Delta \log z_{-jt-\tau}\\[5pt]
    &~\underbrace{-~ \sum_{\tau=0}^{t-1}\delta_{jt} \eta_{j\tau}^{a}}_{\mathclap{\substack{\beta_{j[a_{t-\tau}]}:\text{ bargaining effects}\\\text{of current and past assets}}}} h_{jt-1}^{-1} \Delta \log a_{t-\tau}
    ~\underbrace{-~ \delta_{jt} \eta_{jt}^{\theta_{j}}}_{\mathclap{\substack{\beta_{j[\theta_{j0}]}:\text{ bargaining}\\\text{ effect of }\theta_{j0}}}} h_{jt-1}^{-1} \theta_{j0} 
    ~\underbrace{-~ \delta_{jt} \eta_{jt}^{\theta_{-j}}}_{\mathclap{\substack{\beta_{j[\theta_{-j0}]}:\text{ bargaining}\\\text{ effect of }\theta_{-j0}}}} h_{jt-1}^{-1} \theta_{-j0},
\end{split}
\end{align}
where $-j$ indicates $j$'s spouse. To simplify the discussion, we introduce the parameters $\beta_{j}$ as the reduced form coefficients on wages and distribution factors that enter $j$'s hours. We show in square brackets indexing $\beta_{j}$ which variable each coefficient corresponds to; e.g., $\beta_{j[w_{-jt-\tau}]}$ is the coefficient on $\Delta \log w_{-jt-\tau}$, the partner's wage $\tau$ periods in the past.

Except one's own current wage, all other current and past wages and distribution factors enter the equation exclusively through the Pareto weight. We may thus formulate testable hypotheses on their coefficients in accordance with the alternative models. The own current wage $w_{jt}$ enters \eqref{Eq::EstimableEquation.Final} irrespective of bargaining ($w_{jt}$ is the price of one's own hours), so $\beta_{j[w_{jt}]}$ cannot be part of these hypotheses. 

Distribution factors do not affect behavior in the unitary model, which thus has
\begin{align*}
    {\cal H}_{0}^\text{Unit.}: ~&\beta_{j[w_{-jt}]}=\beta_{j[w_{kt-\tau}]}=\beta_{j[z_{kt}]}=\beta_{j[z_{kt-\tau}]}=\beta_{j[\theta_{k0}]}=0\\
                                &\text{for }\tau\in\{1,\dots,t-1\}\text{ and }k\in\{1,2\}.
\end{align*}
In words, the coefficients on the partner's current wage, all past wages, all current and past distribution factors, and the initial factors at marriage, are zero in the unitary model.

In the full commitment collective model, initial distribution factors affect behavior through the time $t=0$ Pareto weight but later realizations of distribution factors do not. While this seems to suggest that $\beta_{j[\theta_{k0}]}\neq0$, the coefficients on the initial factors be non-zero, this is \emph{not} true in our formulation. Contrasting the reduced form specification of the Pareto weight in \eqref{Eq::Definition.ReducedFormParetoWeight} with its structural counterpart in \eqref{Eq::Definition.ApproxStructuralParetoWeight_Final}, it is clear that $\beta_{j[\theta_{k0}]} \equiv -\delta_{jt} \eta_{jt}^{\theta_{k}}=-\delta_{jt}(e_{\mu_{j},\mu_{jL}})^t e_{\mu_{j},\theta_k}$. Even if $e_{\mu_{j},\theta_k}\neq0$ and the initial factor structurally affects the initial weight, the past does not matter for behavior in full commitment, which has $e_{\mu_{j},\mu_{jL}}=0$ and therefore $\beta_{j[\theta_{k0}]}=0$. In other words, any effect of the initial distribution factor $\theta_{k0}$ on the Pareto weight today is through the recursive structure and the persistence of the weight that are features of limited commitment alone. Consequently, full commitment has
\begin{align*}
    {\cal H}_{0}^\text{FC}:    ~&\beta_{j[w_{-jt}]}=\beta_{j[w_{kt-\tau}]}=\beta_{j[z_{kt}]}=\beta_{j[z_{kt-\tau}]}=\beta_{j[\theta_{k0}]}=0\\
                                &\text{for }\tau\in\{1,\dots,t-1\}\text{ and }k\in\{1,2\},
\end{align*}
which is the same as ${\cal H}_{0}^\text{Unit.}$. This is because our dynamic differences framework revolves around the \emph{growth} of the Pareto weight, $\Delta \log \mu_{jt}$, which is observationally equivalent between full commitment and the unitary model. The main implication is that if we \emph{fail} to reject the common null, this failure cannot inform us about the true underlying structure. 

Under no commitment, past distribution factors do not matter for behavior (this includes the initial factors at marriage for similar reasons as above) while current distribution factors do. No commitment thus has
\begin{align*}
    {\cal H}_{0}^\text{NC}: ~\beta_{j[w_{kt-\tau}]}=\beta_{j[z_{kt-\tau}]}=\beta_{j[\theta_{k0}]}=0 \qquad \text{for }\tau\in\{1,\dots,t-1\}\text{ and }k\in\{1,2\}.
\end{align*}
Moreover, the remaining $\beta_{j}$'s must clearly be of the correct sign: $\beta_{j[w_{-jt}]}>0$ (the partner's wage worsens $j$'s bargaining power and increases $j$'s hours) while $\beta_{j[z_{kt}]}$ is signed according to the assignability of $z_{k}$.\footnote{The partner's wage $w_{-jt}$ worsens $j$'s bargaining power so $\eta^{w_{-j}}_{j0}$, the reduced form coefficient on $w_{-jt}$ in equation \eqref{Eq::Definition.ReducedFormParetoWeight}, must be negative. Then $\beta_{j[w_{-jt}]}\equiv-\delta_{jt} \eta_{j0}^{w_{-j}}>0$ because $\delta_{jt}>0$ by construction.}

Under limited commitment, finally, all current and past distribution factors matter for behavior, so all $\beta_{j}$'s are different from zero. Limited commitment is the most general setting so it is without an alternative hypothesis or a conventional statistical test for it. Nevertheless, limited commitment remains testable in a conceptual sense. Clearly, all bargaining-related $\beta_{j}$'s must be of the correct sign, disciplined by the assignability of the corresponding distribution factors. For example, $\beta_{j[w_{-jt}]}>0$ (as in no commitment), $\beta_{j[w_{jt-\tau}]}<0$ for $\tau\geq1$ (own past wages improve past bargaining power and, through persistence in the Pareto weight, decrease own hours today), $\beta_{j[w_{-jt-\tau}]}>0$ (for opposite reasons), and similarly for the many $\beta_{j[z_{kt-\tau}]}$'s and $\beta_{j[\theta_{k0}]}$'s according to the assignability of $z_{k}$ and $\theta_{k}$. Our test is therefore about the presence of effects from current and past wages and distribution factors, as well as, conceptually, about the sign of such effects. 

The test offers many over-identifying restrictions. The obvious ones concern the \emph{partner's} hours equation. For instance, to reject no commitment, we must reject ${\cal H}_{0}^\text{NC}$ as illustrated above \emph{and} reject the analogous hypothesis for the $\beta_{-j}$'s. Another set of restrictions stems from the presence of multiple distribution factors that give rise to proportionality restrictions as in \citet{Bourguignon2009}. From the structural form of the Pareto weight in \eqref{Eq::Definition.ApproxStructuralParetoWeight_Final}, it is easy to see that the ratio of partial effects of any two concurrent distribution factors is independent of when the factors are timed, e.g. $(\partial \Delta \log \mu_{jt}/\partial \Delta \log w_{1t-\tau})/(\partial \Delta \log \mu_{jt}/ \partial \Delta \log w_{2t-\tau}) = e_{\mu_{j},w_{1}}/e_{\mu_{j},w_{2}}$ independent of $\tau$. This translates into proportionality restrictions on the bargaining effects on hours, that is $\beta_{j[w_{1t-\tau}]}/\beta_{j[w_{2t-\tau}]} = \eta_{j\tau}^{w_{1}}/\eta_{j\tau}^{w_{2}}$ is the same for all $\tau\geq1$.\footnote{The proportionality restrictions extend beyond this illustration. For example, the ratio of partial effects of a given distribution factor $\iota$ periods apart is independent of when the effects are exactly timed.}

Two final remarks are due here. First, the hypotheses do not involve assets even if assets enter the Pareto weight outside of full commitment. Unlike wages or $z_{k}$ and $\theta_{k}$, wealth is non-assignable so it is not clear whom it empowers. Second, although $\beta_{j[w_{jt}]}$ is not useful for testing because the own current wage affects hours irrespective of bargaining, it turns out that we can identify $\eta_{j0}^{w_j}$, the Pareto weight coefficient on $w_{jt}$. This parameter must be positive in no and limited commitment, which offers an additional testable restriction.

\section{Empirical implementation}\label{Section::Implementation}

\subsection{Data requirements}

Our test requires panel data over at least three periods. Two periods are needed to form the concurrent \emph{growth} of hours and distribution factors, thus test whether contemporaneous factors affect behavior; this essentially tests for full commitment as in \citet{Mazzocco2007}. A third period is needed to form the immediately past growth of distribution factors, thus test whether history affects behavior; this in turn separates no and limited commitment. Additional periods strengthen the test with over-identifying restrictions on the role of history. 

The test requires at least one time-varying and one initial distribution factor assignable in the couple in order to assess all parts of ${\cal H}_{0}^\text{FC}$ and ${\cal H}_{0}^\text{NC}$. Assignability enables us to assess the sign of the coefficients. The partner's wage $w_{-jt}$ is the obvious assignable time-varying factor (conditional on wealth and the partner's earnings, $w_{-jt}$ does not affect choices outside of bargaining), so additional $z_{jt}\in Z_{t}$ are not strictly needed.

Given the requirement to have data on the household over a minimum of three periods, it follows that the couple must stay intact, i.e. not divorce, over this course of time. A couple may divorce later (see appendix \ref{Appendix::RecursiveFormulation}) but information from those periods is not part of our test since \eqref{Eq::EstimableEquation.Final} describes behavior for as long as the couple remains married.

\subsection{Sample selection}

We use public data from the Panel Study of Income Dynamics in the US. The PSID started in 1968 as an income and employment survey of a representative sample of households and their split-offs. It was later redesigned to enable the collection also of expenditure and wealth information. As our estimating equation includes consumption and wealth controls, we focus on the period between 1999-2019 when this information is available.\footnote{The PSID is biennial over 1999-2019; we let $\Delta x_{t} = x_{t}-x_{t-1}$ denote the first difference in $x$ between `current' and `previous' periods, even though the `previous' period refers to two calendar years in the past.} 

We focus on the core Survey Research Center (initially representative) sample and we select married households in which the spouses are between 21 and 65 years old. Consistent with our data requirements, we keep couples observed for at least three consecutive periods. We require complete data on earnings, hours, consumption, wealth, and demographics. Given our outcome variable, we restrict the sample to couples in which both partners participate in the labor market.\footnote{Among couples who meet all other selection criteria, 91.6\% of men and 80.1\% of women work for pay.} We return to this last selection in section \ref{Section::Discussion}.

Appendix table \ref{AppTable::SummaryStatistics} summarizes our baseline sample of 13,955 observations that meet these criteria. The sample conforms to expectations about income, hours, and demographics among married couples in the PSID \citep[e.g.][]{BlundellPistaSaporta2016Family_Labour}. For example, women work on average for 1,758 hours/year and earn \$47,843, which is about 79\% of men's average hours (2,231 hours/year) but only 59\% of men's earnings (\$81,742) respectively.

\subsection{Estimation details and econometric issues}\label{SubSection::Implementation_Estimation}

\textbf{Modeling choices.} We must address three final issues before we run the test. The first concerns the marginal utility of wealth. $\Delta \log \lambda_{t}$ reflects the change from $t-1$ to $t$ in the couple's marginal utility over total income and wealth. The couple adjusts wealth endogenously in expectation of future states of the world; this reflects precautionary savings and, in limited commitment, investment in the outside options. As $\Delta \log \lambda_{t}$ is unobserved, we replace it with a function of the growth in household income $\Delta \log y_{t}$ and wealth $\Delta \log a_{t}$, namely $\Delta \log \lambda_{t} = \ell_{\Delta y} \Delta \log y_{t}+\ell_{\Delta a} \Delta \log a_{t}+\ell_{y} \log y_{t-1}+\ell_{a} \log a_{t-1}$. Since $\Delta \log \lambda_{t}$ may depend on the initial values of income and wealth, we also include terms for those.

The second issue concerns the hourly wage. Wages have a lifecycle component that agents typically anticipate and which is unlikely to induce bargaining between spouses. We specify the growth rate of wages as the sum of a deterministic component, anticipated at $t=0$, and a stochastic component \citep[see][for a review of income processes]{MeghirPistaferri2011LifeCycleChoices}. We write $\Delta \log w_{jt} = \boldsymbol{\pi}_{j}^{w\prime}\mathbf{x}^{w}_{jt} + \omega_{jt}$, where $\mathbf{x}^{w}_{jt}$ is a vector of demographics (e.g. age) that enter the deterministic part and $\omega_{jt}$ is the wage shock such that $\mathbb{E}_{t}(\omega_{jt}|\mathbf{x}^{w}_{jt})=0$. We assume that the shock is the only part that induces bargaining between the spouses, that is, the only component of wages that shifts the Pareto weight under no/limited commitment.\footnote{Nevertheless, the deterministic profile of own wages affects the gradient of hours through the own wage term 5 in equation \eqref{Eq::EstimableEquation.BeforeReplacingMu}. We show this point in detail in appendix \ref{Appendix::EstimatingEquation}.}

The third issue concerns our choice of distribution factors in $\Theta_{0}$ and $Z_{t}$. For $\Theta_{0}$, we seek assignable factors set at $t=0$ that remain constant thereafter; but we rarely observe the time of marriage in the PSID, which limits our options. We use two age-gap-at-marriage variables, namely $\Theta_{0} = \{\mathbb{1}[age_1 << age_2], \mathbb{1}[age_1>>age_2]\}$. The first dummy indicates the husband is younger than the wife while the second indicates he is much older.\footnote{We consider the husband older if he is at least 4 years older than the wife.} The underlying premise is that youth empowers oneself as the youngest person's marriage market is typically \emph{more} active, perhaps due to fertility concerns \citep{Low2024ReproductiveCapital}. For $Z_{t}$, we seek assignable factors that vary over time. We follow \citet{ChiapporiOrefficeQuintana2012FatterAttraction} and \citet{DupuyGalichon2014PersonalityTraits} and use anthropometric measures, specifically the spouses' body mass index $Z_{t} = \{BMI_{1t}, BMI_{2t}\}$. Both choices are subject to limitations, which we discuss subsequently. Recall, however, that $z_{jt} \in Z_t$ are not strictly needed for our test (wages \emph{are} time-varying distribution factors) so we present results without $Z_t$ in the text and relegate the results \emph{with} $Z_t$ to appendix \ref{Appendix::Descriptive_Results_Detail}.

\textbf{Specifications and heterogeneity.} After implementing these modeling choices within \eqref{Eq::EstimableEquation.Final}, the final equation for hours of spouse $j\in\{1,2\}$, abstracting from $Z_{t} = \{BMI_{1t}, BMI_{2t}\}$, is given in compact form by
\begin{align}\label{Eq::EstimableEquation.FinalReducedForm}
\notag
\Delta \log h_{jt}
    &= \Big\{b_{j[0]} + \mathbf{b}^\prime_{j[x^w_{jt}]} \mathbf{x}^{w}_{jt}\\
\begin{split}
    &+ \beta_{j[w_{jt}]} \omega_{jt} 
    + \beta_{j[w_{-jt}]} \omega_{-jt}
    + \sum_{\tau=1}^{t-1} \beta_{j[w_{jt-\tau}]} \omega_{jt-\tau} 
    + \sum_{\tau=1}^{t-1} \beta_{j[w_{-jt-\tau}]} \omega_{-jt-\tau}\\[3pt]
    &+ \beta_{j[young_{j}]} \mathbb{1}[age_{j}<<age_{-j}] 
    + \beta_{j[young_{-j}]} \mathbb{1}[age_{j}>>age_{-j}]\\[3pt]
    &+ b_{j[\Delta y_{t}]} \Delta \log y_{t} + b_{j[\Delta a_{t}]} \Delta \log a_{t} + b_{j[y_{t-1}]} \log y_{t-1} + b_{j[a_{t-1}]} \log a_{t-1}
\end{split}\\[3pt]
\notag
    &+ b_{j[\Delta y_{-jt}]} s_{-jt-1} \Delta \log y_{-jt} + b_{j[\Delta q_{t}]} q_{t-1} \Delta \log q_{t} + \sum_{\tau=1}^{t-1}\beta_{j[\Delta a_{t-\tau}]} \Delta \log a_{t-\tau}\Big\}\times h_{jt-1}^{-1}.
\end{align}
Appendix \ref{Appendix::EstimatingEquation} constructs this equation step by step.

We estimate three gradually richer specifications. The \emph{first} is the reduced form linear regression of $\Delta \log h_{jt}$ on the right hand side variables in \eqref{Eq::EstimableEquation.FinalReducedForm}, treating the coefficients as constant in the cross-section/over time. Estimation via OLS delivers estimates of the reduced form bargaining effects of wages and distribution factors, enabling quick testing of the commitment hypotheses. This simplest form of our test thus affirms that the test is easy to implement in reduced form without imposing or estimating preferences.

The pure reduced form neglects the dependence of the coefficients on hours and earnings through $\delta_{jt}=1/(\alpha_{j}^{-1} + \kappa_{t} s_{jt-1} h_{jt-1}^{-1})$. Our \emph{second} specification estimates \eqref{Eq::EstimableEquation.FinalReducedForm} respecting the underlying structure of the coefficients. We fix the tax progressivity parameter $\kappa_{t}=0.185$ \citep[][]{BlundellPistaSaporta2016Family_Labour} and identify $\delta_{jt}$ through $b_{j[y_{-jt}]}$; see appendix \ref{Appendix::EstimatingEquation} for details. This enables us to estimate $\alpha_{j}$ (a scaled Frisch labor supply elasticity) and all bargaining parameters $\eta_{j}$, including $\eta_{j0}^{w_j}$ that describes how the Pareto weight shifts with own current wage $w_{jt}$.

The elasticities of the Pareto weight depend on the immediate past levels of the distribution factors (section \ref{SubSection::Test_Commitment_Dynamics_Pareto}). We reflect this in our \emph{third} specification by letting the $\eta_{j}$'s depend on such past levels. There is no reason to believe that all couples feature the same degree of commitment: in reality, some couples may commit fully while others may not. In principle, we could test for commitment on a household-by-household basis, allowing each household their own degree of commitment, i.e. a household-specific $\eta_{j}$. Albeit theoretically appealing, this requires long time series that we clearly lack. To get around this, we pool households together and estimate a form of aggregate bargaining effect across households, one that depends, however, on the past values of wages and distribution factors. A distribution factor may thus induce \emph{heterogeneous} bargaining effects for given hours and earnings, depending on the distribution factors' historical values.

We estimate all three specifications initially in the baseline sample of households that we observe for at least three consecutive periods. Subsequently, we estimate them again over a smaller sample of households observed for at least four consecutive periods. This strengthens our test with additional restrictions on the role of history. We do not go beyond four periods because we run into small samples. 

\textbf{Empirical strategy.} Unlike the first specification, estimation in the second and third cases requires GMM because of the non-linear structure of $\delta_{jt}$. GMM estimation of the full structure of \eqref{Eq::EstimableEquation.FinalReducedForm} is slow due to the dimensionality of the wage covariates $\mathbf{x}^{w}_{jt}$ and the household observables that we empirically account for (appendix \ref{Appendix::RecursiveFormulation} shows the role of demographics in the model explicitly). To avoid this, we run a first stage regression to net $\Delta \log h_{jt}$ of the demographics. We then estimate the remaining terms in a second stage using residual hours on the left hand side. This two-step estimation is similar to \citet{BlundellPistaSaporta2016Family_Labour}; a single step delivers similar point estimates, albeit much more slowly. 

Our empirical strategy proceeds as follows. First, we regress wage growth $\Delta \log w_{jt}$ on observables to obtain the wage shock $\omega_{jt}$. Second, we regress hours growth $\Delta \log h_{jt}$ on inverse past hours and taste/wage observables (times inverse past hours) to obtain residual hours consistent with the first line in \eqref{Eq::EstimableEquation.FinalReducedForm}.\footnote{The first stage observables for wages and hours include dummies for year, year of birth, education, race, region, number of family members (and its change over time), number of children (and its change over time), the presence of income recipients other than the main couple (present and past), and the presence of outside dependents (present and past). We also include education-year, race-year, and region-year interactions.} Third, we regress residual hours on all other variables in \eqref{Eq::EstimableEquation.FinalReducedForm}, which allows us to estimate the various bargaining effects. Whenever there are over-identifying restrictions, we use multiple moments and a diagonal weighting matrix. We estimate heteroskedasticity-consistent standard errors, with the household serving as a cluster. We discuss measurement error in wages and hours below.

\section{Results}\label{Section::Results}

\textbf{Reduced form specification.} The first results concern the bargaining effects of wages from the reduced form (first) specification of the commitment test. This is the simplest form of our test, one that can quickly inform about commitment in a sample of observations.

Columns 1 \& 2 of table \ref{Table::OLS_Results} show the wage terms in the male and female hours equations in the baseline sample of households observed for at least three periods.\footnote{The number of observations is smaller than 13,955 as the estimating equations are in first difference.} Three points emerge. First, the shock to the partner's wage at $t$ enters significantly positively in the male equation, and it is small, negative, and insignificant in the female equation ($\beta_{j[w_{-jt}]}$).\footnote{The shock to \emph{own} wage at $t$ enters negatively in both equations ($\beta_{j[w_{jt}]}$). This measures an aggregate of substitution and bargaining effects of own wages so $\beta_{j[w_{jt}]}$ is not used as part of the commitment test.} This term should be zero in full commitment but positive in the other regimes. Under no and limited commitment, an increase in the partner's wage empowers the partner, thus worsens one's own relative bargaining position and increases one's own labor supply.\footnote{Recall that we control for family income, wealth, and partner earnings, so the partner's wage effect cannot be interpreted as an income, wealth, or joint taxation effect.} The significant positive effect on the male side is thus indicative of non-full commitment. 

Second, the shocks to wages at $t-1$, both to own and partner wages, enter both equations mostly significantly and with the sign predicted by limited commitment. Under full or no commitment past events do not matter for current behavior so $\beta_{j[w_{jt-1}]}$ and $\beta_{j[w_{-jt-1}]}$ should be zero. By contrast, history matters in limited commitment and past events that shift past bargaining power have lasting effects on behavior in a specific way. This is precisely what we see here: own past shocks reduce own labor supply while the partner's shocks increase it. This is consistent with power shifts under limited commitment, in which favorable shocks improve the bargaining power of the spouse that receives them and, through persistence in the Pareto weight, reduce own future labor supply and increase the partner's.\footnote{Recall that we control for family income and wealth (i.e. realization of the state space), so we account for income and wealth effects that past wage shocks induce on future behavior. Such effects would operate with the same sign across male and female hours, while the coefficient $\beta_{j[w_{kt-1}]}$ clearly flips its sign over $j$.}

\begin{table}[t!]  
\begin{center}
\caption{Commitment test -- summary of reduced form results}\label{Table::OLS_Results}
\begin{tabular}{L{2.1cm} C{0.7cm} C{0.7cm} C{0.7cm} C{0.01cm} C{1.9cm} C{1.9cm} C{0.01cm} C{1.9cm} C{1.9cm}}
\toprule
                                            &&&&                    & \multicolumn{2}{c}{$\geq$ 3 periods}          && \multicolumn{2}{c}{$\geq$ 4 periods (current, }      \\
                                            &&&&                    & \multicolumn{2}{c}{(current \& past shocks)}  && \multicolumn{2}{c}{past, \& older shocks)}           \\
\cmidrule{6-7}\cmidrule{9-10}
\emph{Dependent}                            &&&&                    & (1)                   & (2)                   && (3)                  & (4)                           \\
\emph{variable:}                            &&&&                    & Male                  & Female                && Male                 & Female                        \\
$\Delta \log h_{jt}$                        &FC&NC&LC&              & $j=1$                 & $j=2$                	&& $j=1$                & $j=2$                         \\
\midrule
\multicolumn{10}{l}{\emph{Current shocks} ($t$)}\\
~~$\beta_{j[w_{jt}]}$ & . & . & . &  								& $  -30.674$ 			& $  -10.276$ 			&& $  -34.751$ 			& $  -15.996$ 					\\
 &  &  &  &  														& $(   10.657)$ 		& $(    3.516)$ 		&& $(   19.945)$ 		& $(   12.846)$ 				\\
~~$\beta_{j[w_{-jt}]}$ & $0$ & $+$ & $+$ &  						& $   52.842$ 			& $   -2.564$ 			&& $   39.383$ 			& $  -11.350$ 					\\
 &  &  &  &  														& $(   26.149)$ 		& $(    6.244)$ 		&& $(   30.972)$ 		& $(   18.735)$ 				\\
\multicolumn{10}{l}{\emph{Past shocks} ($t-1$)}\\
~~$\beta_{j[w_{jt-1}]}$ & $0$ & $0$ & $-$ &  						& $  -13.856$ 			& $   -7.338$ 			&& $  -11.018$ 			& $   -5.031$ 					\\
 &  &  &  &  														& $(    5.284)$ 		& $(    3.441)$ 		&& $(   10.115)$ 		& $(   10.305)$ 				\\
~~$\beta_{j[w_{-jt-1}]}$ & $0$ & $0$ & $+$ &  						& $   41.045$ 			& $    6.011$ 			&& $   29.491$ 			& $   -6.877$ 					\\
 &  &  &  &  														& $(   20.351)$ 		& $(    7.113)$ 		&& $(   34.599)$ 		& $(   17.107)$ 				\\
\multicolumn{10}{l}{\emph{Older shocks} ($t-2$)}\\
~~$\beta_{j[w_{jt-2}]}$ & $0$ & $0$ & $-$ &  						&  						&  						&& $  -32.739$ 			& $   -9.643$ 					\\
 &  &  &  &  														&  						&  						&& $(   16.697)$ 		& $(    6.754)$ 				\\
~~$\beta_{j[w_{-jt-2}]}$ & $0$ & $0$ & $+$ &  						&  						&  						&& $    2.417$ 			& $   24.670$ 					\\
 &  &  &  &  														&  						&  						&& $(   28.806)$ 		& $(   15.157)$ 				\\
\multicolumn{10}{l}{\emph{Initial distribution factors} ($t=0$)}\\
~~$\beta_{j[young_{j}]}$ & $0$ & $0$ & $-$ &  						& $  -28.064$ 			& $  -14.680$ 			&& $  -96.972$ 			& $  -47.766$ 					\\
 &  &  &  &  														& $(   26.590)$ 		& $(    6.370)$ 		&& $(   33.136)$ 		& $(   11.893)$ 				\\
~~$\beta_{j[young_{-j}]}$ & $0$ & $0$ & $+$ &  						& $  -80.557$ 			& $   10.376$ 			&& $ -128.468$ 			& $   20.788$ 					\\
 &  &  &  &  														& $(   30.226)$ 		& $(    4.582)$ 		&& $(   40.310)$ 		& $(   10.499)$ 				\\
\noalign{\smallskip}
\mcl{5}{l}{\text{$p$ value for }$ {\cal H}_{0}^\text{FC}$} 			& $    0.000$   		& $    0.000$           && $    0.001$          & $    0.000$ 					\\
\mcl{5}{l}{\text{$p$ value for }$ {\cal H}_{0}^\text{NC}$} 			& $    0.001$   		& $    0.000$           && $    0.001$          & $    0.000$ 					\\
\noalign{\smallskip}
\mcl{5}{l}{Observations}                            				& \mcl{2}{c}{8,513}                             && \mcl{2}{c}{6,028}                    \\
\bottomrule
\end{tabular}
\caption*{\fsz\emph{Notes:} The table reports the coefficients on wages from the reduced form (first) specification of the commitment test. The full results including the terms of the test that we do not explicitly show here appear in appendix table \ref{AppTable::OLS_Results_All}. Standard errors clustered at the household level are in brackets.}
\end{center}
\end{table}

Third, the initial distribution factors enter both equations significantly and mostly with the sign limited commitment postulates. The dummy for being the younger spouse consistently enters negatively ($\beta_{j[young_{j}]}$) while the dummy for being the older spouse enters positively among women, the expected sign under limited commitment, but negatively among men ($\beta_{j[young_{-j}]}$). We introduced these age-gap-at-marriage dummies to reflect the idea that youth empowers oneself through a relatively more active marriage market. Yet, it is unclear if these variables are distribution factors in the true sense or if they also affect tastes. But it is hard to explain from a tastes argument alone why being the youngest reduces hours (the young typically work more), thus explain why $\beta_{j[young_{j}]}<0$ as we find here.

Columns 3 \& 4 show the wage terms in the smaller sample of households observed for at least four consecutive periods. The extra period allows us to introduce the wage shocks at $t-2$ and test additional restrictions on the role of history. While significance in a statistical sense is reduced (we are estimating a more flexible model over a smaller sample), the effects of wage shocks at $t$ and $t-1$ and the effects of the initial factors remain qualitatively unchanged from the baseline. In addition, the shocks at $t-2$ enter both equations consistent with limited commitment. The husband's older shocks reduce his hours and increase the wife's while, symmetrically, the wife's older shocks reduce her hours and raise the husband's. These effects, which are consistent with power shifts under limited commitment as in the case of the $t-1$ shocks above, should be absent in either full or no commitment.\footnote{Wage shocks may be correlated over time or across spouses. This does not jeopardize our test as we explicitly include the male and female shocks over multiple periods as standalone regressors.}

Regardless of the length of history, we reject full commitment (therefore also the unitary model) and no commitment at least at the $0.1\%$ significance level in all cases.\footnote{In principle, we should test the null hypotheses \emph{jointly} across male and female equations. For simplicity, we estimate the equations separately (no cross-equation restrictions) so we also conduct testing separately.} 

The large coefficients of table \ref{Table::OLS_Results} are the result of $h_{jt-1}^{-1}$ multiplying the right hand side of \eqref{Eq::EstimableEquation.FinalReducedForm}. We can thus not interpret the coefficients as the elasticities of hours with respect to wage shocks. It is straightforward to calculate the partial effects of wages but, in the interest of brevity, we only report this in appendix \ref{Appendix::Descriptive_Results_Detail} for our richest specification below.

\textbf{Structural specification.} The second set of results accounts for the underlying structure of the coefficients in \eqref{Eq::EstimableEquation.FinalReducedForm}, which the reduced form results neglect. This allows us to estimate $\alpha_{j}$, a scaled Frisch elasticity of labor supply, and $\eta_{j0}^{w_j}$, the Pareto weight coefficient on \emph{own} \emph{current} wage. It also accounts for the dependence of the coefficients on earnings and hours, thus relaxing the restriction of homogeneity of effects across households.\footnote{The coefficients in \eqref{Eq::EstimableEquation.FinalReducedForm} are functions of $\delta_{jt}=1/(\alpha_{j}^{-1} + \kappa_{t} s_{jt-1} h_{jt-1}^{-1})$, so they depend on the underlying past earnings shares $s_{jt-1}$ and hours $h_{jt-1}$. Appendix \ref{Appendix::EstimatingEquation} provides clarity on this point.}

\begin{table}[t!]  
\begin{center}
\caption{Commitment test -- summary of structural results}\label{Table::Structural_Results}
\begin{tabular}{L{2.1cm} C{0.7cm} C{0.7cm} C{0.7cm} C{0.01cm} C{1.9cm} C{1.9cm} C{0.01cm} C{1.9cm} C{1.9cm}}
\toprule
                                            &&&&                    & \multicolumn{2}{c}{$\geq$ 3 periods}          && \multicolumn{2}{c}{$\geq$ 4 periods (current, }      \\
                                            &&&&                    & \multicolumn{2}{c}{(current \& past shocks)}  && \multicolumn{2}{c}{past, \& older shocks)}           \\
\cmidrule{6-7}\cmidrule{9-10}
\emph{Dependent}                            &&&&                    & (1)                   & (2)                   && (3)                  & (4)                           \\
\emph{variable:}                            &&&&                    & Male                  & Female                && Male                 & Female                        \\
$\Delta \log h_{jt}$                        &FC&NC&LC&              & $j=1$                 & $j=2$                 && $j=1$                & $j=2$                         \\
\midrule
\multicolumn{10}{l}{\emph{Pareto weight elasticities w.r.t current shocks} ($\tau=0$)}\\
~~$\eta_{j0}^{w_{j}}$ & $0$ & $+$ & $+$ &  							& $    1.020$ 			& $    1.007$ 			&& $    1.004$ 			& $    1.006$ 					\\
 &  &  &  &  														& $(    0.010)$ 		& $(    0.010)$ 		&& $(    0.011)$ 		& $(    0.012)$ 				\\
~~$\eta_{j0}^{w_{-j}}$ & $0$ & $-$ & $-$ &  						& $   -0.034$ 			& $   -0.003$ 			&& $   -0.026$ 			& $   -0.006$ 					\\
 &  &  &  &  														& $(    0.022)$ 		& $(    0.012)$ 		&& $(    0.017)$ 		& $(    0.018)$ 				\\
\multicolumn{10}{l}{\emph{Pareto weight elasticities w.r.t shocks 1 period in the past} ($\tau=1$)}\\
~~$\eta_{j1}^{w_{j}}$ & $0$ & $0$ & $+$ &  							& $    0.012$ 			& $    0.009$ 			&& $    0.010$ 			& $    0.002$ 					\\
 &  &  &  &  														& $(    0.006)$ 		& $(    0.008)$ 		&& $(    0.006)$ 		& $(    0.010)$ 				\\
~~$\eta_{j1}^{w_{-j}}$ & $0$ & $0$ & $-$ &  						& $   -0.032$ 			& $    0.000$ 			&& $   -0.031$ 			& $    0.002$ 					\\
 &  &  &  &  														& $(    0.022)$ 		& $(    0.019)$ 		&& $(    0.020)$ 		& $(    0.015)$ 				\\
\multicolumn{10}{l}{\emph{Pareto weight elasticities w.r.t shocks 2 periods in the past} ($\tau=2$)}\\
~~$\eta_{j2}^{w_{j}}$ & $0$ & $0$ & $+$ &  							&  						&  						&& $    0.010$ 			& $    0.009$ 					\\
 &  &  &  &  														&  						&  						&& $(    0.012)$ 		& $(    0.005)$ 				\\
~~$\eta_{j2}^{w_{-j}}$ & $0$ & $0$ & $-$ &  						&  						&  						&& $   -0.023$ 			& $   -0.026$ 					\\
 &  &  &  &  														&  						&  						&& $(    0.018)$ 		& $(    0.016)$ 				\\
\multicolumn{10}{l}{\emph{Pareto weight elasticities w.r.t initial distribution factors}}\\
~~$\eta_{jt}^{young_{j}}$ & $0$ & $0$ & $+$ &  						& $    0.046$ 			& $    0.025$ 			&& $    0.062$ 			& $    0.038$ 					\\
 &  &  &  &  														& $(    0.024)$ 		& $(    0.011)$ 		&& $(    0.024)$ 		& $(    0.014)$ 				\\
~~$\eta_{jt}^{young_{-j}}$ & $0$ & $0$ & $-$ &  					& $    0.096$ 			& $   -0.023$ 			&& $    0.077$ 			& $   -0.021$ 					\\
 &  &  &  &  														& $(    0.035)$ 		& $(    0.010)$ 		&& $(    0.037)$ 		& $(    0.009)$ 				\\
\mcl{5}{l}{\emph{Frisch elasticity} ${\alpha_{j}/\mathbb{E}(h_{jt-1})}$}   
                                                                   	& $    0.545$ 			& $    0.330$ 			&& $    1.017$ 			& $    0.719$ 					\\
 &  &  &  &  														& $(    0.164)$ 		& $(    0.136)$ 		&& $(    0.347)$ 		& $(    0.263)$ 				\\ 
\mcl{5}{l}{\text{$p$ value for }$ {\cal H}_{0}^\text{FC}$}          & $    0.000$           & $    0.000$           && $    0.000$          & $    0.000$   				\\
\mcl{5}{l}{\text{$p$ value for }$ {\cal H}_{0}^\text{NC}$}          & $    0.025$           & $    0.032$           && $    0.052$          & $    0.049$   				\\
\mcl{5}{l}{Observations}                                            & \mcl{2}{c}{8,513}                             && \mcl{2}{c}{6,028}      \\
\bottomrule
\end{tabular}
\caption*{\fsz\emph{Notes:} The table reports the Pareto weight elasticities with respect to wages from the structural (second) specification of the commitment test. The full results including the terms that we do not explicitly show here appear in online appendix table \ref{AppTable::Structural_Results_All}. Standard errors clustered at the household level are in brackets.}
\end{center}
\end{table}

Results across columns 1 \& 2 (three periods) and 3 \& 4 (four periods) of table \ref{Table::Structural_Results} paint a similar picture. We report four main points. First, the own current wage shock improves one's Pareto weight ($\eta_{j0}^{w_{j}}>0$) while the partner's current shock worsens it ($\eta_{j0}^{w_{-j}}<0$). This contradicts full commitment but it is consistent with the other two modes. Second, own past shocks from $t-1$ and $t-2$ empower oneself ($\eta_{j1}^{w_{j}}>0$, $\eta_{j2}^{w_{j}}>0$), while the partner's past shocks either weaken oneself or leave him/her untouched ($\eta_{j1}^{w_{-j}}\leq0$, $\eta_{j2}^{w_{-j}}\leq0$). Under full or no commitment, past shocks should not affect current bargaining power. Statistical significance is generally low but any parameter that \emph{is} significant has the sign that limited commitment requires. Third, being the younger spouse empowers oneself ($\eta_{jt}^{young_{j}}>0$) while being the older spouse produces mixed results as in the reduced form previously. Fourth, $\alpha_{j}/\mathbb{E}(h_{jt-1})$ is approximately equal to the Frisch elasticity of labor supply evaluated at average hours; we estimate it at $0.55$-$1.02$ for men, $0.33$-$0.72$ for women, generally in line with the literature.\footnote{$\alpha_{j}/h_{jt-1}$ is \emph{exactly} equal to the Frisch labor supply elasticity if preferences are separable between hours and consumption. Otherwise, the Frisch elasticity is a function of $\alpha_{j}/h_{jt-1}$ and the extent of non-separability \citep{BlundellPistaSaporta2016Family_Labour}. The larger male elasticity, while not uncommon, e.g. \citet{WuKrueger2021ConsumptionInsurance}, may partly reflect a different degree of consumption-leisure complementarity between men and women.} In all cases, we reject full and no commitment at conventional significance levels.\footnote{It is straightforward to recast the commitment hypotheses in terms of the Pareto weight elasticities $\eta_{j}$.}

The disproportionately large Pareto weight elasticity with respect to own vs the partner's \emph{current} wage ($\eta_{j0}^{w_{j}}$ vs $\eta_{j0}^{w_{-j}}$), in contrast to the similar magnitudes of the analogous elasticities with respect to \emph{past} wages, questions the proportionality properties of distribution factors. To assess this, we impose proportionality in the Pareto weight elasticities, which also enables us to estimate the elasticity of $\mu_{j}$ with respect to its past value.\footnote{We re-estimate the model subject to $\eta_{j\tau}^{w_{k}} = (e_{\mu_{j},\mu_{jL}})^\tau e_{\mu_{j},w_{k}}$, for $j,k\in\{1,2\}$ and $\tau\in\{0,1,2\}$, as per the log-linearized Pareto weight in \eqref{Eq::Definition.ApproxStructuralParetoWeight_Final}.} The results in table \ref{AppTable::DeepStructure_Results_All} are in line with the earlier results without providing a worse overall fit. We reject full and no commitment. We estimate $e_{\mu_{j},\mu_{jL}}$ at about 0.015 (statistically significant at this value), revealing a positive but weak association between current and past Pareto weight.

\textbf{Heterogeneity.} The third set of results allows for heterogeneity in the Pareto weight elasticities $\eta_{j}$ through their dependence on the \emph{past levels} of wages (and distribution factors in general). As each household has its own bundle of wages, we calculate household-specific partial effects of wage shocks, namely $\partial \Delta \log h_{jt} / \partial \omega_{kt-\tau}$ for $j,k\in\{1,2\}$ and $\tau\in\{0,1,2\}$. Appendix figure \ref{AppFigure::PartialEffects} plots these partial effects for all households in the sample. Most households have negative or zero labor supply effects from own past shocks (favorable \emph{own} shocks empower oneself) and positive or zero effects from partner past shocks (favorable \emph{partner} shocks weaken oneself), in line with limited commitment. There is, however, large heterogeneity in commitment. We summarize key moments in table \ref{AppTable::PartialEffects} and we report the parameter estimates in table \ref{AppTable::Heterogeneity_Results_All}, together with a detailed description of the results.

\textbf{Additional time-varying distribution factors.} We use additional distribution factors, specifically $Z_{t-\tau} = \{BMI_{1t-\tau},BMI_{2t-\tau}\}$, $\tau\in\{0,1,2\}$, as a means of over-identification. We report results in appendix tables \ref{AppTable::OLS_Results_All} (reduced form), \ref{AppTable::Structural_Results_All} (structural), and \ref{AppTable::DeepStructure_Results_All} (proportionality restrictions), along with a discussion of the limitations of BMI in this context. The bargaining effects of wages are similar to the baseline, pointing towards limited commitment. The bargaining effects of BMI are also in line with limited commitment.

\textbf{Measurement error.} Measurement error in hours \& wages biases our estimates towards zero (full commitment), but it generally preserves the true sign of the coefficients on the \emph{partner}'s shocks. We consistently reject full and no commitment, even if we focus only on the bargaining effects from the \emph{partner}'s shocks. Appendix \ref{Appendix::Measurement_Error} provides details.

\section{Discussion}\label{Section::Discussion}

We aimed to keep our discussion as simple as possible, so our model purposefully abstracted from several features at the intersection of household decision making and commitment that may matter for the formulation of our test. Appendix \ref{Appendix::Extensions} presents extensions to home production, private leisure, joint leisure, private consumption, endogenous human capital, labor market participation, nonlinear wage effects, and costly renegotiation. We also discuss why we have purposefully not imposed restrictions on the stochastic process governing the wage shocks. We argue (and support with empirical results whenever possible) that none of these issues jeopardize our test for commitment.

\section{Conclusions}\label{Section::Conclusion}

Many policies that entail an intertemporal aspect (from conditional cash transfers, to divorce and property division rules, to child support) cannot be analyzed without reference to some model of household behavior. A particularly important aspect is commitment, the extent to which the spouses commit to a plan that disciplines their behavior in the future. 

In this paper we ask \emph{to what extent spouses commit}. Using a lifecycle collective model, we characterize household behavior in three prominent alternative regimes: full, limited, and no commitment. Current and past news affect the allocation of resources differently in each case. Full commitment is nested within no commitment in terms of the information that matters for such allocation, which in turn is nested within limited commitment. Nesting and natural exclusion restrictions from contemporaneous and historical information allows us to devise a test that distinguishes between the three alternatives. 

We consistently reject full and no commitment in the PSID. By contrast, we find strong evidence for limited commitment. Favorable current and past shocks from 2 ($t-1$) and 4 ($t-2$) years ago decrease one's labor supply and increase their partner's. This occurs even though our model controls for substitution, income, wealth, and tax adjustments that such shocks may induce. We show that these asymmetric effects are consistent with a power shift in the couple, in which favorable news improve the bargaining power of the recipient spouse and, through that, reduce own labor supply and raise the partner's. 

To the best of our knowledge, ours is the first paper that brings the three commitment alternatives together, contrasts their implications for behavior, and proposes a test that distinguishes between them. The insights we provide are applicable to a broad range of cooperative risk sharing arrangements, such as risk sharing among village households.


\begin{singlespace}
\bibliographystyle{chicago}
\bibliography{Bibliography_Commitment}
\end{singlespace}


\pagebreak
\let\origappendix\appendix 
\renewcommand\appendix{\clearpage\pagenumbering{arabic}\origappendix}
\appendix
\renewcommand\appendixtocname{Online Appendix}
\renewcommand\appendixpagename{Online Appendix}
\appendixpage           
\addappheadtotoc        
\numberwithin{equation}{section} 
\numberwithin{table}{section}   
\numberwithin{figure}{section}
\setcounter{footnote}{0}


\section{Recursive formulation of household problem}\label{Appendix::RecursiveFormulation} 

This appendix derives the recursive formulation of the household program across the three commitment alternatives. In contrast to the illustration in the paper, which, for brevity, abstracted from the dependence of preferences or the budget set on individual and household characteristics, this and all following appendices show such dependence explicitly. This is important for completeness and for preparing for the empirical implementation of our test. 

Let individual period utility be $u_{j}(q_{t},h_{jt}; \boldsymbol{\xi}_{jt})$, where $\boldsymbol{\xi}_{jt}$ is a vector of taste shifters such as education or the (possibly stochastic) presence of non decision-making children. Let the mapping from gross to disposable income be $y_{t}^D = \tau(y_{t};\boldsymbol{\psi}_{t})$, which depends on household characteristics $\boldsymbol{\psi}_{t}$ (e.g. presence of young children).\footnote{$\boldsymbol{\psi}_{t}$ may have common elements with the taste shifters $\boldsymbol{\xi}_{jt}$.} Let $X_{t}=\{\boldsymbol{\xi}_{1t},\boldsymbol{\xi}_{2t},\boldsymbol{\psi}_{t}\}$ include all individual and household characteristics that affect preferences or the budget set, some of which may be stochastic (e.g. fertility).

\

\noindent \textbf{Full commitment.} The value function at $t=0$ is given by \eqref{Eq::Definition.FC}. Pooling terms together and setting $t_{0}=0$ yields
\vspace{-1ex}
\begin{equation}\label{AppEq::Definition.Vt0_FC}
    V_{t_0}^\text{FC} (\Omega_{t_0}) = 
    \max_{\{C_{t}\}_{t_0 \leq t \leq \bar{t},\boldsymbol{\omega}_{t}\in\Omega_{t}}} 
    \mathbb{E}_{t_0} \sum_{t=t_0}^{\bar{t}} \beta^{t-t_{0}} \Big( \mu_{1} (\Theta_{0})  u_{1}(q_{t}, h_{1t}) + \mu_{2} (\Theta_{0}) u_{2}(q_{t}, h_{2t})\Big)
\end{equation}
where $C_{t} = \{q_{t}, h_{1t}, h_{2t}, a_{t+1}\}$ is the set of household choice variables in period $t$ and state $\boldsymbol{\omega}_{t}\in\Omega_{t}$. $\boldsymbol{\omega}_{t}$ reflects a particular realization of the state space at $t$, which includes (among other things) the individual/household characteristics $X_{t}$ and wages $W_{t}$ -- more on $\Omega_{t}$ below. Optimal choices out of \eqref{AppEq::Definition.Vt0_FC} are contingent on $\boldsymbol{\omega}_{t}$ but we subsequently drop the explicit conditioning on the state of the world to ease the notation. For similar reasons, we no longer show the dependence of utility on the taste shifters.

Splitting the sum in \eqref{AppEq::Definition.Vt0_FC} in two parts, one for $t_{0}=0$ and another from $t_{1}=1$ through to $\bar{t}$, yields
\begin{align*}
    V_{t_0}^\text{FC} (\Omega_{t_0}) 
    &= \max_{C_{t_0}} ~
    \mu_{1} (\Theta_{0}) u_{1}(q_{t_0}, h_{1t_0}) + \mu_{2} (\Theta_{0}) u_{2}(q_{t_0}, h_{2t_0}) \\
    &+ \beta \max_{\{C_{t}\}_{t_1 \leq t \leq \bar{t}}} 
    \mathbb{E}_{t_0} \sum_{t=t_1}^{\bar{t}} \beta^{t-t_1} \Big( \mu_{1} (\Theta_{0})  u_{1}(q_{t}, h_{1t}) + \mu_{2} (\Theta_{0}) u_{2}(q_{t}, h_{2t})\Big).
\end{align*}
Given \eqref{AppEq::Definition.Vt0_FC}, we may replace the last part with $V_{t_1}^\text{FC}$ to obtain
\begin{equation*}
    V_{t_0}^\text{FC} (\Omega_{t_0}) = 
    \max_{C_{t_0}} ~
    \mu_{1} (\Theta_{0}) u_{1}(q_{t_0}, h_{1t_0}) + \mu_{2} (\Theta_{0}) u_{2}(q_{t_0}, h_{2t_0}) + \beta \mathbb{E}_{t_0} V_{t_1}^\text{FC}(\Omega_{t_1}).
\end{equation*}
For a generic period $t$, this Bellman equation generalizes to become
\begin{equation}\label{AppEq::Definition.RecursiveVt_FC}
    V_{t}^\text{FC} (\Omega_{t}) = 
    \max_{C_{t}} ~
    \mu_{1} (\Theta_{0}) u_{1}(q_{t}, h_{1t}) + \mu_{2} (\Theta_{0}) u_{2}(q_{t}, h_{2t}) + \beta \mathbb{E}_{t} V_{t+1}^\text{FC}(\Omega_{t+1}),
\end{equation}
which is clearly a special case of the general recursive form \eqref{Eq::Definition.RecursiveForm} in the paper if $\mu_{jt}=\mu_{j} (\Theta_{0})$, $j\in\{1,2\}$, i.e. the familiar time-invariance of the Pareto weight under full commitment. 

The policy functions derived from \eqref{AppEq::Definition.RecursiveVt_FC} vary with individual/household characteristics $X_{t}$, wages $W_t$, and assets $a_t$, which affect preferences and the budget set, and, through them, they affect the optimal trade-off between consumption, savings, leisure, and work. The policy functions also vary with the applicable Pareto weight $\mu_{j}(\Theta_{0})$; a relatively larger $\mu_{j}$ implies choices more tailored to spouse $j$, and vice versa. As $\mu_{j}$ only varies with $\Theta_{0}$, it follows that the full state space in full commitment is given by $\Omega_{t}=\{X_{t},W_{t},a_{t},\Theta_{0}\}$.

\

\noindent \textbf{Limited commitment.} The value function at $t=0$ is given by \eqref{Eq::Definition.LC}. We can incorporate the participation constraints into the objective function using a Lagrangian multiplier method \citep[][]{MessnerPavoniSleet2012RecursiveMethods,Marcet2019}, which yields
\vspace{-1ex}
\begin{align*}
    V_{0}^\text{LC}(\Omega_{0}) = 
    \max_{\{C_{t}\}_{0 \leq t \leq \bar{t},\boldsymbol{\omega}_{t}\in\Omega_{t}}} ~
    &\mu_{1} (\Theta_{0}) \left(\mathbb{E}_{0} \sum_{t=0}^{\bar{t}} \beta^{t} u_{1}(q_{t}, h_{1t})\right) + 
     \mu_{2} (\Theta_{0}) \left(\mathbb{E}_{0} \sum_{t=0}^{\bar{t}} \beta^{t} u_{2}(q_{t}, h_{2t})\right)\\
    &+ \sum_{t=1}^{\bar{t}} \beta^{t} \sum_{j} \nu_{jt} 
    \left(\mathbb{E}_{t} \sum_{\tau=t}^{\bar{t}} \beta^{\tau-t} u_{j}(q_{\tau}, h_{j\tau}) - \widetilde{V}_{jt}(X_{jt},w_{jt},Z_{jt},a_{t})\right),
\end{align*}
where the outside option (for example, the continuation value as single) is given by
\begin{align*}
    &\widetilde{V}_{jt}(\Omega_{jt}) = 
    \max_{\{q_{j\tau},h_{j\tau},a_{j\tau+1}\}_{\tau=t,\dots,\bar{t},\boldsymbol{\omega}_{j\tau}\in\Omega_{j\tau}}}
    \mathbb{E}_{t} \sum_{\tau=t}^{\bar{t}} \beta^{\tau-t} \widetilde{u}_{j}(q_{j\tau}, h_{j\tau})\\
    &\text{subject to }~(1+r)a_{j\tau} + \tau(w_{j\tau}h_{j\tau};\boldsymbol{\psi}_{\tau}) = q_{j\tau} + a_{j\tau+1}, ~\forall \tau, \boldsymbol{\omega}_{j\tau}, \text{ and } a_{1t}+a_{2t}=a_{t},
\end{align*}
while $\nu_{jt}$ is the Lagrange multiplier on $j$'s participation constraint at $t$. As such, $\nu_{jt}$ depends on the variables that the underlying constraint depends on, namely $\Omega_{jt} = \{X_{jt}, w_{jt}, Z_{jt}, a_{t}\}$ that affect the \emph{outside} option,\footnote{Individual characteristics $X_{jt} \subseteq X_{t}$ matter for $j$'s life as single through his/her preferences, budget set, and ultimately labor (and possibly remarriage) market future prospects. $X_{jt}$ includes the individual taste shifters $\boldsymbol{\xi}_{jt}$, the single's tax characteristics (in principle a subset of $\boldsymbol{\psi}_{t}$), and possibly some characteristics of their ex-partner that codetermine alimony or child support \citep[e.g.][]{Foerster}. For example, the ex-partner's wage rate may matter for alimony; for simplicity, we do not explicitly show this here but neither the characterization of behavior nor the test for commitment are affected by this.} and several variables below that affect the \emph{inside} option.

The individual and household characteristics $X_{t}$, wages $W_{t}$, and wealth $a_{t}$ affect optimal household choices in limited commitment through their effect on preferences and/or the budget set. The \emph{inside} value of person $j$ reflects such choices, so it varies with $X_{t}$, $W_{t}$, $a_{t}$. The inside value also varies with the applicable Pareto weight in the period/state. However, an important distinction must be made here. Suppose that there is a Pareto weight at the \emph{start} of period $t$, i.e. after the state of the world manifests but before decisions are made in the period, and a potentially different Pareto weight at the \emph{end} of period $t$. Let's call the first weight $\mu_{jt-1}$; this determines what share of the marital surplus accrues to $j$ at the \emph{start} of $t$, i.e. it determines $j$'s inside value at the start of $t$. Whether the participation constraint binds at the \emph{start} of period $t$ depends on $\mu_{jt-1}$. This is ultimately what matters for decision making later on in the period, so the Lagrange multiplier on spouse $j$'s participation constraint depends on $\mu_{jt-1}$.\footnote{The nature of decision making, as shown subsequently, ensures that the participation constraints are satisfied for the applicable weight $\mu_{jt}$ at the end of the period. Moreover, no further updating of the weight takes place before decision making in the following period, therefore $\mu_{jt}$ at the end of $t$ is also the applicable weight at the start of $t + 1$. By deduction, $\mu_{jt-1}$ is therefore the weight that materialized at the end of $t-1$.}

Pooling together the variables that affect $j$'s participation constraint in period $t$, we may express its Lagrange multiplier as $\nu_{jt} = \nu_{j} (X_{t}, W_{t}, Z_{jt}, a_{t}, \mu_{jt-1})$.\footnote{Recall that $X_{jt} \subseteq X_{t}$, so $X_{t}$ summarizes the demographic and tax characteristics of couples and singles.} Assuming for simplicity that $X_{t}$ affects the inside and outside options similarly, it follows that $X_{t}$ does not impact the participation constraints. We thus conclude that $\nu_{jt} = \nu_{j} (W_{t}, Z_{jt}, a_{t}, \mu_{jt-1})$.

Following \citet{Marcet2019}, we use standard algebra and the law of iterated expectations to pool common terms together and write the value function as
\begin{align}\label{AppEq::Definition.Vt0_LC}
\begin{split}
    V_{0}^\text{LC}(\Omega_{0}) = \max_{\{C_{t}\}_{0 \leq t \leq \bar{t},\boldsymbol{\omega}_{t}\in\Omega_{t}}}
    \mathbb{E}_{0} \sum_{t=0}^{\bar{t}} \beta^{t} 
    \Big( 
    &\mu_{1}(W_{t}, Z_{1t}, a_{t}, \mu_{1t-1}) u_{1}(q_{t}, h_{1t}) \\
    + 
    &\mu_{2}(W_{t}, Z_{2t}, a_{t}, \mu_{2t-1}) u_{2}(q_{t}, h_{2t}) + g_{t} (a_{t}) \Big),
\end{split}
\end{align}
where $\mu_{jt} \equiv \mu_{j}(W_{t}, Z_{jt}, a_{t}, \mu_{jt-1}) = \mu_{jt-1} + \nu_{j}(W_{t}, Z_{jt}, a_{t}, \mu_{jt-1})$ with $\mu_{j0} = \mu_{j} (\Theta_{0})$, $j\in\{1,2\}$, and $g_{t} (a_{t}) = -\nu_{1t} \widetilde{V}_{1t}(X_{1t},w_{1t},Z_{1t},a_{t}) - \nu_{2t} \widetilde{V}_{2t}(X_{2t},w_{2t},Z_{2t},a_{t})$ aggregates the spouses' outside options that depend on endogenous assets. 

Following similar algebra to the case of full commitment and admitting that bargaining power is relative in the household (therefore $Z_{t} = \{Z_{1t},Z_{2t}\}$ enters both spouses' Pareto weights through the implicit constraint $\mu_{1t} + \mu_{2t} = \text{\emph{constant}}$), \eqref{AppEq::Definition.Vt0_LC} has a recursive structure which, for a generic period $t$, is given by
\begin{align}\label{AppEq::Definition.RecursiveVt_LC}
\begin{split}
    V_{t}^\text{LC} (\Omega_{t}) = 
    &\max_{C_{t}} ~
    \mu_{1}(W_{t}, Z_{t}, a_{t}, \mu_{1t-1}) u_{1}(q_{t}, h_{1t}) + \mu_{2} (W_{t}, Z_{t}, a_{t}, \mu_{2t-1}) u_{2}(q_{t}, h_{2t}) \\
    &+ g_{t} (a_{t}) + \beta \mathbb{E}_{t} V_{t+1}^\text{LC}(\Omega_{t+1}).
\end{split}
\end{align}
This is a special case of the general recursive form \eqref{Eq::Definition.RecursiveForm} in the paper if $\mu_{jt} \equiv \mu_{j}(W_{t},Z_{t}, a_{t}, \mu_{jt-1})$, subject to the restriction $\mu_{jt} = \mu_{jt-1} + \nu_{jt}$ and $\mu_{j0} = \mu_{j} (\Theta_{0})$, $j\in\{1,2\}$, i.e. the familiar step function of the Pareto weight in limited commitment \citep[e.g.][]{Mazzocco2007,Voena2015}. By similar arguments as in the case of full commitment, the full state space in limited commitment is given by $\Omega_{t}=\{X_{t},W_{t},a_{t},Z_{t},\mu_{jt-1}\}$.

\

\noindent \textbf{No commitment.} The value function at $t=0$ is given by \eqref{Eq::Definition.NC}. Pooling terms together and setting $t_0=0$ yields
\begin{align}\label{AppEq::Definition.Vt0_NC}
    &V_{t_0}^\text{NC} (\Omega_{t_0}) = \\
    \notag &\max_{\{C_{t}\}_{t_0 \leq t \leq \bar{t},\boldsymbol{\omega}_{t}\in\Omega_{t}}} 
    \mathbb{E}_{t_0} \sum_{t=t_0}^{\bar{t}} \beta^{t-t_{0}} 
    \Big( 
    \mu_{1} (\Theta_{0},W_{t},Z_{t},a_{t}) u_{1}(q_{t}, h_{1t}) + 
    \mu_{2} (\Theta_{0},W_{t},Z_{t},a_{t}) u_{2}(q_{t}, h_{2t})\Big).
\end{align}
Following similar algebra to the case of full commitment, \eqref{AppEq::Definition.Vt0_NC} has a recursive structure which, for a generic period $t$, is given by
\begin{align}\label{AppEq::Definition.RecursiveVt_NC}
\begin{split}
    V_{t}^\text{NC} (\Omega_{t}) &= 
    \max_{C_{t}} ~
    \mu_{1} (\Theta_{0},W_{t},Z_{t},a_{t}) u_{1}(q_{t}, h_{1t}) + \mu_{2} (\Theta_{0},W_{t},Z_{t},a_{t}) u_{2}(q_{t}, h_{2t}) \\
    &+ \beta \mathbb{E}_{t} V_{t+1}^\text{NC}(\Omega_{t+1}).
\end{split}
\end{align}
This is a special case of the general recursive form \eqref{Eq::Definition.RecursiveForm} in the paper if $\mu_{jt} = \mu_{j} (\Theta_{0},W_{t},Z_{t},a_{t})$, $j\in\{1,2\}$, i.e. the familiar flexible dependence of the no commitment Pareto weight on current information. The full state space is given by $\Omega_{t}=\{X_{t},W_{t},a_{t},\Theta_{0},Z_{t}\}$.

\

\noindent \textbf{A note on divorce.} We have purposefully abstracted from divorce so far. In all three commitment modes, we can let divorce occur optimally. This will be the case, for instance under limited commitment, when one spouse's participation constraint binds but no feasible allocation can satisfy it without violating the other spouse's constraint. 

Different assumptions exist in the literature about the allocation of assets upon divorce or the spouses' post-divorce welfare. For instance, one may assume that the division of assets is exogenously determined by the legal system \citep[e.g.][]{Voena2015}; alternatively, it may be specified by some prenuptial agreement, which may or may not be optimally designed. Similarly, one may assume that ex-spouses go their separate ways; or they may remain related through future joint decisions \citep[e.g. regarding children, as in][]{ChiapporiostaDiasMeghirXiao2022ChildDevelopmeny}. Remarriage may be considered; in that case, the (expected) Pareto weights within the future (or contingent) union should be taken into account in the definition of each spouse's reservation utility. 

A detailed analysis of these developments is outside the scope of our paper. The crucial aspect, however, is that they remain orthogonal to our main point, which relates to the type of variables that may affect spousal behavior at date $t$. That is, whatever one assumes on the nature and determinants of divorce, the dynamics of Pareto weights \emph{during marriage} for each possible commitment mode remain as described by the previous statements.

\section{Approximation of static optimality conditions}\label{Appendix::DerivationOptimalityConditions}

\noindent \textbf{Baseline model.} The easiest way to derive the general problem's static optimality conditions is to fold \eqref{Eq::Definition.RecursiveForm} back to its non-recursive counterpart and form the Lagrangian function. This is given by 
\begin{equation*}   
    {\cal L} = \mathbb{E}_{0} \sum_{t=0}^{\bar{t}} \beta^t \Big\{ \sum_{j} \mu_{jt} u_{j}(q_{t},h_{jt};\boldsymbol{\xi}_{jt}) + g_{t}(a_{t})+ \lambda_{t} \big((1+r)a_{t} + (1-\chi_{t})y_{t}^{1-\kappa_{t}} - q_{t} - a_{t+1} \big) \Big\},
\end{equation*}
where $\lambda_{t}$ is the multiplier on the sequential budget constraint, in which we have replaced $\tau(y_{t};\boldsymbol{\psi}_{t})$ by $(1-\chi_{t})y_{t}^{1-\kappa_{t}}$. Recall that $\boldsymbol{\xi}_{jt}$ is a vector of taste shifters (such as education or the, possibly stochastic, presence of non decision-making children), while $\boldsymbol{\psi}_{t}$ reflects household tax characteristics (e.g. presence of young children) and, as such, shapes $\chi_{t}$ and $\kappa_{t}$. We leave the restrictions on the Pareto weight in the background. Let period utility be given by $u_{j}(q_{t},h_{jt};\boldsymbol{\xi}_{jt}) = \ddot u_{j}(\ddot q_{t}, \ddot h_{jt})$ where $\ddot q_{t} = q_{t}\exp(-\boldsymbol{\pi}_{j}^{q\prime}\boldsymbol{\xi}_{jt})$, $\ddot h_{jt} = h_{jt}\exp(-\boldsymbol{\pi}_{j}^{h\prime}\boldsymbol{\xi}_{jt})$, and $\boldsymbol{\pi}_{j}^q$ and $\boldsymbol{\pi}_{j}^h$ are parameters that load the taste observables onto consumption and hours. The static first order condition for hours is given by $-\mu_{jt} \ddot u_{j[h]}(\ddot q_{t}, \ddot h_{jt})\exp(-\boldsymbol{\pi}_{j}^{h\prime}\boldsymbol{\xi}_{jt}) = \lambda_{t} (1-\chi_{t})(1-\kappa_{t})y_{t}^{-\kappa_{t}} w_{jt}$ for $j\in\{1,2\}$; this can be alternatively obtained directly from the Bellman equation \eqref{Eq::Definition.RecursiveForm} for $\lambda_{t}$ equal to the costate variable $\beta \mathbb{E}_{t} \partial V_{t+1} / \partial a_{t+1}$ (expected discounted marginal value of wealth). 

Taking logs and a first difference in time yields $\Delta \log \mu_{jt} + \Delta \log (-\ddot u_{j[h]}(\ddot q_{t},\ddot h_{jt})) - \boldsymbol{\pi}_{j}^{h\prime}\Delta\boldsymbol{\xi}_{jt} = \Delta \log \lambda_{t} + \Delta \log (1-\chi_{t})(1-\kappa_{t}) - \Delta \kappa_{t} \log y_{t} + \Delta \log w_{jt}$. Estimation of this expression is impossible outside of a parametrized model because $\ddot u_{j[h]}$ is unknown. To make progress, we follow \citet{BlundellPistaSaporta2016Family_Labour} and expand $\ddot u_{j[h]}$ around its arguments one period ago. A first order Taylor approximation of $\log (-\ddot u_{j[h]}(\ddot q_{t},\ddot h_{jt}))$ around $q_{t-1}$ and $h_{jt-1}$ yields
\begin{equation*}
    \log (-\ddot u_{j[h]}(\ddot q_{t},\ddot h_{jt})) \approx \log (-\ddot u_{j[h]}(\ddot q_{t-1},\ddot h_{jt-1})) 
    +
    \frac{\ddot u_{j[hq]}}{\ddot{u}_{j[h]}} \ddot{q}_{t-1} \Delta \log q_{t} 
    +
    \frac{\ddot u_{j[hh]}}{\ddot{u}_{j[h]}} \ddot{h}_{jt-1} \Delta \log h_{jt},
\end{equation*}
where we use $\Delta x_{t} \approx x_{t-1} \Delta \log x_{t}$ for small changes in $x$. Plugging this into the log differenced optimality condition and rearranging yields 
\begin{align*}
    \Delta \log h_{jt} 
    \approx \alpha_{j} h_{jt-1}^{-1} \big(\boldsymbol{\pi}_{j}^{h\prime}\Delta\boldsymbol{\xi}_{jt} \big.
    &+ \Delta \log (1-\chi_{t})(1-\kappa_{t}) - \Delta \kappa_{t} \log y_{t} \\
    &+ \big. \Delta \log \lambda_{t} - \zeta_{j} q_{t-1} \Delta \log q_{t} + \Delta \log w_{jt} - \Delta \log \mu_{jt} \big),
\end{align*}
where $\alpha_{j} = \ddot{u}_{j[h]}/(\ddot{u}_{j[hh]}\exp(-\boldsymbol{\pi}_{j}^{h\prime}\boldsymbol{\xi}_{jt-1}))$, $\zeta_{j} = (\ddot{u}_{j[hq]}\exp(-\boldsymbol{\pi}_{j}^{q\prime}\boldsymbol{\xi}_{jt-1}))/\ddot{u}_{j[h]}$, and all first and second order partial utilities are timed at $t-1$. $\alpha_{j}$ is strictly positive by our regularity assumptions on period utility, while $\zeta_{j}$ is positive if consumption and leisure are substitutes, negative if they are complements, and zero if preferences are separable between consumption and hours. $\alpha_{j} h_{jt-1}^{-1}$ is approximately equal to spouse $j$'s Frisch elasticity of labor supply and exactly equal if preferences are separable \citep{BlundellPistaSaporta2016Family_Labour}. 

Writing $\Delta \kappa_{t} \log y_{t} \approx \kappa_{t-1}(s_{jt-1} \Delta \log y_{jt} + s_{-jt-1} \Delta \log y_{-jt})$, where $\Delta \log y_{jt}$ is the growth rate of spouse $j$'s earnings and $s_{jt}\geq0$ is $j$'s share of family earnings, replacing own earnings with $y_{jt}=w_{jt}h_{jt}$, assuming for simplicity that $\kappa_{t}=\kappa_{t-1}$ (the progressivity tax parameter does not change between proximate periods), and pooling common terms together yields
\begin{align}\label{AppEq::Approximation_FOC_Baseline}
\begin{split}
\Delta \log h_{jt}
    &= \delta_{jt} \boldsymbol{\pi}_{j}^{h\prime} h_{jt-1}^{-1} \Delta\boldsymbol{\xi}_{jt}
    + \delta_{jt} h_{jt-1}^{-1} \Delta \log (1-\chi_{t})
    -\delta_{jt} \kappa_{t} s_{-jt-1}h_{jt-1}^{-1}\Delta \log y_{-jt} \\
    &+ \delta_{jt} h_{jt-1}^{-1} \Delta \log \lambda_{t}
    - \delta_{jt} \zeta_{j} q_{t-1} h_{jt-1}^{-1} \Delta \log q_{t} \\
    &+ \delta_{jt} (1-\kappa_{t} s_{jt-1}) h_{jt-1}^{-1} \Delta \log w_{jt}
    -\delta_{jt} h_{jt-1}^{-1} \Delta \log \mu_{jt},
\end{split}
\end{align}
where $\delta_{jt} = \left(\alpha_{j}^{-1} + \kappa_{t} s_{jt-1} h_{jt-1}^{-1} \right)^{-1} > 0$, $j\in\{1,2\}$, and $-j$ indicates $j$'s partner. This is expression \eqref{Eq::EstimableEquation.BeforeReplacingMu} in the main text, including an additional term that accounts for individual cross-sectional heterogeneity through the vector of observables $\boldsymbol{\xi}_{jt}$.

\

\noindent \textbf{Extended model with home production (Appendix \ref{Appendix::Extensions}).} The model with home production has the same recursive formulation \eqref{Eq::Definition.RecursiveForm} as the baseline, subject to additional constraints: a home production function for the consumption good and a time budget per spouse. We remove the first constraint replacing consumption $q_{t}$ with $f(x_{t}, d_{1t}, d_{2t})$, where $x_{t}$ is market expenditure and $d_{kt}$ is chores of spouse $k\in\{1,2\}$. We remove the second constraint recasting the utility function to take consumption $q_{t}$ and leisure $l_{jt}=1-h_{jt}-d_{jt}$ as arguments. Now $u_{j[q]}>0$, $u_{j[l]}>0$, $u_{j[qq]}<0$, $u_{j[ll]}<0$, and the signs of $u_{j[ql]}$ and $u_{j[lq]}$ are determined by the nature of the consumption-leisure complementarity. The Lagrangian and the static first order condition for hours are similar to the baseline. Taking logs and a first difference in time yields $\Delta \log \mu_{jt} + \Delta \log \ddot u_{j[l]}(\ddot q_{t},\ddot l_{jt}) - \boldsymbol{\pi}_{j}^{l\prime}\Delta\boldsymbol{\xi}_{jt} = \Delta \log \lambda_{t} + \Delta \log (1-\chi_{t})(1-\kappa_{t}) - \Delta \kappa_{t} \log y_{t} + \Delta \log w_{jt}$, where $\boldsymbol{\pi}_{j}^l$ loads the taste observables onto leisure.

A first order Taylor expansion of $\log \ddot u_{j[l]}(\ddot q_{t},\ddot l_{jt}))$ around $x_{t-1}$, $d_{jt-1}$, $d_{-jt-1}$, $h_{jt-1}$ yields
\begin{align*}
    \log \ddot u_{j[l]}(\ddot q_{t},\ddot l_{jt}) \approx \log \ddot u_{j[l]}(\ddot q_{t-1},\ddot l_{jt-1}) 
    &+
    \frac{\ddot u_{j[lq]}}{\ddot{u}_{j[l]}} \exp(-\boldsymbol{\pi}_{j}^{q\prime}\boldsymbol{\xi}_{jt-1}) f_{x} x_{t-1} \Delta \log x_{t} \\
    &+
    \frac{\ddot u_{j[lq]}}{\ddot{u}_{j[l]}} \exp(-\boldsymbol{\pi}_{j}^{q\prime}\boldsymbol{\xi}_{jt-1}) \sum_{k} f_{d_{k}} d_{kt-1} \Delta \log d_{kt} \\
    &-
    \frac{\ddot u_{j[ll]}}{\ddot{u}_{j[l]}} \exp(-\boldsymbol{\pi}_{j}^{l\prime}\boldsymbol{\xi}_{jt-1}) \left(h_{jt-1} \Delta \log h_{jt} + d_{jt-1} \Delta \log d_{jt}\right),
\end{align*}
where $f_{x} = \partial f / \partial x_{t} >0$ and $f_{d_{k}} = \partial f / \partial d_{kt} >0$, for $k\in\{1,2\}$, are the marginal productivity of market goods and chores. All partial derivatives of the utility and home production functions are timed at $t-1$ given the nature of the log-linearization. Plugging this into the log differenced optimality condition and following the same steps as in the baseline yields 
\begin{align*}
\begin{split}
\Delta \log h_{jt}
    &= \widetilde{\delta}_{jt} \boldsymbol{\pi}_{j}^{l\prime} h_{jt-1}^{-1} \Delta\boldsymbol{\xi}_{jt}
    + \widetilde{\delta}_{jt} h_{jt-1}^{-1} \Delta \log (1-\chi_{t})
    -\widetilde{\delta}_{jt} \kappa_{t} s_{-jt-1}h_{jt-1}^{-1}\Delta \log y_{-jt} 
    + \widetilde{\delta}_{jt} h_{jt-1}^{-1} \Delta \log \lambda_{t} \\
    &- \widetilde{\delta}_{jt} \widetilde{\zeta}_{j} f_{x} x_{t-1} h_{jt-1}^{-1} \Delta \log x_{t}
    - \textstyle \sum_{k} \widetilde{\delta}_{jt} \left(\widetilde{\zeta}_{j} f_{d_{k}} -\boldsymbol{1}[j=k]\widetilde{\alpha}_j^{-1}\right)d_{kt-1} h_{jt-1}^{-1} \Delta \log d_{kt} \\
    &+ \widetilde{\delta}_{jt} (1-\kappa_{t} s_{jt-1}) h_{jt-1}^{-1} \Delta \log w_{jt}
    -\widetilde{\delta}_{jt} h_{jt-1}^{-1} \Delta \log \mu_{jt},
\end{split}
\end{align*}
where $\widetilde{\delta}_{jt} = \left(\kappa_{t} s_{jt-1} h_{jt-1}^{-1} - \widetilde{\alpha}_{j}^{-1} \right)^{-1} > 0$, $\widetilde{\alpha}_{j} = \ddot{u}_{j[l]}/(\ddot{u}_{j[ll]}\exp(-\boldsymbol{\pi}_{j}^{l\prime}\boldsymbol{\xi}_{jt-1}))<0$, $\boldsymbol{1}$ is the indicator function, $\widetilde{\zeta}_{j} = (\ddot{u}_{j[lq]}\exp(-\boldsymbol{\pi}_{j}^{q\prime}\boldsymbol{\xi}_{jt-1}))/\ddot{u}_{j[l]}$, $j,k\in\{1,2\}$, and $-j$ indicates $j$'s partner. $\widetilde{\alpha}_{j} l_{jt-1}^{-1}$ is approximately equal to $j$'s Frisch elasticity of leisure (exactly equal if preferences are separable). The linearized optimality condition with home production is similar to the baseline up to the additional controls for the production inputs.

\

\noindent \textbf{Extended model with private consumption (Appendix \ref{Appendix::Extensions}).} Let individual preferences be given by $u_{j}(q_{t},c_{jt},h_{jt};\boldsymbol{\xi}_{jt})$, where $c_{jt}$ is spouse $j$'s private consumption. In addition to the regularity conditions on utility that were introduced previously, we assume that $u_{j[c]}>0$, $u_{j[cc]}<0$ (concavity), while the signs of $u_{j[cq]}$ and $u_{j[ch]}$ are determined by the nature of the complementarity between private and joint consumption or between private consumption and hours.

The model with private consumption has a recursive formulation analogous to \eqref{Eq::Definition.RecursiveForm}, subject to a modified budget constraint: $(1+r)a_{t} + \tau(y_{t};\boldsymbol{\psi}_{t}) = q_{t} + c_{1t} + c_{2t} + a_{t+1}$. The Lagrangian and the static first order condition for hours are thus analogous to the baseline. Taking logs and a first difference in time yields $\Delta \log \mu_{jt} + \Delta \log (-\ddot u_{j[h]}(\ddot q_{t},\ddot c_{jt},\ddot h_{jt})) - \boldsymbol{\pi}_{j}^{h\prime}\Delta\boldsymbol{\xi}_{jt} = \Delta \log \lambda_{t} + \Delta \log (1-\chi_{t})(1-\kappa_{t}) - \Delta \kappa_{t} \log y_{t} + \Delta \log w_{jt}$, where $\ddot c_{jt} = c_{jt}\exp(-\boldsymbol{\pi}_{j}^{c\prime}\boldsymbol{\xi}_{jt})$ and $\boldsymbol{\pi}_{j}^c$ loads the taste observables onto private consumption.

A first order Taylor expansion of $\log (-\ddot u_{j[h]}(\ddot q_{t},\ddot c_{jt},\ddot h_{jt}))$ around $q_{t-1}$, $c_{jt-1}$, and $h_{jt-1}$ yields
\begin{align*}
    \log (-\ddot u_{j[h]}(\ddot q_{t},\ddot c_{jt},\ddot h_{jt})) \approx \log (-\ddot u_{j[h]}(\ddot q_{t-1},\ddot c_{jt-1},\ddot h_{jt-1})) 
    &+
    \frac{\ddot u_{j[hq]}}{\ddot{u}_{j[h]}} \ddot q_{t-1} \Delta \log q_{t}
    +
    \frac{\ddot u_{j[hc]}}{\ddot{u}_{j[h]}} \ddot c_{jt-1} \Delta \log c_{jt} \\
    &+
    \frac{\ddot u_{j[hh]}}{\ddot{u}_{j[h]}} \ddot h_{jt-1} \Delta \log h_{jt}.
\end{align*}
All partial derivatives are timed at $t-1$ given the nature of the log-linearization. Plugging this into the log differenced optimality condition and following the same steps as in the baseline yields 
\begin{align*}
\begin{split}
\Delta \log h_{jt}
    &= \delta_{jt} \boldsymbol{\pi}_{j}^{h\prime} h_{jt-1}^{-1} \Delta\boldsymbol{\xi}_{jt}
    + \delta_{jt} h_{jt-1}^{-1} \Delta \log (1-\chi_{t})
    -\delta_{jt} \kappa_{t} s_{-jt-1}h_{jt-1}^{-1}\Delta \log y_{-jt} \\
    &+ \delta_{jt} h_{jt-1}^{-1} \Delta \log \lambda_{t}
    - \delta_{jt} \zeta_{j} q_{t-1} h_{jt-1}^{-1} \Delta \log q_{t} 
    - \delta_{jt} \iota_{j} c_{jt-1} h_{jt-1}^{-1} \Delta \log c_{jt} \\
    &+ \delta_{jt} (1-\kappa_{t} s_{jt-1}) h_{jt-1}^{-1} \Delta \log w_{jt}
    -\delta_{jt} h_{jt-1}^{-1} \Delta \log \mu_{jt},
\end{split}
\end{align*}
where $\iota_{j} = (\ddot{u}_{j[hc]}\exp(-\boldsymbol{\pi}_{j}^{c\prime}\boldsymbol{\xi}_{jt-1}))/\ddot{u}_{j[h]}$, while $\delta_{jt}$ and $\zeta_{j}$ are defined like before. The linearized optimality condition with private production is similar to the baseline up to an additional control for the growth in the (empirically unobserved) private good, $\Delta \log c_{jt}$.

\

\noindent \textbf{Extended model with endogenous human capital (Appendix \ref{Appendix::Extensions}).} Let individual hourly wages be given by $w_{jt} = W_{jt}(v_{jt},e_{jt})$, where $v_{jt}$ reflects a stochastic productivity component (which may be subject to persistence) and $e_{jt}$ indicates spouse $j$'s human capital. We think of $e_{jt}$ as a type of experience, e.g., the cumulative number of hours one has spent working up to time $t$, though its definition can be broader. To simplify some of the following statements, we let $W$ be multiplicative, i.e., $w_{jt} = v_{jt}e_{jt}$; this restriction is not important for what follows. 

Human capital accumulates endogenously through the labor supply choices the individual makes. We express this as
\begin{equation*}
	e_{jt+1} = f_{j}(e_{jt},h_{jt}),
\end{equation*}
where $f_{j}$ loads current human capital (perhaps after some depreciation) and labor supply onto future human capital. The linear law of motion that is popular in the literature, namely $e_{jt+1} = (1-depr)e_{jt} + \kappa_j(h_{jt})$, is a special case of this. In that case, $\kappa_{j}$ would be a linear function of $h_{jt}$ (a step function if $h_{jt}$ is discrete), while $depr$ would indicate the period depreciation rate of the stock of human capital.

Endogenous human capital affects the couple's problem through the budget constraint and by changing the dynamic incentives of work in the present versus the future. The couple's state space (incl. observable characteristics) thus becomes $\Omega_{t}=\{X_{t},v_{1t},v_{2t},e_{1t},e_{2t},a_{t},\mu_{jt}\}$. Under limited commitment, individual human capital also enters the singles' state space $\Omega_{jt}=\{X_{jt},v_{jt},e_{jt},Z_{jt},a_{t}\}$ as human capital affects the continuation value as single. Therefore, $e_{jt}$, $j\in\{1,2\}$, is an endogenous state variable that affects both the inside value in marriage and the outside value upon divorce. Unlike financial wealth $a_{t}$, however, individual human capital is in principle not shared between spouses, so $e_{jt}$ is perfectly portable from marriage to divorce.

The Lagrangian function of the household problem is
\begin{align*}   
    {\cal L} = \mathbb{E}_{0} \sum_{t=0}^{\bar{t}} \beta^t \Big\{ \sum_{j} \mu_{jt} u_{j}(q_{t},h_{jt};\boldsymbol{\xi}_{jt}) + g_{t}(a_{t},e_{1t},e_{2t}) 
    &+ \lambda_{t} \big((1+r)a_{t} + (1-\chi_{t})y_{t}^{1-\kappa_{t}} - q_{t} - a_{t+1} \big) \\
    &+ \sum_j \tau_{jt} \big( f_{j}(e_{jt},h_{jt}) - e_{jt+1} \big) \Big\},
\end{align*}
which differs from the Lagrangian of the baseline program in two ways. First, $g_{t} (a_{t},e_{1t},e_{2t}) = -\nu_{1t} \widetilde{V}_{1t}(X_{1t},v_{1t},e_{1t},Z_{1t},a_{t}) - \nu_{2t} \widetilde{V}_{2t}(X_{2t},v_{2t},e_{2t},Z_{2t},a_{t})$ aggregates the spouses' outside options that depend on endogenous assets \emph{and} human capital. Second, the human capital law of motion of each spouse enters the Lagrangian as a standalone constraint in each period $t$, associated with a multiplier $\tau_{jt}\geq0$. 

Following similar steps as previously, the static first order condition for hours is given by $-\mu_{jt} \ddot u_{j[h]}(\ddot q_{t}, \ddot h_{jt})\exp(-\boldsymbol{\pi}_{j}^{h\prime}\boldsymbol{\xi}_{jt}) = \lambda_{t} (1-\chi_{t})(1-\kappa_{t})y_{t}^{-\kappa_{t}} w_{jt} + \tau_{jt}\phi_j$, for $j\in\{1,2\}$, where $\phi_j = \partial f_{j}(e_{jt},h_{jt}) / \partial h_{jt} > 0$. We assume $\phi_j$ is a constant so, for a given stock $e_{jt}$, human capital increases linearly with current work hours, e.g., as implied by the linear human capital law of motion above. This assumption simplifies the next statements but it is not otherwise crucial. Taking logs and a first difference in time yields $\Delta \log \mu_{jt} + \Delta \log (-\ddot u_{j[h]}(\ddot q_{t},\ddot h_{jt})) - \boldsymbol{\pi}_{j}^{h\prime}\Delta\boldsymbol{\xi}_{jt} \approx \varsigma_{t-1}(\Delta \log \lambda_{t} + \Delta \log (1-\chi_{t})(1-\kappa_{t}) - \Delta \kappa_{t} \log y_{t} + \Delta \log w_{jt}) + (1-\varsigma_{t-1})\Delta \log \tau_{jt}$, where $\varsigma_{t-1} = \lambda_{t-1} (1-\chi_{t-1})(1-\kappa_{t-1})y_{t-1}^{-\kappa_{t-1}} w_{jt-1} / (\lambda_{t-1} (1-\chi_{t-1})(1-\kappa_{t-1})y_{t-1}^{-\kappa_{t-1}} w_{jt-1} + \tau_{jt-1}\phi_j)$. Replacing $\log (-\ddot u_{j[h]}(\ddot q_{t},\ddot h_{jt}))$ with its Taylor expansion derived earlier and following the same steps as before, we get
\begin{align*}
\Delta \log h_{jt}
    &= \ddot \delta_{jt} \boldsymbol{\pi}_{j}^{h\prime} h_{jt-1}^{-1} \Delta\boldsymbol{\xi}_{jt}
    + \ddot \delta_{jt} \varsigma_{t-1} h_{jt-1}^{-1} \Delta \log (1-\chi_{t})
    -\ddot \delta_{jt} \varsigma_{t-1} \kappa_{t} s_{-jt-1}h_{jt-1}^{-1}\Delta \log y_{-jt} \\
    &+ \ddot \delta_{jt} \varsigma_{t-1} h_{jt-1}^{-1} \Delta \log \lambda_{t}
    - \ddot  \delta_{jt} \zeta_{j} q_{t-1} h_{jt-1}^{-1} \Delta \log q_{t} \\
    &+ \ddot \delta_{jt} \varsigma_{t-1} (1-\kappa_{t} s_{jt-1}) h_{jt-1}^{-1} \Delta \log w_{jt}
    + \ddot \delta_{jt} (1-\varsigma_{t-1}) h_{jt-1}^{-1} \Delta \log \tau_{jt}
    -\ddot \delta_{jt} h_{jt-1}^{-1} \Delta \log \mu_{jt},
\end{align*}
where $\ddot \delta_{jt} = \left(\alpha_{j}^{-1} + \varsigma_{t-1} \kappa_{t} s_{jt-1} h_{jt-1}^{-1} \right)^{-1} > 0$, and the rest is defined as in the baseline. 

The linearized optimality condition with endogenous human capital differs from the baseline by the term that involves the unobserved growth in the multiplier on $j$'s human capital. To implement this equation empirically, one may replace $\Delta \log \tau_{jt}$ with a reduced-form function of the growth in human capital as, by the definition of Lagrange multipliers, $\Delta \log \tau_{jt}$ is a function of the change in $e_{jt}$ from $t-1$ to $t$.\footnote{In practice, we let $\Delta \log \tau_{jt} = \varrho_{\Delta e} \Delta \log e_{jt} + \varrho_{e} \log e_{jt-1}$ to allow $\Delta \log \tau_{jt}$ to depend on the initial value of human capital. $\varsigma$ is unobserved but poses no problem to the simplest implementation of the test, as $\varsigma$ is subsumed by the reduced form coefficients on the various variables, analogously to the coefficients $b$ in \eqref{Eq::EstimableEquation.FinalReducedForm}.} This is analogous to our implementation of $\Delta \log \lambda_t$, the Lagrange multiplier on the budget constraint, which is similarly unobserved. Importantly, the coefficient on the Pareto weight is signed ($-\ddot \delta_{jt}<0$), which allows us to test for commitment in the exact same way as in the baseline. In other words, if we control for individual human capital similarly to how we control for assets, the test retains its original form based on the current and historical values of the variables that enter the Pareto weight.

\section{Approximation of general Pareto weight}\label{Appendix::ApproximationParetoWeight}

Consider the most general (limited commitment) Pareto weight $\mu_{jt} = \mu_{j}(W_{t},Z_{t},a_{t},\mu_{jt-1})$, with $\mu_{j0} = \mu_{j}(\Theta_0)$, $j\in\{1,2\}$. To simplify the discussion, let $\mu_{jt}$ be a function of one stochastic distribution factor $z_{t}\in Z_{t}$ and the past Pareto weight only, i.e. $\mu_{jt} = \mu_{j}(z_{t},\mu_{jLt})$, and let $\mu_{j0}$ be a function of one initial factor $\theta_{0}\in \Theta_{0}$. Here $Lt$ denotes the first lag before $t$, so $\mu_{jLt}\equiv\mu_{jt-1}$. Suppose that $\ddot{\mu}_{j}(z_{t},\mu_{jLt})$ is the smooth approximation of $\mu_{j}(z_{t},\mu_{jLt})$. A first order approximation of $\log \ddot{\mu}_{j}(z_{t},\mu_{jLt})$ around $z_{t-1}$ and $\mu_{jLt-1}$ yields
\begin{align*}
    \log \ddot{\mu}_{j}(z_{t},\mu_{jLt}) 
        &\approx \log \ddot{\mu}_{j}(z_{t-1},\mu_{jLt-1}) + \ddot{\mu}_{j}(z_{t-1}, \mu_{jLt-1})^{-1} \\
        &\times 
        \left\{ z_{t-1} \frac{\partial \ddot{\mu}_{j}}{\partial z}(z_{t-1}, \mu_{jLt-1}) \Delta \log z_{t}
        + \mu_{jLt-1} \frac{\partial \ddot{\mu}_{j}}{\partial \mu_{jL}}(z_{t-1}, \mu_{jLt-1}) \Delta \log \mu_{jLt}
        \right\}\\
        &\approx \log \ddot{\mu}_{j}(z_{t-1},\mu_{jLt-1}) + e_{\mu_{j},z}(z_{t-1}, \mu_{jLt-1}) \Delta \log z_{t} + e_{\mu_{j},\mu_{jL}}(z_{t-1}, \mu_{jLt-1}) \Delta \log \mu_{jLt},
\end{align*} 
where $\Delta \log \mu_{jLt} = \log \mu_{jLt} - \log \mu_{jLt-1} = \log \mu_{jt-1} - \log \mu_{jt-2}$ and we use again $\Delta x_{t} \approx x_{t-1} \Delta \log x_{t}$ for small changes in the generic variable $x$. The elasticities $e_{\mu_{j},z}$ and $e_{\mu_{j},\mu_{jL}}$ are quasi-structural parameters for the sensitivity of the Pareto weight to the distribution factor and past bargaining power. From the nature of the approximation, these elasticities depend on the \emph{past} levels of \emph{all} the variables that enter $\ddot{\mu}_{j}$, in this case $z_{t-1}$ and $\mu_{jLt-1}$. By recursive substitution, we can remove the right hand side Pareto weight and obtain
\begin{align*}
    \Delta \log \mu_{jt} 
    &\approx \sum_{\tau=0}^{t-1} \left(\prod_{\iota=1}^{\tau} e_{\mu_{j},\mu_{jL}}(z_{t-\iota},\mu_{jLt-\iota})\right) e_{\mu_{j},z}(z_{t-\tau-1},\mu_{jLt-\tau-1}) \Delta \log z_{t-\tau}\\
    &+ \left(\prod_{\iota=1}^{t}e_{\mu_{j},\mu_{jL}}(z_{t-\iota},\mu_{jLt-\iota})\right) e_{\mu_{j},\theta} \theta_{0}\\
    &\approx \sum_{\tau=0}^{t-1} e_{\mu_{j},\mu_{jL}}^\tau (z_{t-1},\mu_{jLt-1}) e_{\mu_{j},z}(z_{t-\tau-1},\mu_{jLt-\tau-1}) \Delta \log z_{t-\tau}\\
    &+ e_{\mu_{j},\mu_{jL}}^t (z_{t-1},\mu_{jLt-1}) e_{\mu_{j},\theta} \theta_{0},
\end{align*}
which is expression \eqref{Eq::Definition.ApproxStructuralParetoWeight_Final} in the main text. For the second approximation above, we consolidate the notation to $\prod_{\iota=1}^{\tau} e_{\mu_{j},\mu_{jL}}(z_{t-\iota},\mu_{jLt-\iota}) = e_{\mu_{j},\mu_{jL}}^\tau(z_{t-1},\mu_{jLt-1})$ and $\prod_{\iota=1}^{t}e_{\mu_{j},\mu_{jL}}(z_{t-\iota},\mu_{jLt-\iota})=e_{\mu_{j},\mu_{jL}}^t(z_{t-1},\mu_{jLt-1})$. Despite the consolidation, the coefficient on $\Delta \log z_{t-\tau}$ still depends on the past levels of the distribution factor through $e_{\mu_{j},z}(z_{t-\tau-1},\mu_{jLt-\tau-1})$.

By the law of motion of the limited commitment weight, namely $\mu_{jt} = \mu_{jt-1} + \nu_{jt}$, it follows that $e_{\mu_{j},\mu_{jL}} = (1 + \partial \nu_{jt} / \partial \mu_{jt-1}) \times (\mu_{jt-1}/\mu_{jt})$. A larger weight at the start of period $t$ (namely $\mu_{jt-1}$) loosens the constraint, so $\partial \nu_{jt} / \partial \mu_{jt-1}<0$. It follows that, if the constraint binds, $1 + \partial \nu_{jt} / \partial \mu_{jt-1}<1$, $\mu_{jt-1}<\mu_{jt}$ (the weight increases), and $e_{\mu_{j},\mu_{jL}}<1$. 

Introducing additional stochastic and initial distribution factors $z_{1t},z_{2t}\in Z_{t}$ and $\theta_{10},\theta_{20}\in \Theta_{0}$, reinstating wages $w_{1t},w_{2t}\in W_{t}$ and assets $a_{t}$ as arguments in the general Pareto weight, and repeating all previous steps, yields the general log-linear formulation
\vspace{-.1cm}
\begin{align*}
    \Delta \log \mu_{jt} 
    &\approx
      \sum_{\tau=0}^{t-1} e_{\mu_{j},\mu_{jL}}^\tau (\Gamma_{t-1}) \Big( e_{\mu_{j},w_{1}}(\Gamma_{t-\tau-1}) \Delta \log w_{1t-\tau}
     + e_{\mu_{j},w_{2}}(\Gamma_{t-\tau-1}) \Delta \log w_{2t-\tau}\Big)\\
    &+\sum_{\tau=0}^{t-1} e_{\mu_{j},\mu_{jL}}^\tau (\Gamma_{t-1}) \Big( e_{\mu_{j},z_{1}}(\Gamma_{t-\tau-1}) \Delta \log z_{1t-\tau}
     + e_{\mu_{j},z_{2}}(\Gamma_{t-\tau-1}) \Delta \log z_{2t-\tau}\Big)\\
    &+\sum_{\tau=0}^{t-1} e_{\mu_{j},\mu_{jL}}^\tau (\Gamma_{t-1}) e_{\mu_{j},a}(\Gamma_{t-\tau-1}) \Delta \log a_{t-\tau} + e_{\mu_{j},\mu_{jL}}^t(\Gamma_{t-1}) \Big(e_{\mu_{j},\theta_1} \theta_{10} + e_{\mu_{j},\theta_2} \theta_{20}\Big),
\end{align*}
where $e_{\mu_{j},w_{k}}$ is the elasticity of $j$'s Pareto weight with respect to $k$'s wage $w_{kt}$, $e_{\mu_{j},z_{k}}$ is the elasticity w.r.t.\ $z_{kt}$, $e_{\mu_{j},a}$ is the elasticity w.r.t.\ assets, and $e_{\mu_{j},\theta_k}$ is the elasticity of $j$'s weight at marriage w.r.t.\ the initial factor $\theta_{k0}$, $k\in\{1,2\}$. From the nature of the approximation, the elasticities depend on the past levels of \emph{all} arguments of the Pareto weight, namely $\Gamma = \{w_{1},w_{2},z_{1},z_{2},a,\mu_{jL}\}$ appropriately timed. The reduced form counterpart is
\vspace{-.1cm}
\begin{align*}
    \Delta \log \mu_{jt} 
    &\approx
      \sum_{\tau=0}^{t-1} \Big( \eta_{j\tau}^{w_{1}} \Delta \log w_{1t-\tau} + \eta_{j\tau}^{w_{2}} \Delta \log w_{2t-\tau} + \eta_{j\tau}^{z_{1}} \Delta \log z_{1t-\tau} + \eta_{j\tau}^{z_{2}} \Delta \log z_{2t-\tau}\Big)\\
    &+\sum_{\tau=0}^{t-1} \eta_{j\tau}^{a} \Delta \log a_{t-\tau} + \eta^{\theta_{1}}_{jt} \theta_{10} + \eta^{\theta_{2}}_{jt} \theta_{20},
\end{align*}
which is the general form of expression \eqref{Eq::Definition.ReducedFormParetoWeight}. The $\eta_{j\tau}$'s are reduced form parameters for the response of $j$'s Pareto weight to the distribution factors $\tau\in\{0,\dots,t\}$ periods in the past.

\section{Derivation of estimating equation for hours}\label{Appendix::EstimatingEquation} 

Let wage growth be $\Delta \log w_{jt} = \boldsymbol{\pi}_{j}^{w\prime}\mathbf{x}^{w}_{jt} + \omega_{jt}$, where $\boldsymbol{\pi}_{j}^{w\prime}\mathbf{x}^{w}_{jt}$ is a deterministic profile and $\omega_{jt}$ is the wage shock. $\omega_{jt}$ should not be confused with the bold case $\boldsymbol{\omega}_{t}$ that we use to describe a realization of the state space. Let the marginal utility of wealth be $\Delta \log \lambda_{t} = \ell_{\Delta y} \Delta \log y_{t}+\ell_{\Delta a} \Delta \log a_{t}+\ell_{y} \log y_{t-1}+\ell_{a} \log a_{t-1}$. Plugging these expressions in \eqref{AppEq::Approximation_FOC_Baseline} yields
\begin{align*}
\Delta \log h_{jt}
    &= \delta_{jt} \boldsymbol{\pi}_{j}^{h\prime} h_{jt-1}^{-1} \Delta\boldsymbol{\xi}_{jt} + \delta_{jt} (1-\kappa_{t} s_{jt-1}) \boldsymbol{\pi}_{j}^{w\prime} h_{jt-1}^{-1} \mathbf{x}^{w}_{jt}
    + \delta_{jt} h_{jt-1}^{-1} \Delta \log (1-\chi_{t}) \\
    &+ \delta_{jt} \ell_{\Delta y} h_{jt-1}^{-1} \Delta \log y_{t} + \delta_{jt} \ell_{\Delta a} h_{jt-1}^{-1} \Delta \log a_{t} + \delta_{jt} \ell_{y} h_{jt-1}^{-1} \log y_{t-1} + \delta_{jt} \ell_{a} h_{jt-1}^{-1} \log a_{t-1} \\
    &-\delta_{jt} \kappa_{t} s_{-jt-1}h_{jt-1}^{-1}\Delta \log y_{-jt}
    - \delta_{jt} \zeta_{j} q_{t-1} h_{jt-1}^{-1} \Delta \log q_{t} \\
    &+ \delta_{jt} (1-\kappa_{t} s_{jt-1}) h_{jt-1}^{-1} \omega_{jt}
    -\delta_{jt} h_{jt-1}^{-1} \Delta \log \mu_{jt}.
\end{align*}

Assume that the deterministic profile of wages does not enter the Pareto weight; and select BMI and the age-gap-at-marriage as the time-varying and initial distribution factors respectively. Replacing $\Delta \log \mu_{jt}$ with the reduced form expression for the dependence of the Pareto weight on its arguments, and pooling common terms together, yields
\begin{align*}
\Delta \log h_{jt}
    &= \delta_{jt} \boldsymbol{\pi}_{j}^{h\prime} h_{jt-1}^{-1} \Delta\boldsymbol{\xi}_{jt} + \delta_{jt} (1-\kappa_{t} s_{jt-1}) \boldsymbol{\pi}_{j}^{w\prime} h_{jt-1}^{-1} \mathbf{x}^{w}_{jt}
    + \delta_{jt} h_{jt-1}^{-1} \Delta \log (1-\chi_{t}) \\
    &+ \delta_{jt} \ell_{\Delta y} h_{jt-1}^{-1} \Delta \log y_{t} + \delta_{jt} (\ell_{\Delta a}-\eta_{j0}^{a}) h_{jt-1}^{-1} \Delta \log a_{t} + \delta_{jt} \ell_{y} h_{jt-1}^{-1} \log y_{t-1} + \delta_{jt} \ell_{a} h_{jt-1}^{-1} \log a_{t-1} \\
    &-\delta_{jt} \kappa_{t} s_{-jt-1}h_{jt-1}^{-1}\Delta \log y_{-jt}
    - \delta_{jt} \zeta_{j} q_{t-1} h_{jt-1}^{-1} \Delta \log q_{t} 
    - \delta_{jt} \sum_{\tau=1}^{t-1} \eta_{j\tau}^{a} h_{jt-1}^{-1} \Delta \log a_{t-\tau} \\
    &+ \delta_{jt} (1-\kappa_{t} s_{jt-1} - \eta_{j0}^{w_{j}}) h_{jt-1}^{-1} \omega_{jt} 
    - \delta_{jt} \eta_{j0}^{w_{-j}} h_{jt-1}^{-1} \omega_{-jt} \\
    &- \delta_{jt} \sum_{\tau=1}^{t-1} \eta_{j\tau}^{w_{j}} h_{jt-1}^{-1} \omega_{jt-\tau} 
    - \delta_{jt} \sum_{\tau=1}^{t-1} \eta_{j\tau}^{w_{-j}} h_{jt-1}^{-1} \omega_{-jt-\tau} \\
    &- \delta_{jt} \sum_{\tau=0}^{t-1} \eta_{j\tau}^{bmi_{j}} h_{jt-1}^{-1} \Delta \log BMI_{jt-\tau} 
    - \delta_{jt} \sum_{\tau=0}^{t-1} \eta_{j\tau}^{bmi_{-j}} h_{jt-1}^{-1} \Delta \log BMI_{-jt-\tau} \\
    &- \delta_{jt} \eta^{young_{j}}_{jt}  h_{jt-1}^{-1} \mathbb{1}[age_{j}<<age_{-j}]
    - \delta_{jt} \eta^{young_{-j}}_{jt} h_{jt-1}^{-1} \mathbb{1}[age_{j}>>age_{-j}].
\end{align*}
Contemporaneous growth in wealth $\Delta \log a_{t}$ appears as part of the marginal utility of wealth (wealth effect) and in the Pareto weight (bargaining effect); we have pooled both effects together in a single term. We replace the parameters by a single reduced form coefficient $b_{j}$ or $\beta_{j}$ associated with each term. We then absorb the tax intercept $\delta_{jt} h_{jt-1}^{-1} \Delta \log (1-\chi_{t})$ in the observables term $b_{j[0]} h_{jt-1}^{-1} + \mathbf{b}^\prime_{j[\Delta \xi_{jt}]} h_{jt-1}^{-1}\Delta\boldsymbol{\xi}_{jt}$. This yields the final compact equation \eqref{Eq::EstimableEquation.FinalReducedForm} in the main text, with the additional terms for $Z_{t-\tau} = \{BMI_{1t-\tau}, BMI_{2t-\tau}\}$ that capture the bargaining effects from current and past BMI changes in the couple.


\noindent \textbf{Identification.} The coefficient on the partner's earnings $b_{j[\Delta y_{-jt}]} = -\delta_{jt} \kappa_{t}$ is identified from a regression of hours growth $\Delta \log h_{jt}$ on the partner's earnings growth $\Delta \log y_{-jt}$ interacted with past earnings shares $s_{-jt-1}$ and inverse past hours $h_{jt-1}^{-1}$, controlling for all other variables that appear in \eqref{Eq::EstimableEquation.FinalReducedForm}. We identify the coefficients on all other terms similarly. Fixing the tax progressivity parameter $\kappa_{t}=0.185$, as in \citet{BlundellPistaSaporta2016Family_Labour}, allows us to identify $\delta_{jt}$ from $b_{j[\Delta y_{-jt}]}$. Identification of $\delta_{jt}$ subsequently enables the identification of the $\eta_{j}$'s, the parameters that describe the reduced form dependence of the Pareto weight on the various distribution factors. For example, $b_{j[w_{-jt}]} = - \delta_{jt} \eta_{j0}^{w_{-j}}$ is the coefficient on the partner's current wage shock; identification of $\delta_{jt}$ permits identification of $\eta_{j0}^{w_{-j}}$, etc.

In one case (table \ref{AppTable::DeepStructure_Results_All}), we replace the reduced form Pareto weight elasticities $\eta$ with their structural counterparts from the log-linearized version of the weight in \eqref{Eq::Definition.ApproxStructuralParetoWeight_Final}. We set $\eta_{j\tau}^{\chi} = (e_{\mu_{j},\mu_{jL}})^\tau e_{\mu_{j},\chi}$ for $\chi\in\{w_{1},w_{2},bmi_{1},bmi_{2},a\}$ and $\tau\in\{0,1,2\}$, and $\eta_{jt}^{\theta} = (e_{\mu_{j},\mu_{jL}})^t e_{\mu_{j},\theta}$ for $\theta=\{\mathbb{1}[age_{j}<<age_{-j}],\mathbb{1}[age_{j}>>age_{-j}]\}$. This effectively imposes proportionality properties over the various distribution factors as per \citet{Bourguignon2009}. In another case (table \ref{AppTable::Heterogeneity_Results_All}), we allow the elasticities $\eta$ depend on the immediately \emph{past levels} of distribution factors. For example, when wages are the sole time-varying distribution factor, we specify the Pareto weight coefficients $\eta_{j0}$ on current wages as
\begin{alignat*}{4}
    \eta_{j0}^{w_j}     =& \eta_{j0}^{w_j\{0\}} &+& \eta_{j0}^{w_j\{1\}}w_{jt-1} &+& \eta_{j0}^{w_j\{2\}}w_{-jt-1} \\
    \eta_{j0}^{w_{-j}}  =& \eta_{j0}^{w_{-j}\{0\}} &+& \eta_{j0}^{w_{-j}\{1\}}w_{jt-1} &+& \eta_{j0}^{w_{-j}\{2\}}w_{-jt-1}.
\end{alignat*}
When BMI serves as an additional time-varying distribution factor, we also add linear terms for male and female BMI. We specify $\eta_{j1}$ similarly as a function of time $t-2$ wages (and BMI) and $\eta_{j2}$ as a function of time $t-3$ wages (and BMI). We also linearly include terms for the age-gap-at-marriage variables. Identification is analogous to the baseline.

\section{Descriptive statistics and detailed results}\label{Appendix::Descriptive_Results_Detail}

This appendix presents descriptive statistics for our estimation sample and provides details on our main empirical results of section \ref{Section::Results}. Specifically, we report the complete set of parameter estimates in each of the three main specifications (reduced form, structural, heterogeneity) and provide an in-depth discussion of heterogeneity and the use of BMI as an additional stochastic distribution factor.

\

\noindent \textbf{Descriptive statistics.} Table \ref{AppTable::SummaryStatistics} presents summary statistics in our baseline estimation sample.

\begin{table}[h]  
\begin{center}
\caption{Summary statistics}\label{AppTable::SummaryStatistics}
\begin{tabular}{L{6.6cm} C{1.7cm} C{1.7cm} C{0.1cm} C{1.7cm} C{1.7cm}}
\toprule
                                                        & \multicolumn{2}{c}{Male $j=1$}    && \multicolumn{2}{c}{Female $j=2$}         \\
\cmidrule{2-3}\cmidrule{5-6}
                                                        & mean              & st.dev.       && mean                 & st.dev.           \\
\midrule
\multicolumn{6}{l}{\emph{Labor market outcomes}}\\
~~Earnings (annual, in \$1000)                          &     81.74         &     91.65     &&     47.84            &     40.25         \\
~~Hours (annual)                                        &     2,231         &       609     &&       1,758          &      660          \\
~~Hours (annual) growth ~$\Delta \log h_{jt}$           &     -0.02         &      0.40     &&     -0.00            &      0.61         \\
~~Hourly wage                                           &     36.95         &     37.97     &&     27.32            &     24.32         \\
~~Hourly wage growth ~$\Delta \log w_{jt}$              &      0.03         &      0.53     &&      0.04            &      0.50         \\
\multicolumn{6}{l}{\emph{Demographics}}\\
~~Age                                                   &     43.76         &     10.46     &&     42.11            &     10.39         \\
~~\% college education                                  &      0.70         &      0.46     &&      0.77            &      0.42         \\
~~BMI                                                   &     27.80         &      4.57     &&     24.91            &      5.49         \\
~~BMI growth ~$\Delta \log BMI_{jt}$                    &      0.01         &      0.07     &&      0.01            &      0.09         \\
                                                        && \multicolumn{3}{c}{Household}    &  \\
\cmidrule{3-5}
                                                        && mean         && st.dev.          &  \\
\midrule   
Consumption (annual, in \$1000)                         &&     45.08    &&     25.84        &  \\
Wealth (annual, in \$1000)                              &&    417.00    &&  1,269.68        &  \\
Number of children                                      &&      1.00    &&      1.10        &  \\
Age gap at marriage (male-female)                       &&      1.65    &&      4.04        &  \\
Survey waves per household                              &&      5.62    &&      2.50        &  \\
Observations [households $\times$ waves]                && \mcl{3}{c}{13,955}               &  \\
\bottomrule
\end{tabular}
\caption*{\fsz\emph{Notes:} The table reports averages and standard deviations of key variables in the baseline estimation sample. All monetary amounts are expressed in 2018 dollars.}
\end{center}
\end{table}

\

\noindent \textbf{Main results in detail.} The following tables present detailed estimation results for our first two main specifications:
\begin{itemize}
    \item Table \ref{AppTable::OLS_Results_All} reports detailed results from the reduced form (first) specification of the commitment test;
    \item Table \ref{AppTable::Structural_Results_All} reports the parameter estimates from the structural (second) specification of the commitment test;
    \item Table \ref{AppTable::DeepStructure_Results_All} reports the parameter estimates from the structural (second) specification of the test, letting $\Delta \log \mu_{jt}$ take the log-linear form in \eqref{Eq::Definition.ApproxStructuralParetoWeight_Final}.
\end{itemize}

\begin{sidewaystable}  
\begin{center}
\caption{Commitment test -- reduced form results in detail}\label{AppTable::OLS_Results_All}
\begin{tabular}{L{2.0cm} C{0.6cm} C{0.7cm} C{0.6cm} C{0.01cm} C{1.5cm} C{1.6cm} C{0.01cm} C{1.5cm} C{1.6cm} C{0.01cm} C{1.7cm} C{1.7cm} C{0.01cm} C{1.7cm} C{1.7cm}}
\toprule
                            &&&&    & \mcl{5}{c}{\textbf{wage shocks}}        && \mcl{5}{c}{\textbf{wage shocks \& BMI}}               \\
\cmidrule{6-10}\cmidrule{12-16}                        
                            &&&&    & \mcl{2}{c}{$\geq$ 3 periods} && \mcl{2}{c}{$\geq$ 4 periods} && \mcl{2}{c}{$\geq$ 3 periods} && \mcl{2}{c}{$\geq$ 4 periods} \\
\cmidrule{6-7}\cmidrule{9-10}\cmidrule{12-13}\cmidrule{15-16}
                            &&&&    & (1)      & (2)      && (3)    & (4)     && (5)    & (6)       && (7)      & (8)   \\
                            &&&&    & Male     & Female   && Male   & Female  && Male   & Female    && Male     & Female   \\
                &FC&NC&LC&          & $j=1$    &  $j=2$   &&  $j=1$ &  $j=2$  && $j=1$  &  $j=2$    &&  $j=1$   &  $j=2$   \\
\midrule
\multicolumn{16}{l}{\emph{Current shocks} ($t$)}\\
~$\beta_{j[w_{jt}]}$ & . & . & . &  & $  -30.674$ & $  -10.276$ &  & $  -34.751$ & $  -15.996$ &  & $  -28.620$ & $  -11.409$ &  & $  -63.576$ & $   -9.979$ \\
 &  &  &  &  & $(   10.657)$ & $(    3.516)$ &  & $(   19.945)$ & $(   12.846)$ &  & $(    9.490)$ & $(    3.868)$ &  & $(   19.756)$ & $(   12.871)$ \\
~$\beta_{j[w_{-jt}]}$ & $0$ & $+$ & $+$ &  & $   52.842$ & $   -2.564$ &  & $   39.383$ & $  -11.350$ &  & $   52.135$ & $    5.605$ &  & $   46.089$ & $   -5.631$ \\
 &  &  &  &  & $(   26.149)$ & $(    6.244)$ &  & $(   30.972)$ & $(   18.735)$ &  & $(   24.542)$ & $(    9.298)$ &  & $(   31.290)$ & $(   19.199)$ \\
~$\beta_{j[bmi_{jt}]}$ & $0$ & $+$ & $+$ &  &  &  &  &  &  &  & $   -7.018$ & $   67.733$ &  & $  157.212$ & $  182.492$ \\
 &  &  &  &  &  &  &  &  &  &  & $(  207.553)$ & $(   54.621)$ &  & $(  251.334)$ & $(   83.683)$ \\
~$\beta_{j[bmi_{-jt}]}$ & $0$ & $-$ & $-$ &  &  &  &  &  &  &  & $   48.146$ & $   17.316$ &  & $  398.058$ & $   38.377$ \\
 &  &  &  &  &  &  &  &  &  &  & $(  136.648)$ & $(   54.047)$ &  & $(  191.855)$ & $(   82.919)$ \\
\multicolumn{16}{l}{\emph{Past shocks} ($t-1$)}\\
~$\beta_{j[w_{jt-1}]}$ & $0$ & $0$ & $-$ &  & $  -13.856$ & $   -7.338$ &  & $  -11.018$ & $   -5.031$ &  & $  -12.662$ & $   -9.340$ &  & $   -1.564$ & $   -0.067$ \\
 &  &  &  &  & $(    5.284)$ & $(    3.441)$ &  & $(   10.115)$ & $(   10.305)$ &  & $(    5.880)$ & $(    4.202)$ &  & $(   11.918)$ & $(   14.176)$ \\
~$\beta_{j[w_{-jt-1}]}$ & $0$ & $0$ & $+$ &  & $   41.045$ & $    6.011$ &  & $   29.491$ & $   -6.877$ &  & $   37.324$ & $    6.338$ &  & $   53.021$ & $  -11.595$ \\
 &  &  &  &  & $(   20.351)$ & $(    7.113)$ &  & $(   34.599)$ & $(   17.107)$ &  & $(   18.307)$ & $(    7.532)$ &  & $(   30.789)$ & $(   17.637)$ \\
~$\beta_{j[bmi_{jt-1}]}$ & $0$ & $0$ & $+$ &  &  &  &  &  &  &  & $  151.789$ & $  111.962$ &  & $   -3.051$ & $  194.269$ \\
 &  &  &  &  &  &  &  &  &  &  & $(  212.888)$ & $(   49.458)$ &  & $(  227.631)$ & $(  110.201)$ \\
~$\beta_{j[bmi_{-jt-1}]}$ & $0$ & $0$ & $-$ &  &  &  &  &  &  &  & $  -23.943$ & $  -35.838$ &  & $ -331.573$ & $   29.395$ \\
 &  &  &  &  &  &  &  &  &  &  & $(  112.992)$ & $(   68.991)$ &  & $(  138.655)$ & $(   85.030)$ \\                      
\multicolumn{16}{l}{\emph{Older shocks} ($t-2$)}\\
~$\beta_{j[w_{jt-2}]}$ & $0$ & $0$ & $-$ &  &  &  &  & $  -32.739$ & $   -9.643$ &  &  &  &  & $  -59.360$ & $   -0.299$ \\
 &  &  &  &  &  &  &  & $(   16.697)$ & $(    6.754)$ &  &  &  &  & $(   17.760)$ & $(    8.106)$ \\
~$\beta_{j[w_{-jt-2}]}$ & $0$ & $0$ & $+$ &  &  &  &  & $    2.417$ & $   24.670$ &  &  &  &  & $   24.988$ & $   20.073$ \\
 &  &  &  &  &  &  &  & $(   28.806)$ & $(   15.157)$ &  &  &  &  & $(   32.180)$ & $(   15.575)$ \\
~$\beta_{j[bmi_{jt-2}]}$ & $0$ & $0$ & $+$ &  &  &  &  &  &  &  &  &  &  & $ -438.600$ & $  139.456$ \\
 &  &  &  &  &  &  &  &  &  &  &  &  &  & $(  247.282)$ & $(  107.369)$ \\
~$\beta_{j[bmi_{-jt-2}]}$ & $0$ & $0$ & $-$ &  &  &  &  &  &  &  &  &  &  & $ -255.479$ & $   40.950$ \\
 &  &  &  &  &  &  &  &  &  &  &  &  &  & $(   85.833)$ & $(   86.437)$ \\                               
\bottomrule
\end{tabular}
\end{center}
\end{sidewaystable}

\begin{sidewaystable}  
\begin{center}
\caption*{Table \ref{AppTable::OLS_Results_All} (continued): Commitment test -- reduced form results in detail}
\begin{tabular}{L{1.9cm} C{0.6cm} C{0.7cm} C{0.6cm} C{0.001cm} C{1.7cm} C{1.7cm} C{0.001cm} C{1.7cm} C{1.7cm} C{0.001cm} C{1.7cm} C{1.6cm} C{0.001cm} C{1.7cm} C{1.7cm}}
\toprule
                            &&&&    & \mcl{5}{c}{\textbf{wage shocks}}        && \mcl{5}{c}{\textbf{wage shocks \& BMI}}               \\
\cmidrule{6-10}\cmidrule{12-16}                        
                            &&&&    & \mcl{2}{c}{$\geq$ 3 periods} && \mcl{2}{c}{$\geq$ 4 periods} && \mcl{2}{c}{$\geq$ 3 periods} && \mcl{2}{c}{$\geq$ 4 periods} \\
\cmidrule{6-7}\cmidrule{9-10}\cmidrule{12-13}\cmidrule{15-16}
                            &&&&    & (1)      & (2)      && (3)    & (4)     && (5)    & (6)       && (7)      & (8)   \\
                            &&&&    & Male     & Female   && Male   & Female  && Male   & Female    && Male     & Female   \\
                &FC&NC&LC&          & $j=1$    &  $j=2$   &&  $j=1$ &  $j=2$  && $j=1$  &  $j=2$    &&  $j=1$   &  $j=2$   \\
\midrule
\multicolumn{16}{l}{\emph{Initial distribution factors} ($t=0$)}\\
~$\beta_{j[young_{j}]}$ & $0$ & $0$ & $-$ &  & $  -28.064$ & $  -14.680$ &  & $  -96.972$ & $  -47.766$ &  & $  -38.991$ & $  -15.300$ &  & $ -119.846$ & $  -43.109$ \\
 &  &  &  &  & $(   26.590)$ & $(    6.370)$ &  & $(   33.136)$ & $(   11.893)$ &  & $(   28.772)$ & $(    7.465)$ &  & $(   36.901)$ & $(   14.170)$ \\
~$\beta_{j[young_{-j}]}$ & $0$ & $0$ & $+$ &  & $  -80.557$ & $   10.376$ &  & $ -128.468$ & $   20.788$ &  & $  -73.021$ & $   10.405$ &  & $  -94.895$ & $   12.586$ \\
 &  &  &  &  & $(   30.226)$ & $(    4.582)$ &  & $(   40.310)$ & $(   10.499)$ &  & $(   30.315)$ & $(    5.645)$ &  & $(   37.749)$ & $(   13.257)$ \\
\multicolumn{16}{l}{\emph{Other controls}}\\
~$ b_{j[\Delta y_{t}]}$ &  &  &  &  & $  250.018$ & $   91.213$ &  & $  320.201$ & $  194.669$ &  & $  245.700$ & $   97.403$ &  & $  420.873$ & $  218.504$ \\
 &  &  &  &  & $(   31.718)$ & $(   32.017)$ &  & $(   49.980)$ & $(   54.148)$ &  & $(   31.770)$ & $(   33.049)$ &  & $(   54.405)$ & $(   49.710)$ \\
~$ b_{j[\Delta a_{t}]}$ &  &  &  &  & $   -0.835$ & $   -4.823$ &  & $   -7.114$ & $   -0.077$ &  & $    1.453$ & $   -2.227$ &  & $  -32.786$ & $    3.751$ \\
 &  &  &  &  & $(   11.700)$ & $(    3.403)$ &  & $(   14.276)$ & $(    5.422)$ &  & $(   12.165)$ & $(    4.582)$ &  & $(   13.840)$ & $(    5.771)$ \\
~$ b_{j[y_{t-1}]}$ &  &  &  &  & $   32.201$ & $    2.047$ &  & $   31.384$ & $    5.729$ &  & $   32.224$ & $    1.359$ &  & $   44.766$ & $    4.844$ \\
 &  &  &  &  & $(    7.304)$ & $(    2.542)$ &  & $(    8.695)$ & $(    6.393)$ &  & $(    6.633)$ & $(    2.557)$ &  & $(    8.238)$ & $(    5.986)$ \\
~$ b_{j[a_{t-1}]}$ &  &  &  &  & $  -33.126$ & $   -2.523$ &  & $  -30.302$ & $   -6.639$ &  & $  -33.096$ & $   -2.115$ &  & $  -43.523$ & $   -6.159$ \\
 &  &  &  &  & $(    7.355)$ & $(    2.247)$ &  & $(    8.509)$ & $(    5.644)$ &  & $(    6.459)$ & $(    2.264)$ &  & $(    7.730)$ & $(    5.425)$ \\
~$ b_{j[\Delta y_{-jt}]}$ &  &  &  &  & $ -186.371$ & $  -85.837$ &  & $ -275.461$ & $ -166.237$ &  & $ -171.123$ & $  -98.992$ &  & $ -334.140$ & $ -197.951$ \\
 &  &  &  &  & $(   40.079)$ & $(   29.805)$ &  & $(   65.870)$ & $(   44.416)$ &  & $(   37.056)$ & $(   32.471)$ &  & $(   63.623)$ & $(   40.809)$ \\
~$ b_{j[\Delta q_{t}]}$ &  &  &  &  & $    0.001$ & $   -0.000$ &  & $    0.001$ & $    0.000$ &  & $    0.000$ & $   -0.000$ &  & $    0.001$ & $   -0.000$ \\
 &  &  &  &  & $(    0.001)$ & $(    0.000)$ &  & $(    0.001)$ & $(    0.000)$ &  & $(    0.001)$ & $(    0.000)$ &  & $(    0.001)$ & $(    0.000)$ \\
~$ b_{j[\Delta a_{t-1}]}$ &  &  &  &  & $   -2.160$ & $   -3.104$ &  & $   -4.693$ & $    0.289$ &  & $   -1.809$ & $   -3.168$ &  & $  -10.137$ & $   -0.424$ \\
 &  &  &  &  & $(    3.933)$ & $(    1.532)$ &  & $(    5.023)$ & $(    2.285)$ &  & $(    4.057)$ & $(    1.222)$ &  & $(    4.754)$ & $(    2.487)$ \\
~$ b_{j[\Delta a_{t-2}]}$ &  &  &  &  &  &  &  & $   -3.201$ & $    1.998$ &  &  &  &  & $   -2.619$ & $    2.839$ \\
 &  &  &  &  &  &  &  & $(    4.008)$ & $(    1.966)$ &  &  &  &  & $(    3.870)$ & $(    2.632)$ \\ 
\noalign{\smallskip}
\mcl{5}{l}{~\text{$ p$ value for }$ {\cal H}_{0}^\text{FC}$} & $    0.000$ & $    0.000$ &  & $    0.001$ & $    0.000$ &  & $    0.012$ & $    0.000$ &  & $    0.000$ & $    0.000$ \\
\mcl{5}{l}{~\text{$ p$ value for }$ {\cal H}_{0}^\text{NC}$} & $    0.001$ & $    0.000$ &  & $    0.001$ & $    0.000$ &  & $    0.010$ & $    0.001$ &  & $    0.000$ & $    0.000$ \\
\noalign{\smallskip}
\mcl{5}{l}{Observations}                       & \mcl{2}{c}{8,513}    && \mcl{2}{c}{6,028} && \mcl{2}{c}{7,616}  &&  \mcl{2}{c}{5,294} \\
\bottomrule 
\end{tabular}
\caption*{\fsz\emph{Notes:} The table reports the coefficient estimates from the reduced form (first) specification of the commitment test with wages (columns 1-4) and wages and BMI (columns 5-8) as time-varying distribution factors. Standard errors clustered at the household level are in brackets.}
\end{center}
\end{sidewaystable}

\begin{sidewaystable}  
\begin{center}
\caption{Commitment test -- structural results in detail}\label{AppTable::Structural_Results_All}
\begin{tabular}{L{1.2cm} C{0.6cm} C{0.7cm} C{0.6cm} C{0.01cm} C{1.5cm} C{1.6cm} C{0.01cm} C{1.5cm} C{1.6cm} C{0.01cm} C{1.7cm} C{1.7cm} C{0.01cm} C{1.7cm} C{1.7cm}}
\toprule
                            &&&&    & \mcl{5}{c}{\textbf{wage shocks}}        && \mcl{5}{c}{\textbf{wage shocks \& BMI}}               \\
\cmidrule{6-10}\cmidrule{12-16}                        
                            &&&&    & \mcl{2}{c}{$\geq$ 3 periods} && \mcl{2}{c}{$\geq$ 4 periods} && \mcl{2}{c}{$\geq$ 3 periods} && \mcl{2}{c}{$\geq$ 4 periods} \\
\cmidrule{6-7}\cmidrule{9-10}\cmidrule{12-13}\cmidrule{15-16}
                            &&&&    & (1)      & (2)      && (3)    & (4)     && (5)    & (6)       && (7)      & (8)   \\
                            &&&&    & Male     & Female   && Male   & Female  && Male   & Female    && Male     & Female   \\
                &FC&NC&LC&          & $j=1$    &  $j=2$   &&  $j=1$ &  $j=2$  && $j=1$  &  $j=2$    &&  $j=1$   &  $j=2$   \\
\midrule
\multicolumn{16}{l}{\emph{Pareto weight elasticities w.r.t current factors} ($\tau=0$)}\\
~$\eta_{j0}^{w_{j}}$ & $0$ & $+$ & $+$ &  & $    1.020$ & $    1.007$ &  & $    1.004$ & $    1.006$ &  & $    1.018$ & $    1.007$ &  & $    1.012$ & $    1.000$ \\
 &  &  &  &  & $(    0.010)$ & $(    0.010)$ &  & $(    0.011)$ & $(    0.012)$ &  & $(    0.009)$ & $(    0.008)$ &  & $(    0.010)$ & $(    0.011)$ \\
~$\eta_{j0}^{w_{-j}}$ & $0$ & $-$ & $-$ &  & $   -0.034$ & $   -0.003$ &  & $   -0.026$ & $   -0.006$ &  & $   -0.036$ & $   -0.019$ &  & $   -0.026$ & $   -0.013$ \\
 &  &  &  &  & $(    0.022)$ & $(    0.012)$ &  & $(    0.017)$ & $(    0.018)$ &  & $(    0.023)$ & $(    0.016)$ &  & $(    0.015)$ & $(    0.018)$ \\
~$\eta_{j0}^{bmi_{j}}$ & $0$ & $-$ & $-$ &  &  &  &  &  &  &  & $   -0.043$ & $   -0.059$ &  & $   -0.104$ & $   -0.073$ \\
 &  &  &  &  &  &  &  &  &  &  & $(    0.214)$ & $(    0.091)$ &  & $(    0.140)$ & $(    0.107)$ \\
~$\eta_{j0}^{bmi_{-j}}$ & $0$ & $+$ & $+$ &  &  &  &  &  &  &  & $    0.050$ & $   -0.016$ &  & $   -0.130$ & $   -0.039$ \\
 &  &  &  &  &  &  &  &  &  &  & $(    0.147)$ & $(    0.085)$ &  & $(    0.100)$ & $(    0.070)$ \\
\multicolumn{16}{l}{\emph{Pareto weight elasticities w.r.t factors 1 period in the past} ($\tau=1$)}\\
~$\eta_{j1}^{w_{j}}$ & $0$ & $0$ & $+$ &  & $    0.012$ & $    0.009$ &  & $    0.010$ & $    0.002$ &  & $    0.012$ & $    0.011$ &  & $    0.005$ & $    0.002$ \\
 &  &  &  &  & $(    0.006)$ & $(    0.008)$ &  & $(    0.006)$ & $(    0.010)$ &  & $(    0.007)$ & $(    0.008)$ &  & $(    0.006)$ & $(    0.012)$ \\
~$\eta_{j1}^{w_{-j}}$ & $0$ & $0$ & $-$ &  & $   -0.032$ & $    0.000$ &  & $   -0.031$ & $    0.002$ &  & $   -0.028$ & $   -0.000$ &  & $   -0.033$ & $    0.001$ \\
 &  &  &  &  & $(    0.022)$ & $(    0.019)$ &  & $(    0.020)$ & $(    0.015)$ &  & $(    0.018)$ & $(    0.017)$ &  & $(    0.015)$ & $(    0.014)$ \\
~$\eta_{j1}^{bmi_{j}}$ & $0$ & $0$ & $-$ &  &  &  &  &  &  &  & $   -0.271$ & $   -0.174$ &  & $   -0.077$ & $   -0.150$ \\
 &  &  &  &  &  &  &  &  &  &  & $(    0.201)$ & $(    0.104)$ &  & $(    0.107)$ & $(    0.114)$ \\
~$\eta_{j1}^{bmi_{-j}}$ & $0$ & $0$ & $+$ &  &  &  &  &  &  &  & $    0.025$ & $    0.051$ &  & $    0.159$ & $    0.062$ \\
 &  &  &  &  &  &  &  &  &  &  & $(    0.108)$ & $(    0.104)$ &  & $(    0.062)$ & $(    0.082)$ \\
\multicolumn{16}{l}{\emph{Pareto weight elasticities w.r.t factors 2 periods in the past} ($\tau=2$)}\\  
~$\eta_{j2}^{w_{j}}$ & $0$ & $0$ & $+$ &  &  &  &  & $    0.010$ & $    0.009$ &  &  &  &  & $    0.019$ & $    0.004$ \\
 &  &  &  &  &  &  &  & $(    0.012)$ & $(    0.005)$ &  &  &  &  & $(    0.012)$ & $(    0.008)$ \\
~$\eta_{j2}^{w_{-j}}$ & $0$ & $0$ & $-$ &  &  &  &  & $   -0.023$ & $   -0.026$ &  &  &  &  & $   -0.029$ & $   -0.022$ \\
 &  &  &  &  &  &  &  & $(    0.018)$ & $(    0.016)$ &  &  &  &  & $(    0.017)$ & $(    0.015)$ \\
~$\eta_{j2}^{bmi_{j}}$ & $0$ & $0$ & $-$ &  &  &  &  &  &  &  &  &  &  & $    0.170$ & $   -0.084$ \\
 &  &  &  &  &  &  &  &  &  &  &  &  &  & $(    0.115)$ & $(    0.124)$ \\
~$\eta_{j2}^{bmi_{-j}}$ & $0$ & $0$ & $+$ &  &  &  &  &  &  &  &  &  &  & $    0.095$ & $    0.126$ \\
 &  &  &  &  &  &  &  &  &  &  &  &  &  & $(    0.042)$ & $(    0.106)$ \\                         
\bottomrule
\end{tabular}
\end{center}
\end{sidewaystable}

\begin{sidewaystable}  
\begin{center}
\caption*{Table \ref{AppTable::Structural_Results_All} (continued): Commitment test -- structural results in detail}
\begin{tabular}{L{1.5cm} C{0.6cm} C{0.7cm} C{0.6cm} C{0.001cm} C{1.6cm} C{1.6cm} C{0.001cm} C{1.6cm} C{1.6cm} C{0.001cm} C{1.6cm} C{1.6cm} C{0.001cm} C{1.6cm} C{1.6cm}}
\toprule
                            &&&&    & \mcl{5}{c}{\textbf{wage shocks}}        && \mcl{5}{c}{\textbf{wage shocks \& BMI}}               \\
\cmidrule{6-10}\cmidrule{12-16}                        
                            &&&&    & \mcl{2}{c}{$\geq$ 3 periods} && \mcl{2}{c}{$\geq$ 4 periods} && \mcl{2}{c}{$\geq$ 3 periods} && \mcl{2}{c}{$\geq$ 4 periods} \\
\cmidrule{6-7}\cmidrule{9-10}\cmidrule{12-13}\cmidrule{15-16}
                            &&&&    & (1)      & (2)      && (3)    & (4)     && (5)    & (6)       && (7)      & (8)   \\
                            &&&&    & Male     & Female   && Male   & Female  && Male   & Female    && Male     & Female   \\
                &FC&NC&LC&          & $j=1$    &  $j=2$   &&  $j=1$ &  $j=2$  && $j=1$  &  $j=2$    &&  $j=1$   &  $j=2$   \\
\midrule
\multicolumn{16}{l}{\emph{Pareto weight elasticities w.r.t initial distribution factors}}\\
~$\eta_{jt}^{young_{j}}$ & $0$ & $0$ & $+$ &  & $    0.046$ & $    0.025$ &  & $    0.062$ & $    0.038$ &  & $    0.060$ & $    0.023$ &  & $    0.064$ & $    0.037$ \\
 &  &  &  &  & $(    0.024)$ & $(    0.011)$ &  & $(    0.024)$ & $(    0.014)$ &  & $(    0.028)$ & $(    0.013)$ &  & $(    0.023)$ & $(    0.012)$ \\
~$\eta_{jt}^{young_{-j}}$ & $0$ & $0$ & $-$ &  & $    0.096$ & $   -0.023$ &  & $    0.077$ & $   -0.021$ &  & $    0.086$ & $   -0.019$ &  & $    0.053$ & $   -0.017$ \\
 &  &  &  &  & $(    0.035)$ & $(    0.010)$ &  & $(    0.037)$ & $(    0.009)$ &  & $(    0.033)$ & $(    0.009)$ &  & $(    0.024)$ & $(    0.010)$ \\
\mcl{5}{l}{\emph{Frisch elasticity} ${\alpha_{j}/\mathbb{E}(h_{jt-1})}^{\#}$}   & $    0.545$ & $    0.330$ &  & $    1.017$ & $    0.719$ &  & $    0.536$ & $    0.382$ &  & $    1.270$ & $    0.817$ \\
 &  &  &  &  & $(    0.164)$ & $(    0.136)$ &  & $(    0.347)$ & $(    0.263)$ &  & $(    0.143)$ & $(    0.150)$ &  & $(    0.372)$ & $(    0.267)$ \\
\multicolumn{16}{l}{\emph{Other terms}}\\
~$\ell_{\Delta y}$ &  &  &  &  & $    0.262$ & $    0.190$ &  & $    0.218$ & $    0.205$ &  & $    0.264$ & $    0.175$ &  & $    0.227$ & $    0.196$ \\
 &  &  &  &  & $(    0.044)$ & $(    0.018)$ &  & $(    0.029)$ & $(    0.013)$ &  & $(    0.041)$ & $(    0.019)$ &  & $(    0.025)$ & $(    0.014)$ \\
\mcl{5}{l}{~$\ell_{\Delta a}-\eta_{j0}^{a}$} & $   -0.002$ & $   -0.012$ &  & $   -0.009$ & $   -0.003$ &  & $    0.000$ & $   -0.008$ &  & $   -0.017$ & $   -0.002$ \\
 &  &  &  &  & $(    0.010)$ & $(    0.009)$ &  & $(    0.007)$ & $(    0.006)$ &  & $(    0.011)$ & $(    0.009)$ &  & $(    0.007)$ & $(    0.007)$ \\
~$\ell_{y}$ &  &  &  &  & $    0.033$ & $    0.007$ &  & $    0.018$ & $    0.005$ &  & $    0.031$ & $    0.005$ &  & $    0.020$ & $    0.005$ \\
 &  &  &  &  & $(    0.010)$ & $(    0.007)$ &  & $(    0.006)$ & $(    0.008)$ &  & $(    0.008)$ & $(    0.006)$ &  & $(    0.006)$ & $(    0.007)$ \\
~$\ell_{a}$ &  &  &  &  & $   -0.033$ & $   -0.008$ &  & $   -0.017$ & $   -0.006$ &  & $   -0.032$ & $   -0.006$ &  & $   -0.020$ & $   -0.006$ \\
 &  &  &  &  & $(    0.010)$ & $(    0.006)$ &  & $(    0.006)$ & $(    0.007)$ &  & $(    0.008)$ & $(    0.005)$ &  & $(    0.006)$ & $(    0.006)$ \\
~$ {\zeta_{j}}^{\#\#}$ &  &  &  &  & $   -0.006$ & $    0.001$ &  & $   -0.006$ & $   -0.001$ &  & $   -0.006$ & $    0.002$ &  & $   -0.005$ & $    0.001$ \\
 &  &  &  &  & $(    0.005)$ & $(    0.002)$ &  & $(    0.004)$ & $(    0.002)$ &  & $(    0.005)$ & $(    0.002)$ &  & $(    0.004)$ & $(    0.002)$ \\
~$\eta_{j1}^{a}$ &  &  &  &  & $   -0.003$ & $   -0.005$ &  & $   -0.005$ & $    0.001$ &  & $   -0.002$ & $   -0.005$ &  & $   -0.006$ & $    0.001$ \\
 &  &  &  &  & $(    0.004)$ & $(    0.003)$ &  & $(    0.003)$ & $(    0.002)$ &  & $(    0.004)$ & $(    0.002)$ &  & $(    0.002)$ & $(    0.002)$ \\
~$\eta_{j2}^{a}$ &  &  &  &  &  &  &  & $   -0.004$ & $    0.003$ &  &  &  &  & $   -0.003$ & $    0.002$ \\
 &  &  &  &  &  &  &  & $(    0.002)$ & $(    0.002)$ &  &  &  &  & $(    0.002)$ & $(    0.002)$ \\
\noalign{\smallskip}
\mcl{5}{l}{~\text{$ p$ value for }$ {\cal H}_{0}^\text{FC}$} & $    0.000$ & $    0.000$ &  & $    0.000$ & $    0.000$ &  & $    0.000$ & $    0.000$ &  & $    0.000$ & $    0.000$ \\
\mcl{5}{l}{~\text{$ p$ value for }$ {\cal H}_{0}^\text{NC}$} & $    0.025$ & $    0.032$ &  & $    0.052$ & $    0.049$ &  & $    0.025$ & $    0.013$ &  & $    0.027$ & $    0.008$ \\
\noalign{\smallskip}
\mcl{5}{l}{Observations}                       & \mcl{2}{c}{8,513}    && \mcl{2}{c}{6,028} && \mcl{2}{c}{7,616}  &&  \mcl{2}{c}{5,294} \\
\bottomrule 
\end{tabular}
\caption*{\fsz\emph{Notes:} The table reports the parameter estimates from the structural (second) specification of the commitment test with wages (columns 1-4) and wages and BMI (columns 5-8) as time-varying distribution factors. Standard errors clustered at the household level are in brackets. See appendix \ref{Appendix::EstimatingEquation} for details on the parameters. $\ell_{\Delta y}$, $\ell_{y}$, $\ell_{a}$ are common across male and female equations in each specification. For simplicity, we estimate the equations separately without imposing cross-equation restrictions. We cannot reject equality of these parameters across equations in most specifications.\\$^\#$We report the Frisch elasticity at the sample average of hours of work; its standard error is calculated with the delta method.\\$^{\#\#}$We multiply $\zeta_{j}$ by $10^4$ for legibility ($\zeta_{j}$ originally multiplies the \emph{level} of consumption so its magnitude is very small).}
\end{center}
\end{sidewaystable}

\begin{sidewaystable}  
\begin{center}
\caption{Commitment test -- structural results with log-linearized Pareto weight}\label{AppTable::DeepStructure_Results_All}
\begin{tabular}{L{1.7cm} C{0.6cm} C{0.7cm} C{0.6cm} C{0.01cm} C{1.5cm} C{1.6cm} C{0.01cm} C{1.5cm} C{1.6cm} C{0.01cm} C{1.7cm} C{1.7cm} C{0.01cm} C{1.7cm} C{1.7cm}}
\toprule
                            &&&&    & \mcl{5}{c}{\textbf{wage shocks}}        && \mcl{5}{c}{\textbf{wage shocks \& BMI}}               \\
\cmidrule{6-10}\cmidrule{12-16}                        
                            &&&&    & \mcl{2}{c}{$\geq$ 3 periods} && \mcl{2}{c}{$\geq$ 4 periods} && \mcl{2}{c}{$\geq$ 3 periods} && \mcl{2}{c}{$\geq$ 4 periods} \\
\cmidrule{6-7}\cmidrule{9-10}\cmidrule{12-13}\cmidrule{15-16}
                            &&&&    & (1)      & (2)      && (3)    & (4)     && (5)    & (6)       && (7)      & (8)   \\
                            &&&&    & Male     & Female   && Male   & Female  && Male   & Female    && Male     & Female   \\
                &FC&NC&LC&          & $j=1$    &  $j=2$   &&  $j=1$ &  $j=2$  && $j=1$  &  $j=2$    &&  $j=1$   &  $j=2$   \\
\midrule
\multicolumn{16}{l}{\emph{Pareto weight elasticities w.r.t distribution factors}}\\
~$ e_{\mu_{j},w_{j}}$ & $0$ & $+$ & $+$ &  & $    1.016$ & $    1.007$ &  & $    0.997$ & $    0.997$ &  & $    1.012$ & $    1.010$ &  & $    0.995$ & $    1.003$ \\
 &  &  &  &  & $(    0.010)$ & $(    0.009)$ &  & $(    0.004)$ & $(    0.006)$ &  & $(    0.007)$ & $(    0.010)$ &  & $(    0.004)$ & $(    0.005)$ \\
~$ e_{\mu_{j},w_{-j}}$ & $0$ & $-$ & $-$ &  & $   -0.030$ & $   -0.003$ &  & $   -0.022$ & $   -0.007$ &  & $   -0.026$ & $    0.003$ &  & $   -0.024$ & $    0.001$ \\
 &  &  &  &  & $(    0.020)$ & $(    0.012)$ &  & $(    0.014)$ & $(    0.018)$ &  & $(    0.018)$ & $(    0.016)$ &  & $(    0.011)$ & $(    0.018)$ \\
~$ e_{\mu_{j},bmi_{j}}$ & $0$ & $-$ & $-$ &  &  &  &  &  &  &  & $    0.017$ & $   -0.000$ &  & $    0.027^a$ & $   -0.076$ \\
 &  &  &  &  &  &  &  &  &  &  & $(    0.189)$ & $(    0.101)$ &  & $(    0.080)$ & $(    0.056)$ \\
~$ e_{\mu_{j},bmi_{-j}}$ & $0$ & $+$ & $+$ &  &  &  &  &  &  &  & $    0.094$ & $    0.061$ &  & $   -0.212^a$ & $   -0.031$ \\
 &  &  &  &  &  &  &  &  &  &  & $(    0.142)$ & $(    0.089)$ &  & $(    0.076)$ & $(    0.055)$ \\
\multicolumn{16}{l}{\emph{Pareto weight elasticity w.r.t past Pareto weight}}\\ 
~$ e_{\mu_{j},\mu_{jL}}$ & $0$ & $0$ & $+$ &  & $    0.015$ & $    0.009$ &  & $    0.001$ & $   -0.008$ &  & $    0.015$ & $    0.015$ &  & $   -0.005^a$ & $   -0.001$ \\
 &  &  &  &  & $(    0.007)$ & $(    0.008)$ &  & $(    0.002)$ & $(    0.006)$ &  & $(    0.007)$ & $(    0.010)$ &  & $(    0.003)$ & $(    0.007)$ \\
\multicolumn{16}{l}{\emph{Pareto weight elasticities w.r.t initial distribution factors}}\\
~$\eta_{jt}^{young_{j}}$ & $0$ & $0$ & $+$ &  & $    0.044$ & $    0.025$ &  & $    0.057$ & $    0.040$ &  & $    0.057$ & $    0.030$ &  & $    0.061$ & $    0.037$ \\
 &  &  &  &  & $(    0.024)$ & $(    0.011)$ &  & $(    0.019)$ & $(    0.012)$ &  & $(    0.027)$ & $(    0.016)$ &  & $(    0.019)$ & $(    0.013)$ \\
~$\eta_{jt}^{young_{-j}}$ & $0$ & $0$ & $-$ &  & $    0.088$ & $   -0.023$ &  & $    0.058$ & $   -0.012$ &  & $    0.102$ & $   -0.011$ &  & $    0.072$ & $   -0.007$ \\
 &  &  &  &  & $(    0.035)$ & $(    0.010)$ &  & $(    0.028)$ & $(    0.007)$ &  & $(    0.037)$ & $(    0.010)$ &  & $(    0.026)$ & $(    0.007)$ \\
\mcl{5}{l}{\emph{Frisch elasticity} ${\alpha_{j}/\mathbb{E}(h_{jt-1})}^{\#}$}  & $    0.584$ & $    0.329$ &  & $    1.239$ & $    0.668$ &  & $    0.526$ & $    0.319$ &  & $    0.998$ & $    0.770$ \\
 &  &  &  &  & $(    0.178)$ & $(    0.133)$ &  & $(    0.371)$ & $(    0.214)$ &  & $(    0.128)$ & $(    0.143)$ &  & $(    0.262)$ & $(    0.207)$ \\
\bottomrule
\end{tabular}
\end{center}
\end{sidewaystable}

\begin{sidewaystable}  
\begin{center}
\caption*{Table \ref{AppTable::DeepStructure_Results_All} (continued): Commitment test -- structural results with log-linearized Pareto weight}
\begin{tabular}{L{1.5cm} C{0.6cm} C{0.7cm} C{0.6cm} C{0.001cm} C{1.6cm} C{1.6cm} C{0.001cm} C{1.6cm} C{1.6cm} C{0.001cm} C{1.6cm} C{1.6cm} C{0.001cm} C{1.6cm} C{1.6cm}}
\toprule
                            &&&&    & \mcl{5}{c}{\textbf{wage shocks}}        && \mcl{5}{c}{\textbf{wage shocks \& BMI}}               \\
\cmidrule{6-10}\cmidrule{12-16}                        
                            &&&&    & \mcl{2}{c}{$\geq$ 3 periods} && \mcl{2}{c}{$\geq$ 4 periods} && \mcl{2}{c}{$\geq$ 3 periods} && \mcl{2}{c}{$\geq$ 4 periods} \\
\cmidrule{6-7}\cmidrule{9-10}\cmidrule{12-13}\cmidrule{15-16}
                            &&&&    & (1)      & (2)      && (3)    & (4)     && (5)    & (6)       && (7)      & (8)   \\
                            &&&&    & Male     & Female   && Male   & Female  && Male   & Female    && Male     & Female   \\
                &FC&NC&LC&          & $j=1$    &  $j=2$   &&  $j=1$ &  $j=2$  && $j=1$  &  $j=2$    &&  $j=1$   &  $j=2$   \\
\midrule
\multicolumn{16}{l}{\emph{Other terms}}\\
~$\ell_{\Delta y}$ &  &  &  &  & $    0.253$ & $    0.190$ &  & $    0.202$ & $    0.193$ &  & $    0.250$ & $    0.191$ &  & $    0.211$ & $    0.198$ \\
 &  &  &  &  & $(    0.038)$ & $(    0.018)$ &  & $(    0.019)$ & $(    0.014)$ &  & $(    0.034)$ & $(    0.019)$ &  & $(    0.016)$ & $(    0.012)$ \\
\mcl{5}{l}{~$\ell_{\Delta a}-e_{\mu_{j},a}$} & $    0.000$ & $   -0.012$ &  & $   -0.007$ & $   -0.002$ &  & $   -0.002$ & $   -0.016$ &  & $   -0.012$ & $    0.002$ \\
 &  &  &  &  & $(    0.010)$ & $(    0.009)$ &  & $(    0.007)$ & $(    0.005)$ &  & $(    0.011)$ & $(    0.012)$ &  & $(    0.007)$ & $(    0.005)$ \\
~$\ell_{y}$ &  &  &  &  & $    0.030$ & $    0.007$ &  & $    0.015$ & $    0.005$ &  & $    0.031$ & $    0.008$ &  & $    0.012$ & $    0.003$ \\
 &  &  &  &  & $(    0.009)$ & $(    0.007)$ &  & $(    0.004)$ & $(    0.006)$ &  & $(    0.007)$ & $(    0.008)$ &  & $(    0.004)$ & $(    0.006)$ \\
~$\ell_{a}$ &  &  &  &  & $   -0.030$ & $   -0.008$ &  & $   -0.014$ & $   -0.005$ &  & $   -0.030$ & $   -0.008$ &  & $   -0.011$ & $   -0.004$ \\
 &  &  &  &  & $(    0.009)$ & $(    0.006)$ &  & $(    0.004)$ & $(    0.006)$ &  & $(    0.007)$ & $(    0.007)$ &  & $(    0.004)$ & $(    0.005)$ \\
~$ {\zeta_{j}}^{\#\#}$ &  &  &  &  & $   -0.008$ & $    0.001$ &  & $   -0.009$ & $    0.001$ &  & $   -0.008$ & $   -0.000$ &  & $   -0.009$ & $    0.001$ \\
 &  &  &  &  & $(    0.004)$ & $(    0.002)$ &  & $(    0.003)$ & $(    0.002)$ &  & $(    0.005)$ & $(    0.002)$ &  & $(    0.003)$ & $(    0.001)$ \\
~$\eta_{j1}^{a}$ &  &  &  &  & $   -0.002$ & $   -0.005$ &  & $   -0.005$ & $    0.001$ &  & $   -0.001$ & $   -0.005$ &  & $   -0.007$ & $   -0.000$ \\
 &  &  &  &  & $(    0.003)$ & $(    0.003)$ &  & $(    0.002)$ & $(    0.002)$ &  & $(    0.004)$ & $(    0.003)$ &  & $(    0.003)$ & $(    0.002)$ \\
~$\eta_{j2}^{a}$ &  &  &  &  &  &  &  & $   -0.002$ & $    0.004$ &  &  &  &  & $   -0.004$ & $    0.003$ \\
 &  &  &  &  &  &  &  & $(    0.002)$ & $(    0.002)$ &  &  &  &  & $(    0.002)$ & $(    0.002)$ \\
\noalign{\smallskip}
\mcl{5}{l}{~\text{$ p$ value for }$ {\cal H}_{0}^\text{FC}$} & $    0.000$ & $    0.000$ &  & $    0.000$ & $    0.000$ &  & $    0.000$ & $    0.000$ &  & $    0.000$ & $    0.000$ \\
\mcl{5}{l}{~\text{$ p$ value for }$ {\cal H}_{0}^\text{NC}$} & $    0.016$ & $    0.015$ &  & $    0.007$ & $    0.012$ &  & $    0.003$ & $    0.015$ &  & $    0.002$ & $    0.004$ \\
\noalign{\smallskip}
\mcl{5}{l}{Observations}                       & \mcl{2}{c}{8,513}    && \mcl{2}{c}{6,028} && \mcl{2}{c}{7,616}  &&  \mcl{2}{c}{5,294} \\
\bottomrule 
\end{tabular}
\caption*{\fsz\emph{Notes:} The table reports the parameter estimates from the structural specification of the commitment test with wages (columns 1-4) and wages and BMI (columns 5-8) as time-varying distribution factors, letting $\Delta \log \mu_{jt}$ take the log-linear form in \eqref{Eq::Definition.ApproxStructuralParetoWeight_Final}. Standard errors clustered at the household level are in brackets. See appendix \ref{Appendix::EstimatingEquation} for details on the parameters. We estimate each of $\eta_{jt}^{young_{j}} = (e_{\mu_{j},\mu_{jL}})^t e_{\mu_{j},young_{j}}$ and $\eta_{jt}^{young_{-j}} = (e_{\mu_{j},\mu_{jL}})^t e_{\mu_{j},young_{-j}}$ as a single parameter because we do not consistently observe the length of marriage $t$. For comparability across specifications, we also estimate each of $\eta_{j1}^{a} = e_{\mu_{j},\mu_{jL}} e_{\mu_{j},a}$ and $\eta_{j2}^{a} = (e_{\mu_{j},\mu_{jL}})^2 e_{\mu_{j},a}$ as a single parameter because $e_{\mu_{j},a}$, the Pareto weight elasticity with respect to assets, is only identified with four periods or more. $\ell_{\Delta y}$, $\ell_{y}$, $\ell_{a}$ are common across male and female equations in each specification. For simplicity, we estimate the equations separately without imposing cross-equation restrictions. We cannot reject equality of these parameters across equations in most specifications.\\
$^\#$We report the Frisch elasticity at the sample average of hours of work; its standard error is calculated with the delta method.\\
$^{\#\#}$We multiply $\zeta_{j}$ by $10^4$ for legibility ($\zeta_{j}$ originally multiplies the \emph{level} of consumption so its magnitude is very small).\\
$^a$In column 7, we obtain a marginally significantly negative $e_{\mu_{j},\mu_{jL}}$, which is not consistent with theory. However, the elasticities with respect to BMI have the opposite sign than expected, so the overall partial effect of past BMI is in line with theory (the partial effect is given by $(e_{\mu_{j},\mu_{jL}})^\tau e_{\mu_{j},bmi_{k}}$ for $\tau\in\{1,2\}$). We attribute $e_{\mu_{j},\mu_{jL}}<0$ to a local minimum which our optimizer on Stata is unable to overcome. }
\end{center}
\end{sidewaystable}
\FloatBarrier

\begin{figure}[]
    \centering
    \caption{Partial effects of wage shocks}\label{AppFigure::PartialEffects}
    \begin{subfigure}[t]{0.99\textwidth}
        \centering
        \includegraphics[width=\textwidth]{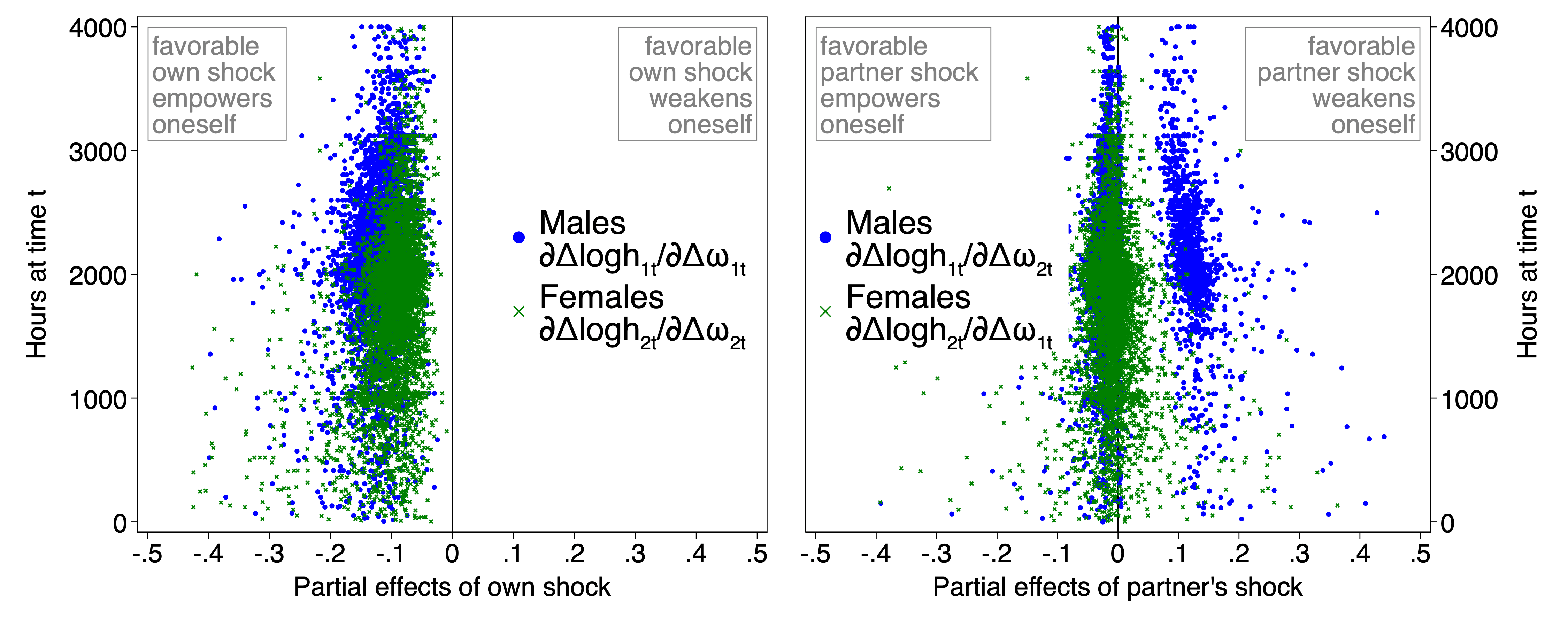}
        \caption{$\partial \Delta \log h_{jt} / \partial \omega_{kt}$, partial effects of current shocks}\label{Figure::PartialEffects_a}
    \end{subfigure}\\[4pt]
    \begin{subfigure}[t]{0.99\textwidth}
        \centering
        \includegraphics[width=\textwidth]{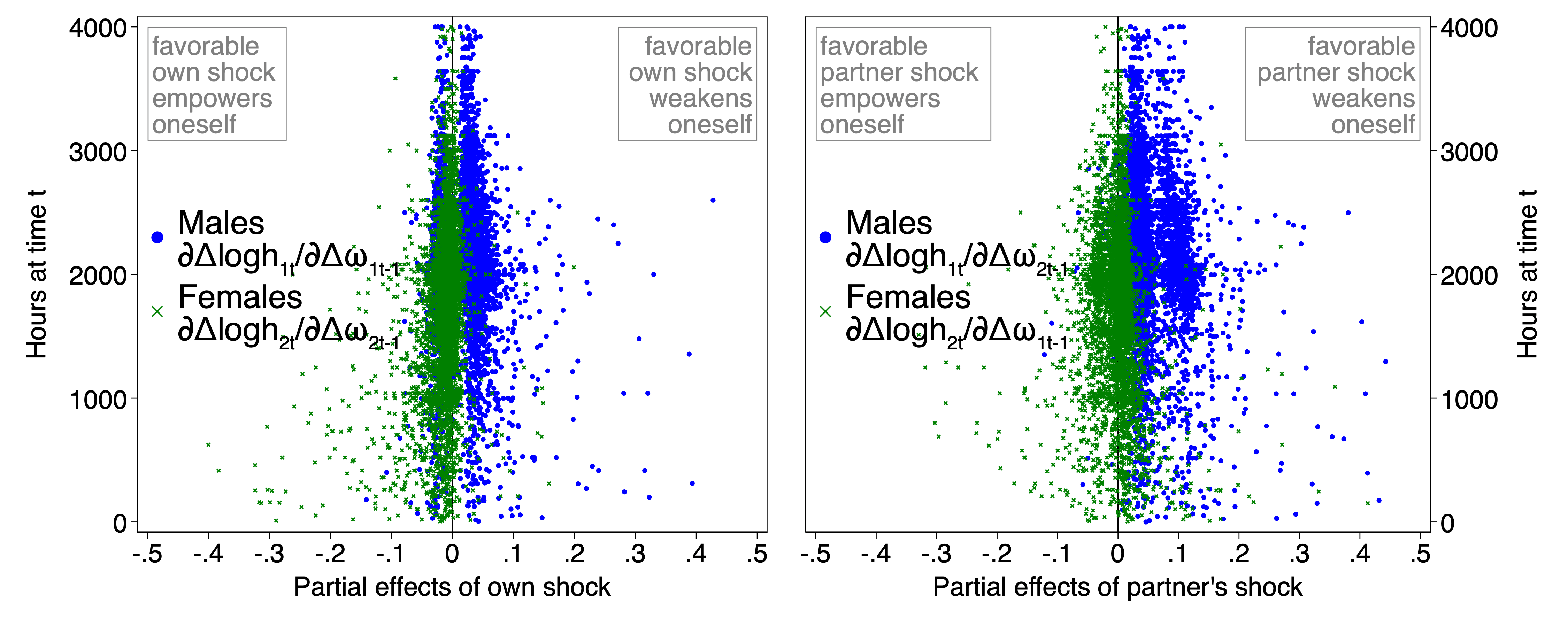}
        \caption{$\partial \Delta \log h_{jt} / \partial \omega_{kt-1}$, partial effects of immediately past shocks}\label{Figure::PartialEffects_b}
    \end{subfigure}\\[4pt]
    \begin{subfigure}[t]{0.99\textwidth}
        \centering
        \includegraphics[width=\textwidth]{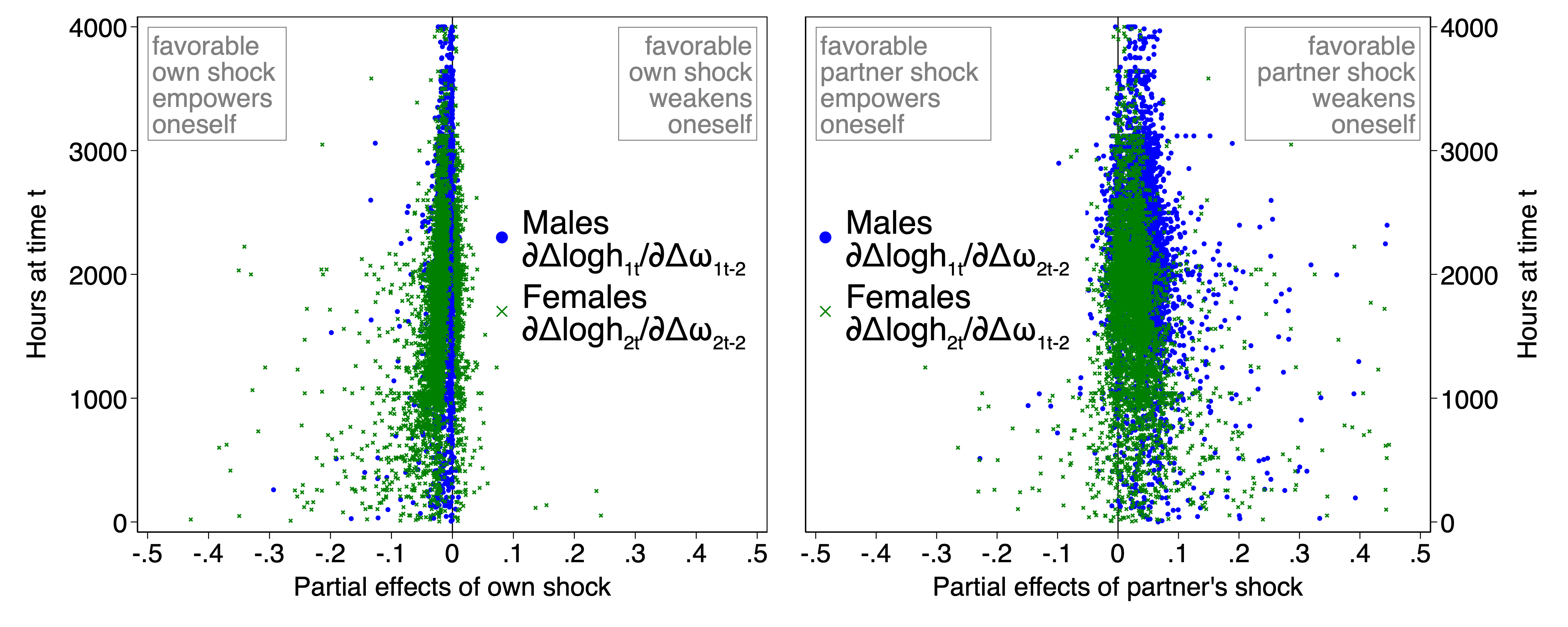}
        \caption{$\partial \Delta \log h_{jt} / \partial \omega_{kt-2}$, partial effects of older shocks}\label{Figure::PartialEffects_c}
    \end{subfigure}%
    \caption*{\fsz\emph{Notes:}
    The figure plots the partial derivative of hours w.r.t wage shocks, $\partial \Delta \log h_{jt} / \partial \omega_{kt-\tau}$, $j,k\in\{1,2\}$, $\tau\in\{0,1,2\}$, across couples observed for $\geq4$ periods. The parameter estimates are in appendix table \ref{AppTable::Heterogeneity_Results_All}.}
\end{figure}

\noindent  \textbf{Heterogeneity.} Our third specification allows the Pareto weight elasticities $\eta_{j}$ to depend on the \emph{past levels} of distribution factors.\footnote{We specify the coefficients $\eta_{j0}$ on current (time $t$) wages as:
\begin{alignat*}{3}
    \eta_{j0}^{w_j}     =& \eta_{j0}^{w_j\{0\}} &+& \eta_{j0}^{w_j\{1\}}w_{jt-1} &+& \eta_{j0}^{w_j\{2\}}w_{-jt-1}, \\
    \eta_{j0}^{w_{-j}}  =& \eta_{j0}^{w_{-j}\{0\}} &+& \eta_{j0}^{w_{-j}\{1\}}w_{jt-1} &+& \eta_{j0}^{w_{-j}\{2\}}w_{-jt-1}. 
\end{alignat*}
We do similarly for $\eta_{j1}$ (as function of $t-2$ wages) and $\eta_{j2}$ (as function of $t-3$ wages). When BMI is an additional distribution factor, we introduce past $Z_{t-\tau} = \{BMI_{1t-\tau}, BMI_{2t-\tau}\}$, $\tau\in\{1,2,3\}$, linearly, e.g. $\eta_{j0}^{w_j} = \eta_{j0}^{w_j\{0\}} + \dots + \eta_{j0}^{w_j\{3\}}BMI_{jt-1} + \eta_{j0}^{w_j\{4\}}BMI_{-jt-1}$, etc. We also include the age-gap-at-marriage dummies, because the bargaining effects of distribution factors depend through $e_{\mu_j,\mu_{jL}}$ on the past Pareto weight and, therefore, recursively on the variables in $\Theta_{0}$ (appendix \ref{Appendix::ApproximationParetoWeight}).} Along with the dependence of $\delta_{jt}$ on earnings and hours (structural specification), the third specification introduces vast amounts of cross-household heterogeneity in the bargaining effects of wages. As each household has its own bundle of earnings, hours, and wages, we use the parameters to calculate household-specific partial effects of wage shocks, namely $\partial \Delta \log h_{jt} / \partial \omega_{kt-\tau}$ for $j,k\in\{1,2\}$ and $\tau\in\{0,1,2\}$. We plot in figure \ref{AppFigure::PartialEffects} the partial effects for all households in the sample; the left graphs show the effects of own shocks (each row corresponds to shocks from a different period) while those on the right show the effects of the partner's shocks. We summarize key moments in table \ref{AppTable::PartialEffects} and we report the parameter estimates in appendix table \ref{AppTable::Heterogeneity_Results_All}.\footnote{For brevity, we only show results from the smaller sample over four time periods, i.e. with shocks from $t-2$ in the equation. Results from the baseline sample over three periods offer similar conclusions.}

The average partial effects of older shocks (shocks from period $t-2$ in figure \ref{Figure::PartialEffects_c}) are $\mathbb{E}(\partial \Delta \log h_{jt} / \partial \omega_{jt-2})<0$ and $\mathbb{E}(\partial \Delta \log h_{jt} / \partial \omega_{-jt-2})>0$. Away from the average, most households exhibit negative or zero labor supply effects from own shocks (so favorable \emph{own} shocks empower oneself) and positive or zero effects from partner shocks (so favorable \emph{partner} shocks weaken oneself). These effects are consistent with power shifts in limited commitment, in which good past shocks improved the bargaining power of the spouse that received them and, through persistence in the Pareto weight, shift future labor supply. It is remarkable that the effects survive 4 calendar years (the time that lapses from $t-2$ to $t$ in the PSID), the controls for all subsequent wage shocks, and the multiple wealth and income controls.

We estimate the average partial effects of immediately past shocks (from $t-1$ in figure \ref{Figure::PartialEffects_b}) at $\mathbb{E}(\partial \Delta \log h_{2t} / \partial \omega_{2t-1})<0$ and $\mathbb{E}(\partial \Delta \log h_{1t} / \partial \omega_{2t-1})>0$, exactly as limited commitment requires. The other two average partial effects seem to contradict limited commitment if taken at face value. $\mathbb{E}(\partial \Delta \log h_{1t} / \partial \omega_{1t-1})$ is weakly positive but the underlying parameters are not statistically significant; $\mathbb{E}(\partial \Delta \log h_{2t} / \partial \omega_{1t-1})$ is weakly negative but the majority of households exhibit \emph{positive} effects exactly as limited commitment postulates. 

\begin{table}[t]  
\begin{center}
\caption{Summary statistics of partial effects of wage shocks}\label{AppTable::PartialEffects}
\begin{tabular}{L{3.7cm} C{0.6cm} C{0.7cm} C{0.6cm} C{1.4cm} C{1.4cm} C{1.4cm} C{1.4cm} C{1.4cm}}
\toprule
                                                        &FC   &NC   &LC     & mean          & st.dev.       & p10           & p50           & p90           \\
\midrule
\multicolumn{9}{l}{\emph{Partial effects of current shocks }$(t)$}\\
~~$\partial \Delta \log h_{1t} / \partial \omega_{1t}$ & $.$ & $.$ & $.$ & $   -0.117$ & $    0.247$ & $   -0.153$ & $   -0.116$ & $   -0.085$ \\
~~$\partial \Delta \log h_{2t} / \partial \omega_{2t}$ & $.$ & $.$ & $.$ & $   -0.111$ & $    0.563$ & $   -0.138$ & $   -0.089$ & $   -0.060$ \\
~~$\partial \Delta \log h_{1t} / \partial \omega_{2t}$ & $0$ & $+$ & $+$ & $    0.011$ & $    0.242$ & $   -0.031$ & $   -0.019$ & $    0.126$ \\
~~$\partial \Delta \log h_{2t} / \partial \omega_{1t}$ & $0$ & $+$ & $+$ & $   -0.007$ & $    0.632$ & $   -0.043$ & $   -0.010$ & $    0.021$ \\
\multicolumn{9}{l}{\emph{Partial effects of immediately past shocks }$(t-1)$}\\
~~$\partial \Delta \log h_{1t} / \partial \omega_{1t-1}$ & $0$ & $0$ & $-$ & $    0.028$ & $    0.136$ & $   -0.022$ & $    0.030$ & $    0.051$ \\
~~$\partial \Delta \log h_{2t} / \partial \omega_{2t-1}$ & $0$ & $0$ & $-$ & $   -0.028$ & $    0.754$ & $   -0.035$ & $   -0.008$ & $    0.005$ \\
~~$\partial \Delta \log h_{1t} / \partial \omega_{2t-1}$ & $0$ & $0$ & $+$ & $    0.058$ & $    0.080$ & $    0.024$ & $    0.042$ & $    0.109$ \\
~~$\partial \Delta \log h_{2t} / \partial \omega_{1t-1}$ & $0$ & $0$ & $+$ & $   -0.008$ & $    0.371$ & $   -0.034$ & $    0.003$ & $    0.019$ \\
\multicolumn{9}{l}{\emph{Partial effects of older shocks }$(t-2)$}\\
~~$\partial \Delta \log h_{1t} / \partial \omega_{1t-2}$ & $0$ & $0$ & $-$ & $   -0.009$ & $    0.020$ & $   -0.021$ & $   -0.004$ & $   -0.001$ \\
~~$\partial \Delta \log h_{2t} / \partial \omega_{2t-2}$ & $0$ & $0$ & $-$ & $   -0.028$ & $    0.158$ & $   -0.043$ & $   -0.021$ & $    0.007$ \\
~~$\partial \Delta \log h_{1t} / \partial \omega_{2t-2}$ & $0$ & $0$ & $+$ & $    0.047$ & $    0.156$ & $    0.002$ & $    0.045$ & $    0.070$ \\
~~$\partial \Delta \log h_{2t} / \partial \omega_{1t-2}$ & $0$ & $0$ & $+$ & $    0.032$ & $    0.340$ & $   -0.004$ & $    0.027$ & $    0.057$ \\
\bottomrule
\end{tabular}
\caption*{\fsz\emph{Notes:} The table reports key summary statistics of the distribution of partial effects of wage shocks across households observed for $\geq4$ periods. The parameter estimates are in appendix table \ref{AppTable::Heterogeneity_Results_All}. }
\end{center}
\end{table}

The partial effects of the partner's current shocks (figure \ref{Figure::PartialEffects_a}) are mostly positive among men, consistent with non-full commitment, but spread around zero among women.

\begin{table}  
\begin{center}
\caption{Commitment test -- structural results with heterogeneity}\label{AppTable::Heterogeneity_Results_All}
\begin{tabular}{L{3.6cm} C{0.01cm} C{2.3cm} C{2.3cm} C{0.01cm} C{2.3cm} C{2.3cm}}
\toprule
                                            && \mcl{5}{c}{\textbf{wage shocks, $\geq$ 4 periods}}                       \\
\cmidrule{3-7}                      
                                            && \mcl{2}{c}{Male $j=1$}           && \mcl{2}{c}{Female $j=2$}             \\
\cmidrule{3-4}\cmidrule{6-7}
                                            && (1) estimate     & (2) st.error  && (3) estimate     & (4) st.error      \\
\midrule
\multicolumn{7}{l}{\emph{Pareto weight elasticities w.r.t current factors} ($\tau=0$)}\\
~$\eta_{j0}^{w_{j}}$                        &&                  &               &&                  &                   \\
~~~constant &  & $    0.999$ & $(    0.022)$ &  & $    1.004$ & $(    0.022)$ \\
~~~$ w_{jt-1}$ &  & $   -0.040$ & $(    0.015)$ &  & $   -0.025$ & $(    0.008)$ \\
~~~$ w_{-jt-1}$ &  & $    0.061$ & $(    0.026)$ &  & $    0.069$ & $(    0.035)$ \\
~~~$\mathbb{1}[age_{j} << age_{-j}]$ &  & $    0.030$ & $(    0.030)$ &  & $   -0.014$ & $(    0.026)$ \\
~~~$\mathbb{1}[age_{j} >> age_{-j}]$ &  & $    0.016$ & $(    0.034)$ &  & $   -0.020$ & $(    0.015)$ \\
\noalign{\smallskip}
~$\eta_{j0}^{w_{-j}}$                       &&                  &               &&                  &                   \\
~~~constant &  & $    0.014$ & $(    0.025)$ &  & $    0.017$ & $(    0.022)$ \\
~~~$ w_{jt-1}$ &  & $    0.021$ & $(    0.023)$ &  & $    0.047$ & $(    0.026)$ \\
~~~$ w_{-jt-1}$ &  & $    0.009$ & $(    0.013)$ &  & $   -0.079$ & $(    0.036)$ \\
~~~$\mathbb{1}[age_{j} << age_{-j}]$ &  & $   -0.023$ & $(    0.038)$ &  & $    0.011$ & $(    0.026)$ \\
~~~$\mathbb{1}[age_{j} >> age_{-j}]$ &  & $   -0.148$ & $(    0.058)$ &  & $    0.040$ & $(    0.045)$ \\
\multicolumn{7}{l}{\emph{Pareto weight elasticities w.r.t factors 1 period in the past} ($\tau=1$)}\\
~$\eta_{j1}^{w_{j}}$                        &&                  &               &&                  &                   \\
~~~constant &  & $   -0.029$ & $(    0.026)$ &  & $    0.008$ & $(    0.015)$ \\
~~~$ w_{jt-2}$ &  & $    0.014$ & $(    0.010)$ &  & $   -0.050$ & $(    0.028)$ \\
~~~$ w_{-jt-2}$ &  & $   -0.054$ & $(    0.066)$ &  & $    0.052$ & $(    0.031)$ \\
~~~$\mathbb{1}[age_{j} << age_{-j}]$ &  & $    0.012$ & $(    0.030)$ &  & $   -0.012$ & $(    0.027)$ \\
~~~$\mathbb{1}[age_{j} >> age_{-j}]$ &  & $    0.059$ & $(    0.045)$ &  & $   -0.003$ & $(    0.024)$ \\
\noalign{\smallskip}
~$\eta_{j1}^{w_{-j}}$                       &&                  &               &&                  &                   \\
~~~constant &  & $   -0.058$ & $(    0.023)$ &  & $   -0.026$ & $(    0.023)$ \\
~~~$ w_{jt-2}$ &  & $    0.005$ & $(    0.027)$ &  & $    0.073$ & $(    0.065)$ \\
~~~$ w_{-jt-2}$ &  & $    0.056$ & $(    0.034)$ &  & $   -0.002$ & $(    0.015)$ \\
~~~$\mathbb{1}[age_{j} << age_{-j}]$ &  & $    0.015$ & $(    0.040)$ &  & $    0.012$ & $(    0.027)$ \\
~~~$\mathbb{1}[age_{j} >> age_{-j}]$ &  & $   -0.063$ & $(    0.064)$ &  & $    0.046$ & $(    0.040)$ \\
\multicolumn{7}{l}{\emph{Pareto weight elasticities w.r.t factors 2 periods in the past} ($\tau=2$)}\\  
~$\eta_{j2}^{w_{j}}$                        &&                  &               &&                  &                   \\
~~~constant &  & $    0.001$ & $(    0.022)$ &  & $    0.029$ & $(    0.008)$ \\
~~~$ w_{jt-3}$ &  & $    0.013$ & $(    0.011)$ &  & $   -0.022$ & $(    0.012)$ \\
~~~$ w_{-jt-3}$ &  & $   -0.009$ & $(    0.050)$ &  & $    0.005$ & $(    0.016)$ \\
~~~$\mathbb{1}[age_{j} << age_{-j}]$ &  & $    0.012$ & $(    0.024)$ &  & $   -0.008$ & $(    0.017)$ \\
~~~$\mathbb{1}[age_{j} >> age_{-j}]$ &  & $    0.018$ & $(    0.030)$ &  & $   -0.034$ & $(    0.011)$ \\
\noalign{\smallskip}
~$\eta_{j2}^{w_{-j}}$                       &&                  &               &&                  &                   \\
~~~constant &  & $   -0.067$ & $(    0.036)$ &  & $   -0.042$ & $(    0.025)$ \\
~~~$ w_{jt-3}$ &  & $    0.047$ & $(    0.034)$ &  & $   -0.047$ & $(    0.045)$ \\
~~~$ w_{-jt-3}$ &  & $   -0.027$ & $(    0.033)$ &  & $    0.052$ & $(    0.045)$ \\
~~~$\mathbb{1}[age_{j} << age_{-j}]$ &  & $    0.061$ & $(    0.040)$ &  & $    0.039$ & $(    0.024)$ \\
~~~$\mathbb{1}[age_{j} >> age_{-j}]$ &  & $    0.032$ & $(    0.039)$ &  & $    0.026$ & $(    0.025)$ \\
\bottomrule
\end{tabular}
\end{center}
\end{table}

\begin{table}[t!]  
\begin{center}
\caption*{Table \ref{AppTable::Heterogeneity_Results_All} (continued): Commitment test -- structural results with heterogeneity}
\begin{tabular}{L{3.6cm} C{0.01cm} C{2.3cm} C{2.3cm} C{0.01cm} C{2.3cm} C{2.3cm}}
\toprule
                                            && \mcl{5}{c}{\textbf{wage shocks, $\geq$ 4 periods}}                       \\
\cmidrule{3-7}                      
                                            && \mcl{2}{c}{Male $j=1$}           && \mcl{2}{c}{Female $j=2$}             \\
\cmidrule{3-4}\cmidrule{6-7}
                                            && (1) estimate     & (2) st.error  && (3) estimate     & (4) st.error      \\
\midrule
\multicolumn{7}{l}{\emph{Pareto weight elasticities w.r.t initial distribution factors}}\\
~$\eta_{jt}^{young_{j}}$ &  & $    0.083$ & $(    0.031)$ &  & $    0.020$ & $(    0.012)$ \\
~$\eta_{jt}^{young_{-j}}$ &  & $    0.080$ & $(    0.042)$ &  & $   -0.019$ & $(    0.008)$ \\
\mcl{2}{l}{\emph{Frisch elast.} ${\alpha_{j}/\mathbb{E}(h_{jt-1})}^{\#}$} & $    1.053$ & $(    0.375)$ &  & $    1.087$ & $(    0.270)$ \\
\noalign{\smallskip}
\multicolumn{7}{l}{\emph{Other terms}}\\
~$\ell_{\Delta y}$ &  & $    0.251$ & $(    0.042)$ &  & $    0.205$ & $(    0.012)$ \\
\mcl{2}{l}{~$\ell_{\Delta a}-\eta_{j0}^{a}$} & $   -0.007$ & $(    0.008)$ &  & $    0.000$ & $(    0.006)$ \\
~$\ell_{y}$ &  & $    0.018$ & $(    0.007)$ &  & $    0.006$ & $(    0.007)$ \\
~$\ell_{a}$ &  & $   -0.017$ & $(    0.006)$ &  & $   -0.007$ & $(    0.007)$ \\
~$ {\zeta_{j}}^{\#\#}$ &  & $   -0.007$ & $(    0.003)$ &  & $    0.000$ & $(    0.002)$ \\
~$\eta_{j1}^{a}$ &  & $   -0.008$ & $(    0.004)$ &  & $   -0.003$ & $(    0.002)$ \\
~$\eta_{j2}^{a}$ &  & $   -0.006$ & $(    0.003)$ &  & $    0.001$ & $(    0.001)$ \\
\noalign{\smallskip}
\mcl{2}{l}{~${\text{$p$ value for }{\cal H}_{0}^\text{FC}}^{\#\#\#}$} & \mcl{2}{c}{$0.042$}  && \mcl{2}{c}{$0.033$}     \\
\mcl{2}{l}{~${\text{$p$ value for }{\cal H}_{0}^\text{NC}}^{\#\#\#}$} & \mcl{2}{c}{$0.044$}  && \mcl{2}{c}{$0.088$}     \\
\noalign{\smallskip}
Observations                                && \mcl{2}{c}{6,028}                && \mcl{2}{c}{6,028}                    \\
\bottomrule 
\end{tabular}
\caption*{\fsz\emph{Notes:} The table reports the parameter estimates from the heterogeneity (third) specification of the commitment test with wages as time-varying distribution factors. The coefficient on each Pareto weight elasticity depends linearly on the immediately past levels of wages and the age-gap-at-marriage dummies. Standard errors clustered at the household level are in brackets. See appendix \ref{Appendix::EstimatingEquation} for details on the parameters. $\ell_{\Delta y}$, $\ell_{y}$, $\ell_{a}$ are common across male and female equations in each specification. For simplicity, we estimate the equations separately without imposing cross-equation restrictions. We cannot reject equality of these parameters across equations. $^\#$We report the Frisch elasticity at the sample average of hours of work; its standard error is calculated with the delta method. $^{\#\#}$We multiply $\zeta_{j}$ by $10^4$ for legibility ($\zeta_{j}$ originally multiplies the \emph{level} of consumption so its magnitude is very small). $^{\#\#\#}$We assess the null hypotheses at the sample average of wages for partners of similar age.}
\end{center}
\end{table}

The partial effects are in fact elasticities of labor supply w.r.t wage shocks. As the effects of past shocks lie on average between $0.01$-$0.05$ in absolute value, a 10\% wage change shifts hours by $0.1\%$-$0.5\%$, with the direction determined by the sign of the elasticity. Past shocks exclusively induce \emph{bargaining} effects, so this gives a sense of the average magnitude of bargaining. However, there is large heterogeneity; clearly, many couples exhibit full commitment (null bargaining effects), while others exhibit limited commitment. 

Despite the theoretical appeal of the heterogeneity specification, almost none of the heterogeneity-specific parameter estimates are statistically significant (table \ref{AppTable::Heterogeneity_Results_All}), which prevents us from making strong assertions. This is to a large extent inevitable. This specification introduces five times as many parameters as the structural (second) specification, so the model is too flexible given our rather limited sample sizes. To improve significance, we thus also explore an indirect approach to heterogeneity by estimating the simplest, reduced form specification of our test over a few well defined subsamples. 

Table \ref{AppTable::OLS_Selective_Heterogeneity} presents results across three different sample splits. In a first split (columns 1-2), we test for commitment among partners who have the same level of education versus those who do not. In a second split (columns 3-4), we compare \emph{older} couples in their \emph{first} marriage to everyone else. In a third split (columns 5-6), we zoom into couples in which the wife is the main earner. In all cases, we focus on households observed for at least three periods, namely our largest sample. To facilitate the interpretation of the results, we interact the current and past wage shocks with a dummy that reflects the applicable sample split in each case; the coefficient on the dummy then picks up how the bargaining effects change along the sample split.

Among couples with \emph{different} levels of education, all past shocks enter with the sign postulated by limited commitment.\footnote{We define four levels of education: high school dropout, high school graduate, (some) college, and more than college.} Specifically, own past shocks reduce future labor supply, while the partner's past shocks increase it; inspect panel B in table \ref{AppTable::OLS_Selective_Heterogeneity}. However, as soon as we interact these wage shocks with the dummy for \emph{same} education, the bargaining effects are pushed towards zero (null bargaining effects). This suggest that those who share the same education behave more in line with full commitment, perhaps due to being a better match than those whose education levels differ. Similar conclusions can be drawn from the bargaining effect from the partner's current shock in panel A. There is evidence for non-full commitment, especially on the men's side, but that bargaining effect is pushed closer to zero among couples who share the same education levels.

Moving to the length of marriage, and focusing on couples in relatively younger marriages, all current and past shocks enter with the sign implied by limited commitment (except $\omega_{1t}$ among women). Statistical significance is low but any parameter that \emph{is} significant has the sign that limited commitment requires. However, as soon as we switch to older couples in their first marriage (namely marriages that have lasted for longer), almost all coefficients become inconsistent with bargaining ($\beta_{1[w_{2t-1}]}$ being the sole exception) and mostly statistically insignificant. This seems to suggest that older couples in their first marriage behave more in line with full commitment, as one would expect.

Finally, we focus on the gender of the main earner. While the coefficients on almost all current and past shocks are consistent with limited commitment in the baseline in which the husband outearns the wife, an interesting split appears when we compare male and female bargaining effects among couples in which the woman (weakly) outearns the husband. In this case, the bargaining effects on male hours are pushed to zero, while the bargaining effects on female hours become more pronounced and more strongly in line with limited commitment. Though the effects remain statistically insignificant, at face value they suggest that women who outearn their husbands engage in renegotiation more than them, likely (and perhaps eventually) leading to the end of certain marriages.

\begin{sidewaystable}
\begin{center}
\caption{Commitment test -- summary reduced form results with selective heterogeneity}\label{AppTable::OLS_Selective_Heterogeneity}
\begin{tabular}{L{2.1cm} C{0.7cm} C{0.7cm} C{0.7cm} C{0.01cm} C{1.9cm} C{1.9cm} C{0.01cm} C{1.9cm} C{1.9cm} C{0.01cm} C{1.9cm} C{1.9cm}}
\toprule
                                            &&&&                    & \multicolumn{8}{c}{\textbf{wage shocks, $\geq$ 3 periods}}  \\
\cmidrule{6-13}
\mcl{5}{l}{\emph{Heterogeneity}}                                    &  \multicolumn{2}{c}{spouses have}  
                                                                    && \multicolumn{2}{c}{older couple}
                                                                    && \multicolumn{2}{c}{wife earnings $\geq$} \\
\emph{event:}                               &&&&                    &  \multicolumn{2}{c}{same education}  
                                                                    && \multicolumn{2}{c}{in first marriage} 
                                                                    && \multicolumn{2}{c}{husband earnings} \\                                                                    
\cmidrule{6-7}\cmidrule{9-10}\cmidrule{12-13}
                                            &&&&                    & (1)     & (2)       && (3)      & (4)       && (5)      & (6)       \\
                                            &&&&                    & Male    & Female    && Male     & Female    && Male     & Female    \\
                                            &FC&NC&LC&              & $j=1$   & $j=2$     && $j=1$    & $j=2$     && $j=1$    & $j=2$     \\
\midrule
\multicolumn{13}{l}{\emph{A. Current shocks} ($t$)}\\
~~$\beta_{j[w_{jt}]}$ & . & . & . &                                 &         &           &&          &           &&          &           \\
~~~constant &  &  &  &  											& $  -24.221$ & $  -10.509$ &  & $  -29.202$ & $   -9.732$ &  & $ -194.086$ & $   -9.710$ \\
 &  &  &  &                                                         & $(   10.324)$ & $(    5.367)$ &  & $(   11.160)$ & $(    3.202)$ &  & $(   62.210)$ & $(    3.534)$ \\
\mcl{5}{l}{~~~$\mathbb{1}[\text{couples with \emph{event} $=$ true}]$}& $  -92.206$ & $   -0.110$ &  & $   -5.100$ & $   15.506$ &  & $  175.603$ & $  -87.495$ \\
 &  &  &  &                                                         & $(   28.042)$ & $(    7.715)$ &  & $(   59.697)$ & $(   27.524)$ &  & $(   62.005)$ & $(   45.776)$ \\
~~$\beta_{j[w_{-jt}]}$ & $0$ & $+$ & $+$ &                          &         &           &&          &           &&          &           \\
~~~constant &  &  &  &  											& $   51.278$ & $  -35.553$ &  & $   44.810$ & $   -7.784$ &  & $   67.600$ & $   -1.758$ \\
 &  &  &  &                                                         & $(   32.890)$ & $(   19.527)$ &  & $(   26.942)$ & $(    6.526)$ &  & $(   32.102)$ & $(    6.360)$ \\
\mcl{5}{l}{~~~$\mathbb{1}[\text{couples with \emph{event} $=$ true}]$}& $  -16.056$ & $   37.500$ &  & $  -48.171$ & $   37.017$ &  & $  -88.329$ & $   27.377$ \\
 &  &  &  &                                                         & $(   44.399)$ & $(   21.301)$ &  & $(   68.809)$ & $(   17.592)$ &  & $(   54.995)$ & $(   28.889)$ \\
\multicolumn{13}{l}{\emph{B. Past shocks} ($t-1$)}\\
~~$\beta_{j[w_{jt-1}]}$ & $0$ & $0$ & $-$ &                         &         &           &&          &           &&          &           \\
~~~constant &  &  &  &  											& $   -9.688$ & $   -8.296$ &  & $  -15.101$ & $   -6.568$ &  & $  -87.945$ & $   -7.844$ \\
 &  &  &  &                                                         & $(    4.983)$ & $(    4.539)$ &  & $(    6.911)$ & $(    3.122)$ &  & $(   26.928)$ & $(    3.593)$ \\
\mcl{5}{l}{~~~$\mathbb{1}[\text{couples with \emph{event} $=$ true}]$}& $    0.062$ & $   10.389$ &  & $   99.646$ & $   50.949$ &  & $   84.647$ & $   -7.275$ \\
 &  &  &  &                                                         & $(   16.732)$ & $(    6.432)$ &  & $(   29.437)$ & $(   49.323)$ &  & $(   27.261)$ & $(   29.754)$ \\
~~$\beta_{j[w_{-jt-1}]}$ & $0$ & $0$ & $+$ &                        &         &           &&          &           &&          &           \\
~~~constant &  &  &  &  											& $   55.615$ & $   24.725$ &  & $   30.082$ & $    6.972$ &  & $   48.430$ & $    6.089$ \\
 &  &  &  &                                                         & $(   29.733)$ & $(   13.679)$ &  & $(   23.571)$ & $(    7.567)$ &  & $(   38.308)$ & $(    7.752)$ \\
\mcl{5}{l}{~~~$\mathbb{1}[\text{couples with \emph{event} $=$ true}]$}& $  -78.030$ & $  -30.818$ &  & $   90.435$ & $  -17.947$ &  & $  -26.773$ & $   23.910$ \\
 &  &  &  &                                                         & $(   39.836)$ & $(   15.523)$ &  & $(   68.195)$ & $(   57.344)$ &  & $(   43.174)$ & $(   48.477)$ \\
\multicolumn{5}{l}{Initial distribution factors ($t=0$)}            & \mcl{2}{c}{yes}     && \mcl{2}{c}{yes}      && \mcl{2}{c}{yes} \\
\multicolumn{5}{l}{Other terms}                                     & \mcl{2}{c}{yes}     && \mcl{2}{c}{yes}      && \mcl{2}{c}{yes} \\
\noalign{\smallskip}
\mcl{5}{l}{Observations}                                            & \mcl{2}{c}{8,513}   && \mcl{2}{c}{8,278}    && \mcl{2}{c}{8,513}  \\
\bottomrule
\end{tabular}
\caption*{\fsz\emph{Notes:} The table reports the coefficient estimates from the reduced form (first) specification of the commitment test implemented over three different sample splits. The subsamples are described in the top of the table. Marital history is not observed for everyone in our baseline sample. Standard errors clustered at the household level are in brackets.}
\end{center}
\end{sidewaystable}

\

\noindent \textbf{Use of BMI as additional time-varying distribution factors.} We use additional distribution factors as a means of over-identification. These factors must be assignable and vary over time. \citet{Voena2015} explores changes in divorce and property division laws; but these changes occur mostly in the 1970s and 1980s, which is outside our time frame.\footnote{See also \cite{Stevenson2007DivorceLaws}. \citet{Chiappori2002} use variation in marital sex ratios across states, which are unlikely to vary much over time. \citet{Blau2016} use the receipt of inheritance as distribution factor; this is assignable and time-varying but we do not observe inheritances in the PSID.} \citet{ChiapporiOrefficeQuintana2012FatterAttraction} and \citet{DupuyGalichon2014PersonalityTraits} use anthropometric measures to determine a person's attractiveness. We take up on this and employ the spouses' body mass index as distribution factor, i.e. $Z_{t} = \{BMI_{1t}, BMI_{2t}\}$. The underlying premise is that one's body mass influences their attractiveness in the marriage market, thus affecting their bargaining power at home. 

We re-estimate all specifications activating $Z_{t-\tau} = \{BMI_{1t-\tau},BMI_{2t-\tau}\}$, $\tau\in\{0,1,2\}$, and report results in columns 5-8 in tables \ref{AppTable::OLS_Results_All} (reduced form), \ref{AppTable::Structural_Results_All} (structural), and \ref{AppTable::DeepStructure_Results_All} (proportionality restrictions).\footnote{Results with heterogeneity lead to similar conclusions. We do not include them for brevity.} The bargaining effects of wages are unchanged from the baseline, pointing again towards limited commitment. The bargaining effects of BMI are also in line with limited commitment: an increase in one's \emph{past} BMI increases his/her hours (as if it weakens one's Pareto weight because that person became relatively less attractive) while, by contrast, an increase in the \emph{partner's past} BMI reduces work (as if the opposite reasons hold). Statistical significance is low but most parameters that \emph{are} significant have the sign that limited commitment postulates. 

The use of BMI as a distribution factor is debatable because shifts in body mass may reflect endogenous choices or affect labor supply directly. But if such shifts exclusively reflected choices, it is unclear why \emph{past} BMI would affect current labor supply, as we find here, beyond its effect on current BMI that we explicitly control for. Moreover, any \emph{direct} effect that BMI has on labor supply would typically be negative (a weight increase limits one's ability to work), which is opposite of what we find here.

\FloatBarrier

\section{Commitment test and measurement error}\label{Appendix::Measurement_Error}

Wages and hours in survey data are subject to measurement error. Typically, one will use an instrument for wages, often lagged wages. This is not possible here because lagged wages directly affect hours in limited commitment (they are in fact standalone regressors in our test), so they do not satisfy the exclusion restriction. 

We take an alternative route and characterize how measurement error biases the main parameter estimates, namely the coefficients on current and past wages. Consider the estimating equation for male hours given by
\begin{equation*}
\Delta h_{1t} = \beta_{1[w_{1t}]} \omega_{1t} + \beta_{1[w_{2t}]} \omega_{2t} + \beta_{1[w_{1t-1}]} \omega_{1t-1} + \beta_{1[w_{2t-1}]} \omega_{2t-1} + \varepsilon_{1t},
\end{equation*}
where $\Delta h_{1t} \equiv \Delta \log h_{1t}$. This corresponds to equation \eqref{Eq::EstimableEquation.FinalReducedForm} for $j=1$ (male hours), abstracting from the non-wage terms (that is, we control for household income, wealth, consumption etc). We focus on current ($t$) and immediate past ($t-1$) wage shocks, as if older shocks do not exist. We maintain that wage shocks and the error term are mean independent. The conclusions we establish for male hours ($j=1$) hold symmetrically for female hours ($j=2$), and they hold also after reintroducing shocks from period $t-2$.

Let shocks not co-vary \emph{between} partners, i.e. $\text{Cov}(\omega_{jt},\omega_{-jt-\tau})=0$, $\forall \tau$. This requires that, first, true male-female shocks are uncorrelated, and second, measurement errors in male-female wages are uncorrelated. We maintain the first assumption to keep this discussion simple but we do \emph{not} impose it in the empirical application. We return to the second assumption below. After these simplifications, the population coefficients are given by
\begin{align}\label{AppEq::IdentifiedParameters}
\begin{split}
\beta_{1[w_{1t}]} 
&= \big\{V(\omega_{1t-1})C(\Delta h_{1t},\omega_{1t}) - C(\omega_{1t},\omega_{1t-1})C(\Delta h_{1t},\omega_{1t-1})\big\}/D_{1} \\
\beta_{1[w_{2t}]} 
&= \big\{V(\omega_{2t-1})C(\Delta h_{1t},\omega_{2t}) - C(\omega_{2t},\omega_{2t-1})C(\Delta h_{1t},\omega_{2t-1})\big\}/D_{2} \\
\beta_{1[w_{1t-1}]} 
&= \big\{V(\omega_{1t})C(\Delta h_{1t},\omega_{1t-1}) - C(\omega_{1t},\omega_{1t-1})C(\Delta h_{1t},\omega_{1t})\big\}/D_{1} \\
\beta_{1[w_{2t-1}]} 
&= \big\{V(\omega_{2t})C(\Delta h_{1t},\omega_{2t-1}) - C(\omega_{2t},\omega_{2t-1})C(\Delta h_{1t},\omega_{2t})\big\}/D_{2},
\end{split}
\end{align}
where $V(\cdot)$ and $C(\cdot,\cdot)$ are the variance and covariance. $D_{j} = V(\omega_{jt})V(\omega_{jt-1}) - (C(\omega_{jt},\omega_{jt-1}))^2$ is the positive determinant of the covariance matrix of shocks of spouse $j=\{1,2\}$. 

We established earlier that $\beta_{1[w_{2t}]}>0$ (coefficient on the partner's current wage) is indicative of no/limited commitment, while $\beta_{1[w_{1t-1}]}<0$ (coefficient on own past wage) and $\beta_{1[w_{2t-1}]}>0$ (coefficient on the partner's past wage) are indicative of limited commitment. We will now assess if measurement error can produce this pattern of signs. 

Let $\Delta h_{1t}^o = \Delta h_{1t} + \Delta e_{t}^{h_1}$ and $\omega_{jt-\tau}^o = \omega_{jt-\tau} + \Delta e_{t-\tau}^{w_j}$ for $j=\{1,2\}$ and $\tau\in\{0,1\}$; that is, the observed variable is equal to the true variable plus measurement error. Let $e^{h_1}$ and $e^{w_j}$ be independent of the true variable and serially uncorrelated. Given the first difference, however, hours/wage growth correlates over time due to mean reversion in the error. Moreover, $e^{w_1} = e^{y_1} - e^{h_1}$, where $e^{y_1}$ is the error in log earnings, as wages are earnings over hours.\footnote{We use the hourly wage rate variable in the PSID (e.g. variable \texttt{ER77414} in 2019), which is calculated as annual earnings over hours, thus is subject to the division bias.} Therefore, error in hours and wages correlate negatively. Assuming time-invariance of the second moments of measurement error, it follows that (i) $V(\Delta e_{t}^{w_j}) = 2\sigma^2_{e^{w_j}}$, (ii) $C(\Delta e_{t}^{w_j},\Delta e_{t-1}^{w_j}) = -\sigma^2_{e^{w_j}}$, (iii) $C(\Delta e_{t}^{h_1},\Delta e_{t}^{w_1}) = -2 \sigma^2_{e^{h_1}}$, (iv) $C(\Delta e_{t}^{h_1},\Delta e_{t-1}^{w_1}) = \sigma^2_{e^{h_1}}$, and (v) $C(\Delta e_{t}^{h_1},\Delta e_{t-\tau}^{w_{2}}) = 0$, for $\tau =\{0,1\}$, where $\sigma^2_{e^{h_1}}$ and $\sigma^2_{e^{w_j}}$, $j\in\{1,2\}$, are the variance of the error in hours and wages.

With measurement error, the estimates $\widehat{\beta}_{1[w_{2t}]}$, $\widehat{\beta}_{1[w_{1t-1}]}$, $\widehat{\beta}_{1[w_{2t-1}]}$ may be biased for the true parameters $\beta_{1[w_{2t}]}$, $\beta_{1[w_{1t-1}]}$, $\beta_{1[w_{2t-1}]}$. The bias may operate separately through the denominators in \eqref{AppEq::IdentifiedParameters}, or through the numerators. We characterize the two in turn.

\

\noindent \textbf{Bias towards full commitment.} With measurement error, the denominator in \eqref{AppEq::IdentifiedParameters} is given by $\widehat{D}_{j} = D_{j} + 3(\sigma^2_{e^{w_j}})^2 + 2\sigma^2_{e^{w_j}}\left(V(\omega_{jt})+V(\omega_{jt-1})+C(\omega_{jt},\omega_{jt-1})\right)$, which is larger than $D_j$.\footnote{The covariance between consecutive residual wages is typically negative (e.g. in a permanent-transitory process, it is negative due to mean reversion of the transitory shock). However, $|C(\omega_{jt},\omega_{jt-1})|$ is always smaller than the largest standalone variance, so the bracketed term is strictly positive.} Measurement error thus inflates the positive denominator, which, ceteris paribus, biases all coefficients towards zero. Null bargaining effects reflect full commitment, so measurement error biases the parameter estimates towards full commitment.

On top of this bias towards full commitment, measurement error biases the numerators in \eqref{AppEq::IdentifiedParameters}. Below we discuss this bias starting from the coefficients on the partner's wages.

\

\noindent \textbf{Coefficient on partner's current wage}. The numerator of $\widehat{\beta}_{1[w_{2t}]}$ is given by $\text{num}(\widehat{\beta}_{1[w_{2t}]}) = \text{num}(\beta_{1[w_{2t}]}) + \sigma^2_{e^{w_{2}}}\left(2C(\Delta h_{1t},\omega_{2t}) + C(\Delta h_{1t},\omega_{2t-1})\right)$, where $\text{num}(\beta_{1[w_{2t}]})$ is shown in \eqref{AppEq::IdentifiedParameters}. How $\text{num}(\widehat{\beta}_{1[w_{2t}]})$ compares with $\text{num}(\beta_{1[w_{2t}]})$ depends on the signs of $C(\Delta h_{1t},\omega_{2t})$ and $C(\Delta h_{1t},\omega_{2t-1})$, so we distinguish four cases:
\begin{enumerate}
    \item Let $C(\Delta h_{1t},\omega_{2t})>0$ and $C(\Delta h_{1t},\omega_{2t-1})>0$.\\
    Then $\text{num}(\beta_{1[w_{2t}]})>0$, so true $\beta_{1[w_{2t}]}>0$, in line with no/limited commitment.\footnote{We treat the first-order covariance between residual wages as negative, as is typically found empirically; e.g. in a permanent-transitory process, it is negative due to mean reversion of the transitory shock.} It follows that $\text{num}(\widehat{\beta}_{1[w_{2t}]})>\text{num}(\beta_{1[w_{2t}]})$; that is, measurement error preserves the sign of the numerator and increases its magnitude. However, measurement error inflates the denominator $D_{2}$ by much more so measurement error overall biases the estimate downwards towards zero (null bargaining effects, full commitment).\footnote{\label{AppFootnote::Measurement_error}Note that $\sigma^2_{e^{w_{2}}}\left(2V(\omega_{2t}) + V(\omega_{2t-1})\right)$, which is only part of the increase in the denominator due to the error, is larger than $\sigma^2_{e^{w_{2}}}\left(2C(\Delta h_{1t},\omega_{2t}) + C(\Delta h_{1t},\omega_{2t-1})\right)$ in the numerator if the covariances between hours and partner shocks are smaller than the variance of the shocks. This is generally true: a regression of hours on partner shocks, controlling for income, wealth, etc., produces coefficients $\leq1$ in absolute value.} 

    \item Let $C(\Delta h_{1t},\omega_{2t})\leq0$ and $C(\Delta h_{1t},\omega_{2t-1})\leq0$.\\
    Then $\text{num}(\beta_{1[w_{2t}]})\leq0$ and $\beta_{1[w_{2t}]}\leq0$, which is inconsistent with non-full commitment. The question is then whether measurement error biases $\widehat{\beta}_{1[w_{2t}]}$ towards positive values, making it artificially consistent with no/limited commitment. Clearly not as $\text{num}(\widehat{\beta}_{1[w_{2t}]})\leq\text{num}(\beta_{1[w_{2t}]})$ in this case, so $\widehat{\beta}_{1[w_{2t}]}$ remains inconsistent with bargaining. 

    \item Let $C(\Delta h_{1t},\omega_{2t})>0$ and $C(\Delta h_{1t},\omega_{2t-1})\leq0$.
    	\begin{itemize}
     		\item[(a)] Let $\beta_{1[w_{2t}]}>0$, in line with no/limited commitment. The covariance of concurrent hours-wages is empirically larger than the covariance one period apart, i.e. $|C(\Delta h_{1t},\omega_{2t})|>|C(\Delta h_{1t},\omega_{2t-1})|$, so $\text{num}(\widehat{\beta}_{1[w_{2t}]})>\text{num}(\beta_{1[w_{2t}]})$. Measurement error, however, inflates the denominator by more than the numerator (see footnote \ref{AppFootnote::Measurement_error}), so $\widehat{\beta}_{1[w_{2t}]}$ is biased towards zero (null bargaining effects; full commitment).

    		\item[(b)] Let $\beta_{1[w_{2t}]}\leq0$, contradicting no/limited commitment. It must be $|C(\Delta h_{1t},\omega_{2t})|<<|C(\Delta h_{1t},\omega_{2t-1})|$, which ensures $\text{num}(\beta_{1[w_{2t}]})\leq0$. Depending on the sign of $2C(\Delta h_{1t},\omega_{2t}) + C(\Delta h_{1t},\omega_{2t-1})$, measurement error could bias the numerator downwards ($\widehat{\beta}_{1[w_{2t}]}$ remains inconsistent with bargaining) or towards zero (full commitment). It would flip its sign to positive only if $|C(\Delta h_{1t},\omega_{2t})|>|C(\Delta h_{1t},\omega_{2t-1})|$, which is a contradiction as $\beta_{1[w_{2t}]}$ would have been positive in that case.
    	\end{itemize}

    \item Let $C(\Delta h_{1t},\omega_{2t})\leq0$ and $C(\Delta h_{1t},\omega_{2t-1})>0$. 
    	\begin{itemize}
     		\item[(a)] Let $\beta_{1[w_{2t}]}>0$, in line with no/limited commitment. In this case $|C(\Delta h_{1t},\omega_{2t})|<<|C(\Delta h_{1t},\omega_{2t-1})|$, which ensures that $\text{num}(\beta_{1[w_{2t}]})>0$. Depending on the sign of $2C(\Delta h_{1t},\omega_{2t}) + C(\Delta h_{1t},\omega_{2t-1})$, measurement error could bias the numerator downwards (thus $\widehat{\beta}_{1[w_{2t}]}$ becomes inconsistent with no/limited commitment) or upwards (remains correctly in line with no/limited commitment).

     		\item[(b)] Let $\beta_{1[w_{2t}]}\leq0$, contradicting no/limited commitment. In this case $|C(\Delta h_{1t},\omega_{2t})|\geq|C(\Delta h_{1t},\omega_{2t-1})|$, which ensures $\text{num}(\beta_{1[w_{2t}]})\leq0$. Measurement error then implies $\text{num}(\widehat{\beta}_{1[w_{2t}]})<\text{num}(\beta_{1[w_{2t}]})$, so $\widehat{\beta}_{1[w_{2t}]}$ remains inconsistent with bargaining.
     	\end{itemize}
\end{enumerate}
In brief: when the true coefficient $\beta_{1[w_{2t}]}$ on the partner's current wage is positive, in line with no/limited commitment, measurement error either preserves the sign or biases the estimate towards zero (full commitment); when it is weakly negative, contradicting no/limited commitment, measurement error does \emph{not} flip its sign. In no case does measurement error erroneously move $\widehat{\beta}_{1[w_{2t}]}$ towards positive values (no/limited commitment) when the true $\beta_{1[w_{2t}]}$ is incompatible with bargaining. Any evidence for no/limited commitment through $\widehat{\beta}_{1[w_{2t}]}$, as we find in practice, is thus not the byproduct of measurement error.

\

\noindent \textbf{Coefficient on partner's past wage}. An analogous discussion applies to the coefficient on the partner's \emph{past} wage. The numerator of $\widehat{\beta}_{1[w_{2t-1}]}$ is given by $\text{num}(\widehat{\beta}_{1[w_{2t-1}]}) = \text{num}(\beta_{1[w_{2t-1}]}) + \sigma^2_{e^{w_{2}}}\left(C(\Delta h_{1t},\omega_{2t}) + 2C(\Delta h_{1t},\omega_{2t-1})\right)$, where $\text{num}(\beta_{1[w_{2t-1}]})$ is shown in \eqref{AppEq::IdentifiedParameters}. We distinguish four cases:
\begin{enumerate}
    \item Let $C(\Delta h_{1t},\omega_{2t})>0$ and $C(\Delta h_{1t},\omega_{2t-1})>0$.\\
    Then $\text{num}(\beta_{1[w_{2t-1}]})>0$, so true $\beta_{1[w_{2t-1}]}>0$, in line with limited commitment. Then $\text{num}(\widehat{\beta}_{1[w_{2t-1}]})>\text{num}(\beta_{1[w_{2t-1}]})$; that is, measurement error preserves the sign of the numerator and increases its magnitude. However, measurement error inflates the denominator $D_{2}$ by much more (see footnote \ref{AppFootnote::Measurement_error}), so measurement error overall biases the estimate downwards towards zero (null bargaining effects, full commitment).

    \item Let $C(\Delta h_{1t},\omega_{2t})\leq0$ and $C(\Delta h_{1t},\omega_{2t-1})\leq0$.\\
    Then $\text{num}(\beta_{1[w_{2t-1}]})\leq0$, so $\beta_{1[w_{2t-1}]}\leq0$, inconsistent with limited commitment. The question is then whether measurement error biases $\widehat{\beta}_{1[w_{2t-1}]}$ towards positive values, making it artificially consistent with limited commitment. Clearly not as $\text{num}(\widehat{\beta}_{1[w_{2t-1}]})\leq\text{num}(\beta_{1[w_{2t-1}]})$ in this case, so $\widehat{\beta}_{1[w_{2t-1}]}$ remains inconsistent with bargaining. 

    \item Let $C(\Delta h_{1t},\omega_{2t})>0$ and $C(\Delta h_{1t},\omega_{2t-1})\leq0$.
        \begin{itemize}
            \item[(a)] Let $\beta_{1[w_{2t-1}]}>0$, as in limited commitment. Depending on the sign of $C(\Delta h_{1t},\omega_{2t}) + 2C(\Delta h_{1t},\omega_{2t-1})$, measurement error could bias the numerator towards zero (full commitment) or upwards ($\widehat{\beta}_{1[w_{2t-1}]}$ remains in line with limited commitment). 

            \item[(b)] Let $\beta_{1[w_{2t-1}]}\leq0$, incompatible with limited commitment. In this case $|C(\Delta h_{1t},\omega_{2t})|<<|C(\Delta h_{1t},\omega_{2t-1})|$, which ensures $\text{num}(\beta_{1[w_{2t-1}]})\leq0$. Measurement error implies $\text{num}(\widehat{\beta}_{1[w_{2t-1}]})<\text{num}(\beta_{1[w_{2t-1}]})$, so $\widehat{\beta}_{1[w_{2t-1}]}$ remains inconsistent with bargaining.
        \end{itemize}

    \item Let $C(\Delta h_{1t},\omega_{2t})\leq0$ and $C(\Delta h_{1t},\omega_{2t-1})>0$. 
        \begin{itemize}
            \item[(a)] Let $\beta_{1[w_{2t-1}]}>0$, in line with limited commitment. In this case $|C(\Delta h_{1t},\omega_{2t})|<<|C(\Delta h_{1t},\omega_{2t-1})|$, which ensures $\text{num}(\beta_{1[w_{2t-1}]})>0$. Measurement error implies $\text{num}(\widehat{\beta}_{1[w_{2t-1}]})>\text{num}(\beta_{1[w_{2t-1}]})$, so $\widehat{\beta}_{1[w_{2t-1}]}$ remains consistent with bargaining.

            \item[(b)] Let $\beta_{1[w_{2t-1}]}\leq0$, incompatible with limited commitment. In this case $|C(\Delta h_{1t},\omega_{2t})|\geq|C(\Delta h_{1t},\omega_{2t-1})|$, which ensures $\text{num}(\beta_{1[w_{2t}]})\leq0$. Depending on the sign of $C(\Delta h_{1t},\omega_{2t}) + 2C(\Delta h_{1t},\omega_{2t-1})$, measurement error could bias the numerator downwards ($\widehat{\beta}_{1[w_{2t-1}]}$ remains incompatible with limited commitment) or towards zero (full commitment). It may flip its sign to positive if $|C(\Delta h_{1t},\omega_{2t})|<<|C(\Delta h_{1t},\omega_{2t-1})|$; however, this is a contradiction as $\beta_{1[w_{2t-1}]}$ would be positive in such case.
        \end{itemize}
\end{enumerate}
In brief: as in the case of $\widehat{\beta}_{1[w_{2t}]}$, in no case does measurement error erroneously move $\widehat{\beta}_{1[w_{2t-1}]}$ towards positive values (limited commitment) when the true $\beta_{1[w_{2t-1}]}$ is incompatible with bargaining. Any evidence for limited commitment through $\widehat{\beta}_{1[w_{2t-1}]}$, as we find in practice, is thus not the byproduct of measurement error.

\

\noindent \textbf{Coefficient on own past wage.} The numerator of $\widehat{\beta}_{1[w_{1t-1}]}$ is given by $\text{num}(\widehat{\beta}_{1[w_{1t-1}]}) = \text{num}(\beta_{1[w_{1t-1}]}) + \sigma^2_{e^{w_{1}}}(C(\Delta h_{1t},\omega_{1t}) + 2C(\Delta h_{1t},\omega_{1t-1})) + \sigma^2_{e^{h_{1}}}(V(\omega_{1t})+2C(\omega_{1t},\omega_{1t-1}))$, where $\text{num}(\beta_{1[w_{1t-1}]})$ is shown in \eqref{AppEq::IdentifiedParameters}. The second bracketed term is positive -- under e.g. a permanent-transitory process for $\omega_{1t}$, this term is equal to the variance of the permanent wage shock. By contrast, the first bracketed term may be positive or negative depending on the signs of $C(\Delta h_{1t},\omega_{1t})$ and $C(\Delta h_{1t},\omega_{1t-1})$, for which we distinguish four cases: 
\begin{enumerate}
    \item Let $C(\Delta h_{1t},\omega_{1t})>0$ and $C(\Delta h_{1t},\omega_{1t-1})>0$.\\
    Then $\text{num}(\beta_{1[w_{1t-1}]})$ is positive, and $\beta_{1[w_{1t-1}]}>0$, contradicting limited commitment. It follows that $\text{num}(\widehat{\beta}_{1[w_{1t-1}]})$ is biased upwards, so measurement error preserves the true sign of the parameter, and the estimate remains inconsistent with limited commitment. 

    \item Let $C(\Delta h_{1t},\omega_{1t})\leq0$ and $C(\Delta h_{1t},\omega_{1t-1})\leq0$.\\
    Then $\text{num}(\beta_{1[w_{1t-1}]})$ is negative, so $\beta_{1[w_{1t-1}]}<0$, in line with limited commitment. Whether $\text{num}(\widehat{\beta}_{1[w_{1t-1}]})$ is biased upwards (towards null bargaining effects; full commitment) or downwards (exaggerating the bargaining effect of past wages in limited commitment) depends on the relative magnitude of the first vs the second (positive) bracketed term, in absolute terms. The overall effect is such that we either erroneously \emph{fail} to reject full/no commitment (upward bias) or correctly reject it in favor of limited commitment. In either case, measurement error does not make us erroneously reject full/no commitment in favor of limited commitment.

    \item Let $C(\Delta h_{1t},\omega_{1t})>0$ and $C(\Delta h_{1t},\omega_{1t-1})\leq0$.\\
    Measurement error may induce upwards or downwards bias, depending on the relative magnitude of the covariances and the relative magnitude of the first vs second bracketed terms in $\text{num}(\widehat{\beta}_{1[w_{1t-1}]})$. Therefore the overall effect is ambiguous. 

    \item Let $C(\Delta h_{1t},\omega_{1t})\leq0$ and $C(\Delta h_{1t},\omega_{1t-1})>0$: Same as case \#3 above.
\end{enumerate}
In brief: in contrast to the coefficients on the partner's wages, the bias induced by measurement error on $\widehat{\beta}_{1[w_{1t-1}]}$ cannot always be signed.

\

\noindent \textbf{A note on the coefficient on own current wage.} While $\beta_{1[w_{1t}]}$ is not part of our test for commitment (see section \ref{SubSec::Test_Commitment_Formulation}), it is interesting to note that $\text{num}(\widehat{\beta}_{1[w_{1t}]}) = \text{num}(\beta_{1[w_{1t}]}) + \sigma^2_{e^{w_{1}}}(2C(\Delta h_{1t},\omega_{1t}) + C(\Delta h_{1t},\omega_{1t-1})) - \sigma^2_{e^{h_{1}}}(2V(\omega_{1t-1})+C(\omega_{1t},\omega_{1t-1})) - 3\sigma^2_{e^{w_{1}}}\sigma^2_{e^{h_{1}}}$, with $\text{num}(\beta_{1[w_{1t}]})$ shown in \eqref{AppEq::IdentifiedParameters}. For reasonable values of the hours/wage moments, $\text{num}(\widehat{\beta}_{1[w_{1t}]})$ is biased downwards and becomes negative if the variance of measurement error is large. $\widehat{\beta}_{1[w_{1t}]}$ is thus downwards biased, or even negative as we often estimate in the data, due to the negative correlation between error in concurrent hours and wages. This may also help explain the large values for $\widehat{\eta_{j0}^{w_{j}}}$ (which underlies $\widehat{\beta}_{1[w_{1t}]}$) that we find throughout the paper. 

\

\noindent \textbf{Correlation between error in spousal wages.} The characterization so far assumes uncorrelated errors in spousal wages. If error in male and female wages is positively correlated \citep[as is often implied given that wage shocks typically covary positively between spouses;][]{BlundellPistaSaporta2016Family_Labour}, the effects of concurrent spousal wages on labor supply are biased towards having similar signs. Limited commitment requires \emph{asymmetric} effects from the time $t-1$ spousal wages on one's labor supply, as we often find in the empirical application, so measurement error dampens this asymmetry, i.e. it operates against limited commitment.

\

\noindent \textbf{Summary.} Measurement error in wages and hours inflates the denominators in \eqref{AppEq::IdentifiedParameters}, pushing all parameters towards zero (null bargaining effects, full commitment). This is a classical consequence of measurement error. In the numerators, measurement error either preserves the true sign of the coefficients on the \emph{partner}'s wages, or it further biases them towards zero (full commitment). By contrast, measurement error can induce arbitrary bias on the coefficients on \emph{own} past wages, potentially erroneously flipping their true sign.

With measurement error, the test for commitment should rely primarily on the coefficients on the \emph{partner's} wage shocks. The results in the paper reveal that the cross-effects from the \emph{partner}'s current and past shocks are strongly in line with limited commitment. Any evidence for limited commitment through those parameters is not the byproduct of measurement error as in no case does measurement error erroneously flip their sign making them falsely consistent with limited commitment. Conducting the commitment test utilizing only those latter parameters (as opposed to also including the parameters pertaining to own shocks), we still strongly reject full and no commitment in favor of limited commitment.\footnote{Testing in this way, the average $p$ value for ${\cal H}_{0}^\text{FC}$ (${\cal H}_{0}^\text{NC}$) is $0.002$ ($0.003$) over the columns of table \ref{Table::OLS_Results}.}

\section{Discussion and extensions}\label{Appendix::Extensions}

\textbf{Home production.} Suppose that, as in \citet{LiseYamada2019}, the public good $q_{t}$ is produced domestically via the home production function $q_{t} = f(x_{t}, d_{1t}, d_{2t})$, where $x_{t}$ is money expenditure and $d_{kt}$ is spouse $k$'s housework time. We show in appendix \ref{Appendix::DerivationOptimalityConditions} that the estimating equation for hours is analogous to the baseline \eqref{Eq::EstimableEquation.Final}, namely
\begin{align}\tag{\ref{Eq::EstimableEquation.Final}.hp}\label{Eq::EstimableEquation.Final.HP}
\begin{split}
\Delta \log h_{jt}
    = ~\dots~
    &- \widetilde{\delta}_{jt} \widetilde{\zeta}_{j} f_{x} x_{t-1} h_{jt-1}^{-1} \Delta \log x_{t} \\
    &- \textstyle \sum_{k} \widetilde{\delta}_{jt} \left(\widetilde{\zeta}_{j} f_{d_{k}} -\boldsymbol{1}[j=k]\widetilde{\alpha}_j^{-1}\right)d_{kt-1} h_{jt-1}^{-1} \Delta \log d_{kt},\\
\end{split}
\end{align}
with the additive terms for money expenditure and the spouses' housework times replacing the term for $q_t$. Slightly redefined compared to the baseline, $\widetilde{\delta}_{jt}>0$ is given by one over $\kappa_{t} s_{jt-1} h_{jt-1}^{-1} - \widetilde{\alpha}_{j}^{-1}$, where $\widetilde{\alpha}_{j}<0$ is approximately equal to $j$'s Frisch elasticity of leisure scaled by his/her leisure hours. $f_{x}$ and $f_{d_{k}}$ denote the marginal productivities of money and time, $\widetilde{\zeta}_{j}$ reflects the nature of the consumption-leisure complementarity, and $\boldsymbol{1}$ is the indicator function. In essence, after controlling for shifts in expenditure and spousal chores, in addition to our controls for income, wealth, etc., any effect on hours of current and past distribution factors, including wages, operate exclusively through the Pareto weight. The exclusion restrictions on current and past distribution factors thus remain intact and our test retains its original form. 

We re-estimate all specifications allowing for home production and report results in appendix table \ref{AppTable::OLS_Results_All_ExtendedModel}.\footnote{We report results from the reduced form specification only; results from the other specifications lead to similar conclusions but we do not show them for brevity. Statistical significance is lower than in baseline because we estimate a more flexible model over a smaller sample (we do not observe housework for everyone).} Our main finding remains: own past wage shocks reduce labor supply while the partner's current and past shocks increase it, in line with limited commitment.

In obtaining the augmented equation in \eqref{Eq::EstimableEquation.Final.HP}, we put no restrictions on the spouses' relative productivities at home or on how strongly each of them likes the domestic good (captured by $\widetilde{\zeta}_j$) relative to leisure (captured by $\widetilde{\alpha}_j$). Chores will shift with wages/distribution factors reflecting preferences, productivities, and bargaining -- as such, additional predictions arise about the relationship between chores and wage/distribution factor histories. Unlike labor hours, however, we cannot sign these predictions without restricting the sign of $\widetilde{\zeta}_j$ and its magnitude relative to $\widetilde{\alpha}_j$. By using labor supply as outcome variable, our test maintains a level of generality that the use of chores cannot afford: by controlling for income, wealth, expenditure, and chores, all of which would otherwise adjust to the economic environment in a complicated way, we can sign the relationship between hours and the Pareto weight (and the variables therein), based solely on our regularity conditions on period utility. 

\begin{sidewaystable}  
\begin{center}
\caption{Commitment test -- reduced form results from model with home production}\label{AppTable::OLS_Results_All_ExtendedModel}
\begin{tabular}{L{2.0cm} C{0.6cm} C{0.7cm} C{0.6cm} C{0.01cm} C{1.5cm} C{1.6cm} C{0.01cm} C{1.5cm} C{1.6cm} C{0.01cm} C{1.7cm} C{1.7cm} C{0.01cm} C{1.7cm} C{1.7cm}}
\toprule
                            &&&&    & \mcl{5}{c}{\textbf{wage shocks}}        && \mcl{5}{c}{\textbf{wage shocks \& BMI}}               \\
\cmidrule{6-10}\cmidrule{12-16}                        
                            &&&&    & \mcl{2}{c}{$\geq$ 3 periods} && \mcl{2}{c}{$\geq$ 4 periods} && \mcl{2}{c}{$\geq$ 3 periods} && \mcl{2}{c}{$\geq$ 4 periods} \\
\cmidrule{6-7}\cmidrule{9-10}\cmidrule{12-13}\cmidrule{15-16}
                            &&&&    & (1)      & (2)      && (3)    & (4)     && (5)    & (6)       && (7)      & (8)   \\
                            &&&&    & Male     & Female   && Male   & Female  && Male   & Female    && Male     & Female   \\
                &FC&NC&LC&          & $j=1$    &  $j=2$   &&  $j=1$ &  $j=2$  && $j=1$  &  $j=2$    &&  $j=1$   &  $j=2$   \\
\midrule
\multicolumn{16}{l}{\emph{Current shocks} ($t$)}\\
~$\beta_{j[w_{jt}]}$ & . & . & . &  & $  -25.549$ & $  -11.152$ &  & $  -30.775$ & $  -12.895$ &  & $  -30.152$ & $  -11.763$ &  & $  -61.977$ & $   -6.541$ \\
 &  &  &  &  & $(   10.023)$ & $(    5.021)$ &  & $(   19.661)$ & $(   12.542)$ &  & $(    7.169)$ & $(    5.008)$ &  & $(   20.434)$ & $(   11.797)$ \\
~$\beta_{j[w_{-jt}]}$ & $0$ & $+$ & $+$ &  & $   56.635$ & $  -14.447$ &  & $   39.899$ & $  -22.833$ &  & $   54.620$ & $  -10.944$ &  & $   57.254$ & $  -16.832$ \\
 &  &  &  &  & $(   31.405)$ & $(   10.707)$ &  & $(   31.646)$ & $(   19.293)$ &  & $(   23.810)$ & $(   10.689)$ &  & $(   34.128)$ & $(   19.705)$ \\
~$\beta_{j[bmi_{jt}]}$ & $0$ & $+$ & $+$ &  &  &  &  &  &  &  & $  -67.533$ & $   70.908$ &  & $  114.434$ & $  197.030$ \\
 &  &  &  &  &  &  &  &  &  &  & $(  244.464)$ & $(   58.790)$ &  & $(  245.185)$ & $(   84.167)$ \\
~$\beta_{j[bmi_{-jt}]}$ & $0$ & $-$ & $-$ &  &  &  &  &  &  &  & $  461.408$ & $   11.030$ &  & $  480.751$ & $  -57.693$ \\
 &  &  &  &  &  &  &  &  &  &  & $(  171.107)$ & $(   66.830)$ &  & $(  190.153)$ & $(  103.215)$ \\
\multicolumn{16}{l}{\emph{Past shocks} ($t-1$)}\\
~$\beta_{j[w_{jt-1}]}$ & $0$ & $0$ & $-$ &  & $   -7.427$ & $   -2.800$ &  & $   -9.589$ & $   -2.917$ &  & $    5.743$ & $   -3.573$ &  & $    6.232$ & $   -0.214$ \\
 &  &  &  &  & $(    6.937)$ & $(    3.642)$ &  & $(   11.364)$ & $(    9.693)$ &  & $(    7.922)$ & $(    4.404)$ &  & $(   12.667)$ & $(   13.724)$ \\
~$\beta_{j[w_{-jt-1}]}$ & $0$ & $0$ & $+$ &  & $   31.111$ & $    4.650$ &  & $   33.291$ & $  -14.557$ &  & $   43.115$ & $    6.801$ &  & $   52.354$ & $  -20.876$ \\
 &  &  &  &  & $(   21.521)$ & $(   12.139)$ &  & $(   35.108)$ & $(   18.498)$ &  & $(   18.871)$ & $(   11.788)$ &  & $(   32.750)$ & $(   18.765)$ \\
~$\beta_{j[bmi_{jt-1}]}$ & $0$ & $0$ & $+$ &  &  &  &  &  &  &  & $  165.042$ & $  120.685$ &  & $    8.008$ & $  201.795$ \\
 &  &  &  &  &  &  &  &  &  &  & $(  217.238)$ & $(   54.733)$ &  & $(  233.759)$ & $(  111.799)$ \\
~$\beta_{j[bmi_{-jt-1}]}$ & $0$ & $0$ & $-$ &  &  &  &  &  &  &  & $ -217.787$ & $   22.307$ &  & $ -299.162$ & $   12.967$ \\
 &  &  &  &  &  &  &  &  &  &  & $(  106.018)$ & $(   61.415)$ &  & $(  146.144)$ & $(   94.813)$ \\
\multicolumn{16}{l}{\emph{Older shocks} ($t-2$)}\\
~$\beta_{j[w_{jt-2}]}$ & $0$ & $0$ & $-$ &  &  &  &  & $  -26.658$ & $   -6.506$ &  &  &  &  & $  -56.900$ & $    4.310$ \\
 &  &  &  &  &  &  &  & $(   16.885)$ & $(    7.124)$ &  &  &  &  & $(   18.159)$ & $(    8.107)$ \\
~$\beta_{j[w_{-jt-2}]}$ & $0$ & $0$ & $+$ &  &  &  &  & $    0.310$ & $   14.647$ &  &  &  &  & $   33.901$ & $   14.740$ \\
 &  &  &  &  &  &  &  & $(   30.527)$ & $(   18.766)$ &  &  &  &  & $(   33.748)$ & $(   19.389)$ \\
~$\beta_{j[bmi_{jt-2}]}$ & $0$ & $0$ & $+$ &  &  &  &  &  &  &  &  &  &  & $ -506.887$ & $  147.853$ \\
 &  &  &  &  &  &  &  &  &  &  &  &  &  & $(  265.048)$ & $(  103.282)$ \\
~$\beta_{j[bmi_{-jt-2}]}$ & $0$ & $0$ & $-$ &  &  &  &  &  &  &  &  &  &  & $ -244.176$ & $   42.594$ \\
 &  &  &  &  &  &  &  &  &  &  &  &  &  & $(   84.383)$ & $(   87.121)$ \\
\bottomrule
\end{tabular}
\end{center}
\end{sidewaystable}

\begin{sidewaystable}  
\begin{center}
\caption*{Table \ref{AppTable::OLS_Results_All_ExtendedModel} (continued): Commitment test -- reduced form results from model with home production}
\begin{tabular}{L{1.9cm} C{0.6cm} C{0.7cm} C{0.6cm} C{0.001cm} C{1.7cm} C{1.7cm} C{0.001cm} C{1.7cm} C{1.7cm} C{0.001cm} C{1.7cm} C{1.6cm} C{0.001cm} C{1.7cm} C{1.7cm}}
\toprule
                            &&&&    & \mcl{5}{c}{\textbf{wage shocks}}        && \mcl{5}{c}{\textbf{wage shocks \& BMI}}               \\
\cmidrule{6-10}\cmidrule{12-16}                        
                            &&&&    & \mcl{2}{c}{$\geq$ 3 periods} && \mcl{2}{c}{$\geq$ 4 periods} && \mcl{2}{c}{$\geq$ 3 periods} && \mcl{2}{c}{$\geq$ 4 periods} \\
\cmidrule{6-7}\cmidrule{9-10}\cmidrule{12-13}\cmidrule{15-16}
                            &&&&    & (1)      & (2)      && (3)    & (4)     && (5)    & (6)       && (7)      & (8)   \\
                            &&&&    & Male     & Female   && Male   & Female  && Male   & Female    && Male     & Female   \\
                &FC&NC&LC&          & $j=1$    &  $j=2$   &&  $j=1$ &  $j=2$  && $j=1$  &  $j=2$    &&  $j=1$   &  $j=2$   \\
\midrule
\multicolumn{16}{l}{\emph{Initial distribution factors} ($t=0$)}\\
~$\beta_{j[young_{j}]}$ & $0$ & $0$ & $-$ &  & $  -49.788$ & $  -28.257$ &  & $  -95.829$ & $  -65.286$ &  & $  -59.983$ & $  -26.883$ &  & $ -126.686$ & $  -60.522$ \\
 &  &  &  &  & $(   32.410)$ & $(    9.332)$ &  & $(   37.009)$ & $(   14.616)$ &  & $(   33.217)$ & $(   10.036)$ &  & $(   39.893)$ & $(   17.209)$ \\
~$\beta_{j[young_{-j}]}$ & $0$ & $0$ & $+$ &  & $  -64.736$ & $   11.741$ &  & $ -120.887$ & $   10.585$ &  & $  -44.975$ & $    8.848$ &  & $  -97.870$ & $   -0.621$ \\
 &  &  &  &  & $(   37.702)$ & $(    5.023)$ &  & $(   39.494)$ & $(   11.562)$ &  & $(   33.140)$ & $(    5.924)$ &  & $(   38.654)$ & $(   14.652)$ \\
\multicolumn{16}{l}{\emph{Other controls}}\\
~$ b_{j[\Delta y_{t}]}$ &  &  &  &  & $  310.452$ & $  113.351$ &  & $  310.220$ & $  192.609$ &  & $  346.924$ & $  113.656$ &  & $  429.084$ & $  232.623$ \\
 &  &  &  &  & $(   47.264)$ & $(   45.884)$ &  & $(   49.267)$ & $(   54.192)$ &  & $(   43.983)$ & $(   45.847)$ &  & $(   53.484)$ & $(   50.342)$ \\
~$ b_{j[\Delta a_{t}]}$ &  &  &  &  & $    5.151$ & $   -9.304$ &  & $   -2.575$ & $   -3.966$ &  & $    4.384$ & $   -7.786$ &  & $  -31.149$ & $   -0.976$ \\
 &  &  &  &  & $(   12.928)$ & $(    4.393)$ &  & $(   15.728)$ & $(    6.417)$ &  & $(   12.234)$ & $(    4.980)$ &  & $(   15.617)$ & $(    7.270)$ \\
~$ b_{j[y_{t-1}]}$ &  &  &  &  & $   23.440$ & $    5.443$ &  & $   33.936$ & $    9.473$ &  & $   23.362$ & $    1.449$ &  & $   48.163$ & $    6.718$ \\
 &  &  &  &  & $(    7.386)$ & $(    3.273)$ &  & $(    8.768)$ & $(    7.075)$ &  & $(    6.684)$ & $(    3.750)$ &  & $(    9.293)$ & $(    6.280)$ \\
~$ b_{j[a_{t-1}]}$ &  &  &  &  & $  -24.879$ & $   -5.510$ &  & $  -33.031$ & $   -9.323$ &  & $  -25.429$ & $   -2.144$ &  & $  -46.819$ & $   -7.319$ \\
 &  &  &  &  & $(    7.487)$ & $(    2.920)$ &  & $(    8.513)$ & $(    6.053)$ &  & $(    6.529)$ & $(    3.324)$ &  & $(    8.666)$ & $(    5.540)$ \\
~$ b_{j[\Delta y_{-jt}]}$ &  &  &  &  & $ -300.183$ & $  -86.503$ &  & $ -267.806$ & $ -161.406$ &  & $ -314.290$ & $  -90.450$ &  & $ -339.960$ & $ -207.910$ \\
 &  &  &  &  & $(   67.033)$ & $(   39.956)$ &  & $(   63.662)$ & $(   45.959)$ &  & $(   53.737)$ & $(   40.753)$ &  & $(   62.721)$ & $(   43.294)$ \\
~$ b_{j[\Delta x_{t}]}$ &  &  &  &  & $    0.001$ & $   -0.000$ &  & $    0.001$ & $    0.000$ &  & $    0.001$ & $   -0.000$ &  & $    0.001$ & $    0.000$ \\
 &  &  &  &  & $(    0.001)$ & $(    0.000)$ &  & $(    0.001)$ & $(    0.000)$ &  & $(    0.001)$ & $(    0.000)$ &  & $(    0.001)$ & $(    0.000)$ \\
~$ b_{j[\Delta d_{jt}]}$ &  &  &  &  & $   -0.278$ & $    0.320$ &  & $   -0.207$ & $    0.963$ &  & $   -0.099$ & $    0.182$ &  & $   -1.528$ & $    0.952$ \\
 &  &  &  &  & $(    0.939)$ & $(    0.167)$ &  & $(    1.171)$ & $(    0.473)$ &  & $(    0.924)$ & $(    0.171)$ &  & $(    1.303)$ & $(    0.568)$ \\
~$ b_{j[\Delta d_{-jt}]}$ &  &  &  &  & $    0.474$ & $    0.439$ &  & $   -0.185$ & $   -0.341$ &  & $    1.287$ & $    0.525$ &  & $    1.420$ & $    0.429$ \\
 &  &  &  &  & $(    0.889)$ & $(    0.642)$ &  & $(    1.142)$ & $(    1.259)$ &  & $(    1.035)$ & $(    0.697)$ &  & $(    1.269)$ & $(    1.243)$ \\
~$ b_{j[\Delta a_{t-1}]}$ &  &  &  &  & $   -6.867$ & $   -7.170$ &  & $   -5.028$ & $   -1.096$ &  & $  -10.163$ & $   -5.008$ &  & $  -13.208$ & $    0.214$ \\
 &  &  &  &  & $(    3.933)$ & $(    2.085)$ &  & $(    5.664)$ & $(    3.917)$ &  & $(    4.078)$ & $(    2.399)$ &  & $(    5.893)$ & $(    4.318)$ \\
~$ b_{j[\Delta a_{t-2}]}$ &  &  &  &  &  &  &  & $   -3.949$ & $    2.850$ &  &  &  &  & $   -3.159$ & $    3.804$ \\
 &  &  &  &  &  &  &  & $(    4.073)$ & $(    2.286)$ &  &  &  &  & $(    4.030)$ & $(    2.932)$ \\
\noalign{\smallskip}
\mcl{5}{l}{~\text{$ p$ value for }$ {\cal H}_{0}^\text{FC}$} & $    0.028$ & $    0.001$ &  & $    0.002$ & $    0.000$ &  & $    0.000$ & $    0.005$ &  & $    0.000$ & $    0.000$ \\
\mcl{5}{l}{~\text{$ p$ value for }$ {\cal H}_{0}^\text{NC}$} & $    0.055$ & $    0.002$ &  & $    0.005$ & $    0.000$ &  & $    0.004$ & $    0.004$ &  & $    0.000$ & $    0.000$ \\
\noalign{\smallskip}
\mcl{5}{l}{Observations}                       		& \mcl{2}{c}{7,673}    	&& \mcl{2}{c}{5,461} 	&& \mcl{2}{c}{6,862}  	&&  \mcl{2}{c}{4,785} \\
\bottomrule 
\end{tabular}
\caption*{\fsz\emph{Notes:} The table reports the coefficient estimates from the reduced form (first) specification of the commitment test with wages (columns 1-4) and wages and BMI (columns 5-8) as time-varying distribution factors. Standard errors clustered at the household level are in brackets.}
\end{center}
\end{sidewaystable}

\

\noindent \textbf{Private leisure.} The regularity conditions on period utility follow in fact from our implicit assumption that labor supply and leisure are negatively related. This allows us to sign the bargaining effects of wages/distribution factors on hours and implement our test empirically using readily available data on labor supply. Any true improvement in bargaining power, however, must be accompanied by an increase in own leisure and a simultaneous decrease in the partner's. While the characterization of the effects on leisure is straightforward, their implementation is not, because leisure is not observed in our data. 

We take an indirect route to define leisure as total hours minus labor supply, commuting, and chores.\footnote{We set leisure $l_{jt}=5840-h_{jt}-m_{jt}-d_{jt}$, where 5840 is total annual time based on 16 productive hours per day. Commuting $m_{jt}$ is not observed before 2011, so we set $m_{jt}=0.1 \times h_{jt}$ reflecting the average commuting time in the PSID whenever we observe it.} We run the reduced form and structural specifications of our test using leisure as the outcome variable and report results in tables \ref{AppTable::OLS_Results_Leisure_All}-\ref{AppTable::Structural_Results_Leisure_All}. Own past shocks \emph{increase} leisure while the partner's past shocks \emph{reduce} it, in line with limited commitment. We reject full and no commitment based on female leisure, but we fail to reject no commitment using male leisure as our estimates there lack statistical significance. Yet, any effect of past shock that \emph{is} significant has the sign that limited commitment requires. Current partner shocks, by contrast, often produce counterintuitive results (such shocks seem to increase own leisure although they should reduce it), though those results may partly reflect omitted \emph{joint} leisure.

\begin{table}  
\begin{center}
\caption{Commitment test based on leisure -- reduced form results in detail}\label{AppTable::OLS_Results_Leisure_All}
\resizebox{0.85\columnwidth}{!}{
\begin{tabular}{L{2.0cm} C{0.6cm} C{0.7cm} C{0.6cm} C{0.01cm} C{2.0cm} C{2.0cm} C{0.01cm} C{2.0cm} C{2.0cm}}
\toprule
                            &&&&    & \mcl{5}{c}{\textbf{wage shocks}}        \\
\cmidrule{6-10}                        
                            &&&&    & \mcl{2}{c}{$\geq$ 3 periods} && \mcl{2}{c}{$\geq$ 4 periods} \\
\cmidrule{6-7}\cmidrule{9-10}
\emph{Dependent}            &&&&    & (1)      & (2)      && (3)    & (4)     \\
\emph{variable:}            &&&&    & Male     & Female   && Male   & Female  \\
$\Delta \log l_{jt}$   &FC&NC&LC&   & $j=1$    &  $j=2$   &&  $j=1$ &  $j=2$  \\
\midrule
\multicolumn{10}{l}{\emph{Current shocks} ($t$)}\\
~~$\beta_{j[w_{jt}]}$ & . & . & . &  & $  459.178$ & $  569.196$ &  & $  592.072$ & $  599.834$ \\
 &  &  &  &  & $(   72.891)$ & $(   33.432)$ &  & $(   94.998)$ & $(   36.953)$ \\
~~$\beta_{j[w_{-jt}]}$ & $0$ & $-$ & $-$ &  & $   46.103$ & $   96.173$ &  & $   39.169$ & $   72.344$ \\
 &  &  &  &  & $(   34.099)$ & $(   24.157)$ &  & $(   30.023)$ & $(   30.430)$ \\
\multicolumn{10}{l}{\emph{Past shocks} ($t-1$)}\\
~~$\beta_{j[w_{jt-1}]}$ & $0$ & $0$ & $+$ &  & $   35.242$ & $   91.268$ &  & $  -10.409$ & $   72.237$ \\
 &  &  &  &  & $(   40.048)$ & $(   20.035)$ &  & $(   45.968)$ & $(   26.648)$ \\
~~$\beta_{j[w_{-jt-1}]}$ & $0$ & $0$ & $-$ &  & $  -22.435$ & $  -24.545$ &  & $   -6.118$ & $  -84.716$ \\
 &  &  &  &  & $(   24.346)$ & $(   16.485)$ &  & $(   29.420)$ & $(   22.299)$ \\                    
\multicolumn{10}{l}{\emph{Older shocks} ($t-2$)}\\
~~$\beta_{j[w_{jt-2}]}$ & $0$ & $0$ & $+$ &  &  &  &  & $  -22.233$ & $   19.368$ \\
 &  &  &  &  &  &  &  & $(   29.324)$ & $(   20.637)$ \\
~~$\beta_{j[w_{-jt-2}]}$ & $0$ & $0$ & $-$ &  &  &  &  & $    0.996$ & $  -59.648$ \\
 &  &  &  &  &  &  &  & $(   29.428)$ & $(   17.649)$ \\                         
\multicolumn{10}{l}{\emph{Initial distribution factors} ($t=0$)}\\
~~$\beta_{j[young_{j}]}$ & $0$ & $0$ & $+$ &  & $  -13.108$ & $   29.098$ &  & $  -79.712$ & $   22.037$ \\
 &  &  &  &  & $(   52.345)$ & $(   21.910)$ &  & $(   65.734)$ & $(   24.147)$ \\
~~$\beta_{j[young_{-j}]}$ & $0$ & $0$ & $-$ &  & $   -5.751$ & $   31.320$ &  & $  -24.987$ & $  -20.103$ \\
 &  &  &  &  & $(   26.747)$ & $(   26.097)$ &  & $(   24.430)$ & $(   30.514)$ \\
\multicolumn{10}{l}{\emph{Other controls}}\\
~$ b_{j[\Delta y_{t}]}$ &  &  &  &  & $ -729.919$ & $ -806.207$ &  & $-1087.301$ & $-1098.749$ \\
 &  &  &  &  & $(  135.413)$ & $(   78.293)$ &  & $(  118.935)$ & $(   81.506)$ \\
~$ b_{j[\Delta a_{t}]}$ &  &  &  &  & $   -8.522$ & $   39.493$ &  & $    1.672$ & $   28.742$ \\
 &  &  &  &  & $(   14.501)$ & $(   10.007)$ &  & $(   14.792)$ & $(   10.076)$ \\
~$ b_{j[y_{t-1}]}$ &  &  &  &  & $   -3.398$ & $  -28.430$ &  & $   18.362$ & $  -19.162$ \\
 &  &  &  &  & $(   18.312)$ & $(    8.337)$ &  & $(   10.405)$ & $(    9.901)$ \\
~$ b_{j[a_{t-1}]}$ &  &  &  &  & $    1.710$ & $   24.813$ &  & $  -18.205$ & $   18.728$ \\
 &  &  &  &  & $(   17.396)$ & $(    7.768)$ &  & $(    9.981)$ & $(    9.185)$ \\
~$ b_{j[\Delta y_{-jt}]}$ &  &  &  &  & $  452.830$ & $  546.818$ &  & $  692.852$ & $  791.918$ \\
 &  &  &  &  & $(   97.529)$ & $(   69.182)$ &  & $(   94.012)$ & $(   79.334)$ \\
~$ b_{j[\Delta q_{t}]}$ &  &  &  &  & $   -0.001$ & $   -0.001$ &  & $    0.000$ & $    0.000$ \\
 &  &  &  &  & $(    0.001)$ & $(    0.000)$ &  & $(    0.001)$ & $(    0.000)$ \\
~$ b_{j[\Delta a_{t-1}]}$ &  &  &  &  & $    0.105$ & $   -4.320$ &  & $   15.964$ & $   12.856$ \\
 &  &  &  &  & $(    6.650)$ & $(    6.121)$ &  & $(    4.175)$ & $(    4.370)$ \\
~$ b_{j[\Delta a_{t-2}]}$ &  &  &  &  &  &  &  & $    7.785$ & $    2.907$ \\
 &  &  &  &  &  &  &  & $(    2.912)$ & $(    2.901)$ \\
\noalign{\smallskip}
\mcl{5}{l}{~\text{$ p$ value for }$ {\cal H}_{0}^\text{FC}$} & $    0.589$ & $    0.000$ &  & $    0.121$ & $    0.000$ \\
\mcl{5}{l}{~\text{$ p$ value for }$ {\cal H}_{0}^\text{NC}$} & $    0.734$ & $    0.000$ &  & $    0.579$ & $    0.001$ \\
\noalign{\smallskip}
\mcl{5}{l}{Observations}                       & \mcl{2}{c}{8,411}    && \mcl{2}{c}{5,960} \\
\bottomrule 
\end{tabular}}
\caption*{\fsz\emph{Notes:} The table reports the coefficient estimates from the reduced form (first) specification of the commitment test with wages as time-varying distribution factors, using leisure as the outcome variable. Leisure  $l_{jt}$ is defined as 5840 hours (total annual time based on 16 productive hours per day), minus labor supply, chores, and commuting. The sample sizes are smaller than in table \ref{Table::OLS_Results} because we remove observations with negative leisure. Standard errors clustered at the household level are in brackets.}
\end{center}
\end{table}

\begin{table}  
\begin{center}
\caption{Commitment test based on leisure -- structural results in detail}\label{AppTable::Structural_Results_Leisure_All}
\resizebox{0.81\columnwidth}{!}{
\begin{tabular}{L{1.8cm} C{0.6cm} C{0.7cm} C{0.6cm} C{0.01cm} C{1.8cm} C{1.8cm} C{0.01cm} C{1.8cm} C{1.8cm}}
\toprule
                            &&&&    & \mcl{5}{c}{\textbf{wage shocks}}        \\
\cmidrule{6-10}                        
                            &&&&    & \mcl{2}{c}{$\geq$ 3 periods} && \mcl{2}{c}{$\geq$ 4 periods} \\
\cmidrule{6-7}\cmidrule{9-10}
\mcl{5}{l}{\emph{Dependent}}        & (1)      & (2)      && (3)    & (4)     \\
\mcl{5}{l}{\emph{variable:}}        & Male     & Female   && Male   & Female  \\
$\Delta \log l_{jt}$  &FC&NC&LC&    & $j=1$    &  $j=2$   &&  $j=1$ &  $j=2$  \\
\midrule
\multicolumn{10}{l}{\emph{Pareto weight elasticities w.r.t current factors} ($\tau=0$)}\\
~~$\eta_{j0}^{w_{j}}$ & $0$ & $+$ & $+$ &  & $    1.070$ & $    1.113$ &  & $    1.047$ & $    1.060$ \\
 &  &  &  &  & $(    0.028)$ & $(    0.022)$ &  & $(    0.020)$ & $(    0.012)$ \\
~~$\eta_{j0}^{w_{-j}}$ & $0$ & $-$ & $-$ &  & $    0.014$ & $    0.029$ &  & $    0.009$ & $    0.015$ \\
 &  &  &  &  & $(    0.015)$ & $(    0.009)$ &  & $(    0.008)$ & $(    0.007)$ \\
\multicolumn{10}{l}{\emph{Pareto weight elasticities w.r.t factors 1 period in the past} ($\tau=1$)}\\
~~$\eta_{j1}^{w_{j}}$ & $0$ & $0$ & $+$ &  & $   -0.002$ & $    0.026$ &  & $   -0.014$ & $    0.012$ \\
 &  &  &  &  & $(    0.011)$ & $(    0.008)$ &  & $(    0.012)$ & $(    0.007)$ \\
~~$\eta_{j1}^{w_{-j}}$ & $0$ & $0$ & $-$ &  & $   -0.007$ & $   -0.006$ &  & $   -0.003$ & $   -0.017$ \\
 &  &  &  &  & $(    0.009)$ & $(    0.005)$ &  & $(    0.006)$ & $(    0.005)$ \\
\multicolumn{10}{l}{\emph{Pareto weight elasticities w.r.t factors 2 periods in the past} ($\tau=2$)}\\  
~~$\eta_{j2}^{w_{j}}$ & $0$ & $0$ & $+$ &  &  &  &  & $   -0.009$ & $    0.002$ \\
 &  &  &  &  &  &  &  & $(    0.008)$ & $(    0.004)$ \\
~~$\eta_{j2}^{w_{-j}}$ & $0$ & $0$ & $-$ &  &  &  &  & $   -0.004$ & $   -0.012$ \\
 &  &  &  &  &  &  &  & $(    0.008)$ & $(    0.004)$ \\                    
\multicolumn{10}{l}{\emph{Pareto weight elasticities w.r.t initial distribution factors}}\\
~~$\eta_{jt}^{young_{j}}$ & $0$ & $0$ & $+$ &  & $   -0.005$ & $    0.009$ &  & $   -0.023$ & $    0.005$ \\
 &  &  &  &  & $(    0.020)$ & $(    0.008)$ &  & $(    0.018)$ & $(    0.005)$ \\
~~$\eta_{jt}^{young_{-j}}$ & $0$ & $0$ & $-$ &  & $   -0.002$ & $    0.008$ &  & $   -0.007$ & $   -0.007$ \\
 &  &  &  &  & $(    0.009)$ & $(    0.009)$ &  & $(    0.006)$ & $(    0.007)$ \\ 
\mcl{5}{l}{\emph{Frisch elasticity} ${\alpha_{j}/\mathbb{E}(l_{jt-1})}^{\#}$} & $   -0.960$ & $   -1.046$ &  & $   -1.372$ & $   -1.547$ \\
 &  &  &  &  & $(    0.210)$ & $(    0.138)$ &  & $(    0.197)$ & $(    0.165)$ \\
\multicolumn{10}{l}{\emph{Other terms}}\\
~$\ell_{\Delta y}$ &  &  &  &  & $    0.285$ & $    0.270$ &  & $    0.288$ & $    0.256$ \\
 &  &  &  &  & $(    0.044)$ & $(    0.017)$ &  & $(    0.031)$ & $(    0.014)$ \\
\mcl{5}{l}{~$\ell_{\Delta a}-\eta_{j0}^{a}$} & $    0.001$ & $   -0.011$ &  & $   -0.000$ & $   -0.005$ \\
 &  &  &  &  & $(    0.005)$ & $(    0.004)$ &  & $(    0.004)$ & $(    0.002)$ \\
~$\ell_{y}$ &  &  &  &  & $    0.004$ & $    0.008$ &  & $   -0.004$ & $    0.004$ \\
 &  &  &  &  & $(    0.006)$ & $(    0.003)$ &  & $(    0.003)$ & $(    0.002)$ \\
~$\ell_{a}$ &  &  &  &  & $   -0.004$ & $   -0.007$ &  & $    0.004$ & $   -0.004$ \\
 &  &  &  &  & $(    0.006)$ & $(    0.003)$ &  & $(    0.003)$ & $(    0.002)$ \\
~$ {\zeta_{j}}^{\#\#}$ &  &  &  &  & $   -0.002$ & $   -0.002$ &  & $    0.000$ & $    0.001$ \\
 &  &  &  &  & $(    0.002)$ & $(    0.002)$ &  & $(    0.002)$ & $(    0.001)$ \\
~$\eta_{j1}^{a}$ &  &  &  &  & $   -0.001$ & $    0.001$ &  & $   -0.004$ & $   -0.003$ \\
 &  &  &  &  & $(    0.002)$ & $(    0.002)$ &  & $(    0.001)$ & $(    0.001)$ \\
~$\eta_{j2}^{a}$ &  &  &  &  &  &  &  & $   -0.002$ & $   -0.001$ \\
 &  &  &  &  &  &  &  & $(    0.001)$ & $(    0.001)$ \\
\noalign{\smallskip}
\mcl{5}{l}{~\text{$ p$ value for }$ {\cal H}_{0}^\text{FC}$} & $    0.000$ & $    0.000$ &  & $    0.000$ & $    0.000$ \\
\mcl{5}{l}{~\text{$ p$ value for }$ {\cal H}_{0}^\text{NC}$} & $    0.937$ & $    0.024$ &  & $    0.331$ & $    0.014$ \\
\noalign{\smallskip}
\mcl{5}{l}{Observations}                       & \mcl{2}{c}{8,411}    && \mcl{2}{c}{5,960} \\
\bottomrule 
\end{tabular}}
\caption*{\fsz\emph{Notes:} The table reports the parameter estimates from the structural (second) specification of the commitment test with wages as time-varying distribution factors, using leisure as the outcome variable. See notes in table \ref{AppTable::OLS_Results_Leisure_All} for comments on leisure and sample sizes. $^\#$We report the Frisch elasticity of leisure at the sample average of leisure hours; its standard error is calculated with the delta method. $^{\#\#}$We multiply $\zeta_{j}$ by $10^4$ for legibility.}
\end{center}
\end{table}

\

\noindent \textbf{Private consumption, joint leisure.} Extensions to private consumption \citep[][]{LiseYamada2019} and joint leisure \citep[][]{CosaertTheloudisVerheyden2022Togetherness} are straightforward: the estimating equation includes terms for own private consumption and joint leisure, but it is otherwise similar to the baseline. The commitment test thus retains its original form based on the (current and past values of the) variables that enter the Pareto weight. 

As an illustration, appendix \ref{Appendix::DerivationOptimalityConditions} derives the estimating equation for hours when individual preferences are given by $u_{j}(q_{t},c_{jt},h_{jt};\boldsymbol{\xi}_{jt})$, where $c_{jt}$ is spouse $j$'s private consumption. The estimating equation is then given by
\begin{equation*}
	\Delta \log h_{jt} = ~\dots~ - \delta_{jt} \iota_{j} c_{jt-1} h_{jt-1}^{-1} \Delta \log c_{jt},
\end{equation*}
which is analogous to the baseline \eqref{Eq::EstimableEquation.Final} up to an additive term for the growth in own private consumption $\Delta \log c_{jt}$. $\iota_{j}$ denotes the nature of the complementarity between hours and private consumption. What is important is that, conditional on private and joint consumption, income, wealth etc., any remaining effects from current and past distribution factors operate \emph{exclusively} through the Pareto weight. This is true without restricting how preferences over private versus public consumption compare between spouses, exactly as we did not restrict how preferences over leisure/joint consumption compared between spouses in the baseline. The case of joint leisure is similar. In addition, private consumption may be used as an additional outcome variable, thus offering additional predictions for the role of current and historical shocks. We do not take these features to the data because the PSID lacks the necessary information on private expenditures or joint leisure.

\

\noindent \textbf{Endogenous human capital.} Suppose that, as in \citet{EcksteinWolpin1989DynamicLabourForceParticipation}, wages depend on human capital that accumulates endogenously over time. Specifically, let individual hourly wages (abstracting from a deterministic profile) be given by $w_{jt} = v_{jt}e_{jt}$, where $v_{jt}$ reflects a stochastic productivity component and $e_{jt}$ indicates spouse $j$'s human capital. We think of $e_{jt}$ as a type of work experience, e.g. the cumulative number of hours one has spent working up to time $t$, though its definition can be broadened. Human capital accumulates endogenously through the labor supply choices the individual makes. We express this as
\begin{equation*}
	e_{jt+1} = f_{j}(e_{jt},h_{jt}),
\end{equation*}
where $f_{j}$ loads current human capital (perhaps after some depreciation) and labor supply onto future human capital; we assume the partial derivative of $f_{j}$ with respect to each of its arguments to be weakly positive. The linear law of motion for $e_{jt+1}$ that is popular in the literature is a special case of this. 

Endogenous human capital affects the couple's problem through the budget constraint and by changing the dynamic incentives of work in the present versus the future. Under limited commitment, individual human capital also affects the continuation value as single. Therefore, $e_{jt}$ is an endogenous state variable that affects both the inside value in marriage and the outside value upon divorce. Moreover, and unlike financial wealth, human capital is in principle not directly transferable between spouses.

We show in appendix \ref{Appendix::DerivationOptimalityConditions} that, in this case, the estimating equation for hours at $t$ is similar to the baseline \eqref{Eq::EstimableEquation.Final} up to two additive terms for the growth in individual human capital between $t-1$ and $t$ and for its level at $t-1$. These terms capture the growth in the unobserved multiplier on $j$'s human capital law of motion, which enters the household program as a standalone constraint. What is important is that the coefficient on the Pareto weight remains signed, which allows us to test for commitment in the exact same way as in the baseline. That is, if we control for human capital similarly to how we control for assets, the test retains its original form based on the current and historical values of the variables that enter the Pareto weight. In the case of wages, the relevant distribution factor is the shock to the \emph{exogenous} part of wages, namely to $v_{jt}$ in the above notation.

\begin{table}  
\begin{center}
\caption{Commitment test with endogenous human capital -- reduced form results in detail}\label{AppTable::OLS_Results_Endogenous_Human_Capital}
\resizebox{0.82\columnwidth}{!}{
\begin{tabular}{L{2.0cm} C{0.6cm} C{0.7cm} C{0.6cm} C{0.01cm} C{2.0cm} C{2.0cm} C{0.01cm} C{2.0cm} C{2.0cm}}
\toprule
                            &&&&    & \mcl{5}{c}{\textbf{wage shocks}}        \\
\cmidrule{6-10}                        
                            &&&&    & \mcl{2}{c}{$\geq$ 3 periods} && \mcl{2}{c}{$\geq$ 4 periods} \\
\cmidrule{6-7}\cmidrule{9-10}
                            &&&&    & (1) Male     & (2) Female   && (3) Male   & (4) Female  \\
                            &FC&NC&LC&   & $j=1$    &  $j=2$   &&  $j=1$ &  $j=2$  \\
\midrule
\multicolumn{10}{l}{\emph{Current shocks} ($t$)}\\
~$\beta_{j[w_{jt}]}$ & . & . & . &  & $  -24.419$ & $  -10.167$ &  & $  -25.316$ & $  -20.853$ \\
 &  &  &  &  & $(   11.964)$ & $(    3.677)$ &  & $(   24.684)$ & $(   13.279)$ \\
~$\beta_{j[w_{-jt}]}$ & $0$ & $+$ & $+$ &  & $   35.257$ & $    0.483$ &  & $  -22.749$ & $   -0.129$ \\
 &  &  &  &  & $(   27.717)$ & $(    6.467)$ &  & $(   29.808)$ & $(   20.705)$ \\
\multicolumn{10}{l}{\emph{Past shocks} ($t-1$)}\\
~$\beta_{j[w_{jt-1}]}$ & $0$ & $0$ & $-$ &  & $  -12.167$ & $   -6.270$ &  & $   -0.329$ & $  -12.100$ \\
 &  &  &  &  & $(    6.543)$ & $(    3.477)$ &  & $(   10.800)$ & $(   11.227)$ \\
~$\beta_{j[w_{-jt-1}]}$ & $0$ & $0$ & $+$ &  & $   42.610$ & $    7.310$ &  & $    7.678$ & $   -5.619$ \\
 &  &  &  &  & $(   23.131)$ & $(    5.881)$ &  & $(   33.231)$ & $(   19.064)$ \\                     
\multicolumn{10}{l}{\emph{Older shocks} ($t-2$)}\\
~$\beta_{j[w_{jt-2}]}$ & $0$ & $0$ & $-$ &  &  &  &  & $  -37.894$ & $  -16.565$ \\
 &  &  &  &  &  &  &  & $(   19.728)$ & $(    8.625)$ \\
~$\beta_{j[w_{-jt-2}]}$ & $0$ & $0$ & $+$ &  &  &  &  & $  -17.231$ & $   17.156$ \\
 &  &  &  &  &  &  &  & $(   27.125)$ & $(   14.508)$ \\                       
\multicolumn{10}{l}{\emph{Initial distribution factors} ($t=0$)}\\
~$\beta_{j[young_{j}]}$ & $0$ & $0$ & $-$ &  & $  -20.543$ & $  -13.395$ &  & $  -68.994$ & $  -45.443$ \\
 &  &  &  &  & $(   26.756)$ & $(    6.566)$ &  & $(   31.762)$ & $(   12.657)$ \\
~$\beta_{j[young_{-j}]}$ & $0$ & $0$ & $+$ &  & $  -85.861$ & $    6.523$ &  & $ -109.423$ & $   16.438$ \\
 &  &  &  &  & $(   29.068)$ & $(    5.469)$ &  & $(   41.651)$ & $(   11.041)$ \\
\multicolumn{10}{l}{\emph{Other controls}}\\
~$ b_{j[\Delta y_{t}]}$ &  &  &  &  & $  289.839$ & $  114.454$ &  & $  411.689$ & $  234.134$ \\
 &  &  &  &  & $(   37.516)$ & $(   45.863)$ &  & $(   56.714)$ & $(   51.533)$ \\
~$ b_{j[\Delta a_{t}]}$ &  &  &  &  & $    2.323$ & $   -4.338$ &  & $    9.911$ & $    2.311$ \\
 &  &  &  &  & $(   13.248)$ & $(    3.252)$ &  & $(   15.406)$ & $(    5.304)$ \\
~$ b_{j[y_{t-1}]}$ &  &  &  &  & $   66.091$ & $   10.792$ &  & $   97.551$ & $   22.962$ \\
 &  &  &  &  & $(   12.527)$ & $(    5.575)$ &  & $(   20.762)$ & $(   11.452)$ \\
~$ b_{j[a_{t-1}]}$ &  &  &  &  & $  -35.506$ & $   -2.091$ &  & $  -17.629$ & $   -5.973$ \\
 &  &  &  &  & $(   10.626)$ & $(    2.573)$ &  & $(   12.542)$ & $(    5.148)$ \\
~$ b_{j[\Delta y_{-jt}]}$ &  &  &  &  & $ -181.052$ & $ -107.510$ &  & $ -266.163$ & $ -201.590$ \\
 &  &  &  &  & $(   42.416)$ & $(   44.870)$ &  & $(   59.545)$ & $(   42.696)$ \\
~$ b_{j[\Delta q_{t}]}$ &  &  &  &  & $    0.000$ & $   -0.000$ &  & $    0.001$ & $    0.000$ \\
 &  &  &  &  & $(    0.001)$ & $(    0.000)$ &  & $(    0.001)$ & $(    0.000)$ \\
~$ b_{j[\Delta a_{t-1}]}$ &  &  &  &  & $    2.965$ & $   -3.125$ &  & $    6.118$ & $    0.907$ \\
 &  &  &  &  & $(    4.059)$ & $(    1.799)$ &  & $(    4.715)$ & $(    3.302)$ \\
~$ b_{j[\Delta a_{t-2}]}$ &  &  &  &  &  &  &  & $   -1.644$ & $   -0.044$ \\
 &  &  &  &  &  &  &  & $(    4.717)$ & $(    2.271)$ \\
~$ b_{j[\Delta e_{jt}]}$ &  &  &  &  & $-1329.544$ & $ -151.080$ &  & $-2996.015$ & $ -429.334$ \\
 &  &  &  &  & $(  327.759)$ & $(   94.107)$ &  & $(  622.340)$ & $(  189.228)$ \\
~$ b_{j[e_{jt-1}]}$ &  &  &  &  & $  -76.846$ & $  -31.467$ &  & $ -211.880$ & $  -55.622$ \\
 &  &  &  &  & $(   38.857)$ & $(   18.804)$ &  & $(   46.915)$ & $(   25.279)$ \\
\noalign{\smallskip}
\mcl{5}{l}{~\text{$ p$ value for }$ {\cal H}_{0}^\text{FC}$} & $    0.000$ & $    0.018$ &  & $    0.049$ & $    0.013$ \\
\mcl{5}{l}{~\text{$ p$ value for }$ {\cal H}_{0}^\text{NC}$} & $    0.001$ & $    0.009$ &  & $    0.030$ & $    0.008$ \\
\noalign{\smallskip}
\mcl{5}{l}{Observations}                       & $8,513$ & $8,505$    && $        6,028$ & $        6,028$ \\
\bottomrule 
\end{tabular}}
\caption*{\fsz\emph{Notes:} The table reports estimates from the reduced form (first) specification of the test with wages as distribution factors, accounting for endogenous human capital. $b_{j[\Delta e_{jt}]}$ is the coefficient on $\Delta \log e_{jt}$ and $b_{j[e_{jt-1}]}$ is the coefficient on $\log e_{jt-1}$. The sample sizes are smaller than in table \ref{Table::OLS_Results} because we remove a few women with negative human capital. Standard errors clustered at the household level are in brackets.}
\end{center}
\end{table}

We estimate the reduced form specification of the test accounting for endogenous human capital. We do not fully observe individuals' work history in the PSID, let alone other forms of human capital, so we resort to defining human capital (in years) as age minus years of education minus 6. We report the results in table \ref{AppTable::OLS_Results_Endogenous_Human_Capital}. 

Own past shocks from $t-1$ and $t-2$ consistently reduce own labor supply in the future. The effects from the partner's shocks are mixed but mostly non-different from zero. Statistical significance is lower than in the baseline (the model is more flexible), but any coefficient that \emph{is} significant has the sign that limited commitment postulates. Consequently, we reject both full and no commitment. Results for the structural specification are similar but we do not show them in the interest of brevity.

\

\noindent \textbf{Labor market participation.} An extension to the extensive margin of labor supply is straightforward; see \citet{BlundellChiapporiMagnacMeghir2007NonParticipation} and \citet{Voena2015}. We do not model it here because our estimating equations rely on the log-linearization of the agents' marginal utility, which is not possible with discrete outcomes. We would have to resort to a numerical solution to a fully specified model, subject to the limitation of parameterizing preferences and expectations, thus conducting a joint test of commitment \emph{and} the specification used. 

The exclusion restrictions of distribution factors are, however, unrelated to the log-linearization. Consequently, our test based on the role of current and past shocks remains conceptually intact by the lack of extensive margin. Empirically, our results are sensitive to this choice only to the extent that those \emph{not} participating operate in a different commitment mode than the rest. While full commitment facilitates specialization (so those not working may operate in full commitment), it is hard to believe that this would meaningfully change our results in light of the high participation rates (among couples who meet all our other selection criteria, 91.6\% of men and 80.1\% of women work for pay). 

\

\noindent \textbf{Stochastic process for wages.} We have conducted the characterization of behavior and proposed a test for commitment based (mostly) on shocks to individual hourly wages, without taking a stance on the process governing those shocks. This is because, as we argue below, the test does not require us to take a stance on the stochastic process for $\omega_{jt}$. There are two somewhat related issues that require discussion in this regard. 

The first issue concerns the fact that $\omega_{jt}$ may have distinct permanent and transitory components. Unless directly observed, it is impossible to separate the two without imposing some structure. Empirically, our results depend on the extent to which wage shocks truly induce bargaining under no/limited commitment, which may be debatable for transitory or small permanent shocks. As $\omega_{jt}$ may partly reflect such `unimportant' shocks, the test may \emph{fail} to reject full/no commitment, even if limited commitment is the true underlying mode. In practice, we strongly reject full and no commitment despite any `unimportant' transitory components. This suggests that the rejection would have been even stronger if we had isolated the `important' persistent/large shocks underlying $\omega_{jt}$.

A second issue is that wages may be correlated over time, reflecting persistence, path dependence, or mean reversion. It is hard to discipline these dynamics without specifying a stochastic process for $\omega_{jt}$. The important question, however, is to what extent such dynamics affect our test. To put differently, if past wages affect future wages, we must make sure that our test correctly picks up bargaining/commitment, rather than these wage dynamics.

Assuming joint taxation and taste observables away, the first-order condition for hours is given by $-\mu_{jt} u_{j[h]}(q_{t}, h_{jt}) = \lambda_{t} w_{jt}$ (appendix \ref{Appendix::DerivationOptimalityConditions}). This condition is derived for \emph{any} process underlying $w_{jt}$ (endogenous wages being an exception; see human capital above); it fully characterizes optimal behavior and forms the basis for our estimating equations in the paper. Past wages may affect behavior through three channels: the Pareto weight $\mu_{jt}$ (bargaining; our main point), the current wage $w_{jt}$ (wage persistence), and the marginal utility of wealth $\lambda_t$ (wealth effects and expectations about future wages). To interpret any effects from past wages as bargaining effects through $\mu_{jt}$, we must control for $w_{jt}$ and $\lambda_t$ -- which we clearly do in \eqref{Eq::EstimableEquation.Final}. In other words, as long as we control for $w_{jt}$ and $\lambda_t$, the past plays no other role in the optimality condition for hours, except through $\mu_{jt}$. Clearly the way we control for the unobserved $\lambda_t$ (polynomial in assets and income -- see section \ref{SubSection::Implementation_Estimation}) may be open to debate, but we tried multiple richer specifications (higher-order polynomials, controls for wealth and income in additional time periods) and our results have not been sensitive.

\

\noindent \textbf{Nonlinear wage effects.} A related issue concerns the possibility of nonlinear effects from wages on hours, through bargaining or other channels. The idea is that large wage shocks may induce large shifts in bargaining and/or in wealth, while smaller shocks may leave the Pareto weight and the budget set unchanged. This introduces bias: even if limited commitment is the true underlying mode, our test may \emph{fail} to reject full/no commitment when $\omega_{jt}$ consists of small shocks that leave the Pareto weight mostly intact. While, in practice, we strongly reject full and no commitment despite the presence of any `unimportant' small shocks underlying $\omega_{jt}$, we also take an alternative route to explicitly allow for nonlinear wage effects.

Assuming joint taxation and taste observables away, the first-order condition for hours after a first difference in logs is given by $\Delta \log \mu_{jt} + \Delta \log(-u_{j[h]}(q_{t}, h_{jt})) = \Delta \log \lambda_{t} + \Delta \log w_{jt}$. This equation holds \emph{exactly}, and wage growth $\Delta \log w_{jt}$ enters it \emph{linearly}. However, shocks may enter the Pareto weight $\mu_{jt}$ nonlinearly and hours may relate nonlinearly to the marginal utility $u_{j[h]}$. We have carried out log-linear approximations to $\mu_{jt}$ and $-u_{j[h]}$, and it is precisely these linearizations that give rise to the bias we mentioned above. 

We maintain the log-linear approximation to $-u_{j[h]}$ (the estimating equation is otherwise not in closed-form for hours) but we carry out a \emph{second}-order Taylor approximation to $\log \mu_{jt}$. This is a straightforward generalization of the approximations in appendix \ref{Appendix::ApproximationParetoWeight}. It results in writing $\Delta \log \mu_{jt}$ as a quadratic polynomial in wage shocks and all other distribution factors. The estimating equation for hours is similar to the baseline \eqref{Eq::EstimableEquation.Final}, up to additive terms for the square of current and past wage shocks, the square of current and past distribution factors (e.g. BMI), and all pair-wise interaction terms. Wage shocks in this case do not affect hours uniformly; instead, their effect depends on their size and sign.

\begin{table}  
\begin{center}
\caption{Commitment test with nonlinear wage effects -- reduced form results in detail}\label{AppTable::OLS_Results_NonlinearWageEffects}
\resizebox{0.95\columnwidth}{!}{
\begin{tabular}{L{2.0cm} C{0.6cm} C{0.7cm} C{0.7cm} C{0.01cm} C{2.0cm} C{2.0cm} C{0.01cm} C{2.0cm} C{2.0cm}}
\toprule
                            &&&&    & \mcl{5}{c}{\textbf{wage shocks}}        \\
\cmidrule{6-10}                        
                            &&&&    & \mcl{2}{c}{$\geq$ 3 periods} && \mcl{2}{c}{$\geq$ 4 periods} \\
\cmidrule{6-7}\cmidrule{9-10}
                            &&&&    & (1)      & (2)      && (3)    & (4)     \\
                            &&&&    & Male     & Female   && Male   & Female  \\
                            &FC&NC&LC&   & $j=1$    &  $j=2$   &&  $j=1$ &  $j=2$  \\
\midrule
\multicolumn{10}{l}{\emph{Current shocks} ($t$)}\\
~~$\beta_{j[w_{jt}]}$ & . & . & . &  & $  -46.574$ & $  -36.954$ &  & $  -49.674$ & $  -44.847$ \\
 &  &  &  &  & $(   14.179)$ & $(    8.250)$ &  & $(   22.150)$ & $(   17.840)$ \\
~~$\beta_{j[w_{jt}^2]}$ & . & . & . &  & $  -18.017$ & $  -11.310$ &  & $  -13.032$ & $  -15.173$ \\
 &  &  &  &  & $(    8.237)$ & $(    3.062)$ &  & $(   15.103)$ & $(    4.642)$ \\
~~$\beta_{j[w_{-jt}]}$ & $0$ & $(+)$ & $(+)$ &  & $   24.700$ & $  -10.609$ &  & $   -1.665$ & $   -9.495$ \\
 &  &  &  &  & $(   29.470)$ & $(    8.003)$ &  & $(   32.952)$ & $(   19.723)$ \\
~~$\beta_{j[w_{-jt}^2]}$ & $0$ &  &  &  & $  -21.449$ & $  -11.623$ &  & $  -42.064$ & $   -7.381$ \\
 &  &  &  &  & $(   13.382)$ & $(    4.512)$ &  & $(   17.411)$ & $(   10.658)$ \\
\multicolumn{10}{l}{\emph{Past shocks} ($t-1$)}\\
~~$\beta_{j[w_{jt-1}]}$ & $0$ & $0$ & $(-)$ &  & $    0.310$ & $   -7.225$ &  & $   -1.200$ & $  -12.060$ \\
 &  &  &  &  & $(    4.072)$ & $(    7.063)$ &  & $(   11.708)$ & $(   11.318)$ \\
~~$\beta_{j[w_{jt-1}^2]}$ & $0$ & $0$ &  &  & $  -14.393$ & $   -2.118$ &  & $  -17.420$ & $   -3.432$ \\
 &  &  &  &  & $(    5.048)$ & $(    3.846)$ &  & $(   14.144)$ & $(    4.538)$ \\
~~$\beta_{j[w_{-jt-1}]}$ & $0$ & $0$ & $(+)$ &  & $   49.496$ & $   11.162$ &  & $   37.794$ & $   -4.299$ \\
 &  &  &  &  & $(   22.699)$ & $(    7.236)$ &  & $(   35.211)$ & $(   14.516)$ \\
~~$\beta_{j[w_{-jt-1}^2]}$ & $0$ & $0$ &  &  & $   11.062$ & $   -2.815$ &  & $   10.049$ & $   19.748$ \\
 &  &  &  &  & $(   10.779)$ & $(    2.057)$ &  & $(   20.405)$ & $(    8.351)$ \\                     
\multicolumn{10}{l}{\emph{Older shocks} ($t-2$)}\\
~~$\beta_{j[w_{jt-2}]}$ & $0$ & $0$ & $(-)$ &  &  &  &  & $  -27.638$ & $  -12.827$ \\
 &  &  &  &  &  &  &  & $(   18.882)$ & $(    7.139)$ \\
~~$\beta_{j[w_{jt-2}^2]}$ & $0$ & $0$ &  &  &  &  &  & $    2.680$ & $    2.570$ \\
 &  &  &  &  &  &  &  & $(    7.837)$ & $(    2.451)$ \\
~~$\beta_{j[w_{-jt-2}]}$ & $0$ & $0$ & $(+)$ &  &  &  &  & $   36.626$ & $   48.005$ \\
 &  &  &  &  &  &  &  & $(   30.802)$ & $(   18.696)$ \\
~~$\beta_{j[w_{-jt-2}^2]}$ & $0$ & $0$ &  &  &  &  &  & $  -32.217$ & $   -5.251$ \\
 &  &  &  &  &  &  &  & $(   23.404)$ & $(    4.660)$ \\                        
\bottomrule 
\end{tabular}}
\end{center}
\end{table}

\begin{table}  
\begin{center}
\caption*{Table \ref{AppTable::OLS_Results_NonlinearWageEffects} (continued): Commitment test with nonlinear wage effects -- reduced form results in detail}
\resizebox{0.95\columnwidth}{!}{
\begin{tabular}{L{2.0cm} C{0.6cm} C{0.7cm} C{0.7cm} C{0.01cm} C{2.0cm} C{2.0cm} C{0.01cm} C{2.0cm} C{2.0cm}}
\toprule
                            &&&&    & \mcl{5}{c}{\textbf{wage shocks}}        \\
\cmidrule{6-10}                        
                            &&&&    & \mcl{2}{c}{$\geq$ 3 periods} && \mcl{2}{c}{$\geq$ 4 periods} \\
\cmidrule{6-7}\cmidrule{9-10}
                            &&&&    & (1)      & (2)      && (3)    & (4)     \\
                            &&&&    & Male     & Female   && Male   & Female  \\
                            &FC&NC&LC&   & $j=1$    &  $j=2$   &&  $j=1$ &  $j=2$  \\
\midrule                
\multicolumn{10}{l}{\emph{Initial distribution factors} ($t=0$)}\\
~~$\beta_{j[young_{j}]}$ & $0$ & $0$ & $-$ &  & $  -69.599$ & $  -37.408$ &  & $ -133.636$ & $  -44.005$ \\
 &  &  &  &  & $(   27.910)$ & $(    9.711)$ &  & $(   36.621)$ & $(   12.826)$ \\
~~$\beta_{j[young_{-j}]}$ & $0$ & $0$ & $+$ &  & $ -107.822$ & $   10.812$ &  & $ -152.297$ & $   18.435$ \\
 &  &  &  &  & $(   31.655)$ & $(    6.507)$ &  & $(   40.714)$ & $(    9.949)$ \\
\multicolumn{10}{l}{\emph{Other controls}}\\
~$ b_{j[\Delta y_{t}]}$ &  &  &  &  & $  272.385$ & $  122.285$ &  & $  336.983$ & $  195.537$ \\
 &  &  &  &  & $(   37.924)$ & $(   46.312)$ &  & $(   57.365)$ & $(   61.871)$ \\
~$ b_{j[\Delta a_{t}]}$ &  &  &  &  & $   -2.060$ & $   -0.594$ &  & $   -4.220$ & $    7.221$ \\
 &  &  &  &  & $(   13.083)$ & $(    5.111)$ &  & $(   14.021)$ & $(    8.538)$ \\
~$ b_{j[y_{t-1}]}$ &  &  &  &  & $   34.834$ & $    5.598$ &  & $   31.902$ & $    2.439$ \\
 &  &  &  &  & $(    9.669)$ & $(    3.568)$ &  & $(    9.988)$ & $(    7.673)$ \\
~$ b_{j[a_{t-1}]}$ &  &  &  &  & $  -32.307$ & $   -5.392$ &  & $  -27.494$ & $   -3.418$ \\
 &  &  &  &  & $(    9.044)$ & $(    3.122)$ &  & $(    9.179)$ & $(    6.894)$ \\
~$ b_{j[\Delta y_{-jt}]}$ &  &  &  &  & $ -241.594$ & $ -103.130$ &  & $ -311.108$ & $ -161.300$ \\
 &  &  &  &  & $(   55.143)$ & $(   43.989)$ &  & $(   86.238)$ & $(   50.358)$ \\
~$ b_{j[\Delta q_{t}]}$ &  &  &  &  & $    0.001$ & $   -0.000$ &  & $    0.000$ & $    0.000$ \\
 &  &  &  &  & $(    0.001)$ & $(    0.000)$ &  & $(    0.001)$ & $(    0.000)$ \\
~$ b_{j[\Delta a_{t-1}]}$ &  &  &  &  & $    2.706$ & $  -12.297$ &  & $   -3.987$ & $  -10.644$ \\
 &  &  &  &  & $(   12.318)$ & $(    3.752)$ &  & $(   14.427)$ & $(    8.348)$ \\
~$ b_{j[\Delta a_{t-2}]}$ &  &  &  &  &  &  &  & $   -4.847$ & $    1.086$ \\
 &  &  &  &  &  &  &  & $(    6.356)$ & $(    2.718)$ \\
~$ b_{j[(\Delta a_{t})^2]}$ &  &  &  &  & $    2.641$ & $    4.290$ &  & $    5.264$ & $    7.381$ \\
 &  &  &  &  & $(    3.271)$ & $(    2.551)$ &  & $(    3.040)$ & $(    2.978)$ \\
~$ b_{j[(\Delta a_{t-1})^2]}$ &  &  &  &  & $   -0.777$ & $    1.056$ &  & $   -0.516$ & $    0.643$ \\
 &  &  &  &  & $(    1.231)$ & $(    0.431)$ &  & $(    1.486)$ & $(    0.797)$ \\
~$ b_{j[(\Delta a_{t-2})^2]}$ &  &  &  &  &  &  &  & $    0.333$ & $    0.461$ \\
 &  &  &  &  &  &  &  & $(    0.597)$ & $(    0.242)$ \\
\noalign{\smallskip}
\mcl{5}{l}{~\text{$ p$ value for }$ {{\cal H}_{0}^\text{FC}}^{\#}$} & $    0.000$ & $    0.000$ &  & $    0.000$ & $    0.000$ \\
\mcl{5}{l}{~\text{$ p$ value for }$ {{\cal H}_{0}^\text{NC}}^{\#}$} & $    0.000$ & $    0.000$ &  & $    0.000$ & $    0.000$ \\
\noalign{\smallskip}
\mcl{5}{l}{Observations}                       & \mcl{2}{c}{8,513}    && \mcl{2}{c}{6,028} \\
\bottomrule 
\end{tabular}}
\caption*{\fsz\emph{Notes:} The table reports the coefficient estimates from the reduced form (first) specification of the commitment test with wages as time-varying distribution factors, allowing for nonlinear bargaining effects of wages through a second-order Taylor approximation of the Pareto weight. $\beta_{j[w_{kt-\tau}^2]}$ is the coefficient on $\omega_{kt-\tau}^2$ for $k\in\{1,2\}$ and $\tau=\{0,1,2\}$. The sign of the typical bargaining effects under no/limited commitment appears in brackets because, with nonlinear terms, what matters is the sign of the partial effect $\partial \Delta \log h_{jt} / \partial \omega_{kt-\tau}$ for a given level of the wage shock, which we report in table \ref{AppTable::PartialEffects_NonlinearWageEffects}. Standard errors clustered at the household level are in brackets. $^{\#}$We assess the null hypotheses at the average wage shock of zero.}
\end{center}
\end{table}

\begin{sidewaystable}[t]  
\begin{center}
\caption{Average partial effects of wage shocks along the distribution of shocks}\label{AppTable::PartialEffects_NonlinearWageEffects}
\begin{tabular}{L{3.7cm} C{0.6cm} C{0.7cm} C{0.6cm} C{1.5cm} C{1.5cm} C{1.5cm} C{1.5cm} c C{1.5cm} C{1.5cm} C{1.5cm} C{1.5cm}}
\toprule
                                                        &&&&\multicolumn{4}{c}{\textbf{wage cut}}&&\multicolumn{4}{c}{\textbf{wage rise}}\\
                                                        &&&&\multicolumn{4}{c}{(negative shock to $\Delta \log w_{k\tau}$)}&&\multicolumn{4}{c}{(positive shock to $\Delta \log w_{k\tau}$)}\\
\cmidrule{5-8}\cmidrule{10-13}                                                   
                                                        &&&&\multicolumn{4}{c}{wage shock $\omega =$}&&\multicolumn{4}{c}{wage shock $\omega =$}\\
                                                        &FC   &NC   &LC     & $-0.5$ & $-0.25$ & $-0.1$ & $-0.01$ && $0.01$ & $0.1$ & $0.25$ & $0.5$\\
\midrule
\multicolumn{13}{l}{\emph{Partial effects of current shocks }$(\tau=t)$}\\
~~$\partial \Delta \log h_{1t} / \partial \omega_{1t}$ & $.$ & $.$ & $.$ & $   -0.020$ & $   -0.024$ & $   -0.026$ & $   -0.027$ &  & $   -0.027$ & $   -0.029$ & $   -0.031$ & $   -0.035$ \\
~~$\partial \Delta \log h_{2t} / \partial \omega_{2t}$ & $.$ & $.$ & $.$ & $   -0.028$ & $   -0.035$ & $   -0.040$ & $   -0.042$ &  & $   -0.043$ & $   -0.045$ & $   -0.050$ & $   -0.057$ \\
~~$\partial \Delta \log h_{1t} / \partial \omega_{2t}$ & $0$ & $+$ & $+$ & $    0.022$ & $    0.011$ & $    0.004$ & $   -0.000$ &  & $   -0.001$ & $   -0.006$ & $   -0.012$ & $   -0.024$ \\
~~$\partial \Delta \log h_{2t} / \partial \omega_{1t}$ & $0$ & $+$ & $+$ & $   -0.002$ & $   -0.005$ & $   -0.008$ & $   -0.009$ &  & $   -0.009$ & $   -0.010$ & $   -0.012$ & $   -0.016$ \\
\multicolumn{13}{l}{\emph{Partial effects of immediately past shocks }$(\tau=t-1)$}\\
~~$\partial \Delta \log h_{1t} / \partial \omega_{1t-1}$ & $0$ & $0$ & $-$ & $    0.009$ & $    0.004$ & $    0.001$ & $   -0.000$ &  & $   -0.001$ & $   -0.003$ & $   -0.005$ & $   -0.010$ \\
~~$\partial \Delta \log h_{2t} / \partial \omega_{2t-1}$ & $0$ & $0$ & $-$ & $   -0.008$ & $   -0.010$ & $   -0.011$ & $   -0.011$ &  & $   -0.011$ & $   -0.012$ & $   -0.013$ & $   -0.015$ \\
~~$\partial \Delta \log h_{1t} / \partial \omega_{2t-1}$ & $0$ & $0$ & $+$ & $    0.015$ & $    0.018$ & $    0.020$ & $    0.021$ &  & $    0.021$ & $    0.022$ & $    0.024$ & $    0.026$ \\
~~$\partial \Delta \log h_{2t} / \partial \omega_{1t-1}$ & $0$ & $0$ & $+$ & $   -0.023$ & $   -0.013$ & $   -0.008$ & $   -0.004$ &  & $   -0.004$ & $   -0.000$ & $    0.005$ & $    0.015$ \\
\multicolumn{13}{l}{\emph{Partial effects of older shocks }$(\tau=t-2)$}\\
~~$\partial \Delta \log h_{1t} / \partial \omega_{1t-2}$ & $0$ & $0$ & $-$ & $   -0.017$ & $   -0.016$ & $   -0.016$ & $   -0.015$ &  & $   -0.015$ & $   -0.015$ & $   -0.014$ & $   -0.014$ \\
~~$\partial \Delta \log h_{2t} / \partial \omega_{2t-2}$ & $0$ & $0$ & $-$ & $   -0.015$ & $   -0.013$ & $   -0.013$ & $   -0.012$ &  & $   -0.012$ & $   -0.012$ & $   -0.011$ & $   -0.010$ \\
~~$\partial \Delta \log h_{1t} / \partial \omega_{2t-2}$ & $0$ & $0$ & $+$ & $    0.038$ & $    0.029$ & $    0.024$ & $    0.021$ &  & $    0.020$ & $    0.017$ & $    0.011$ & $    0.002$ \\
~~$\partial \Delta \log h_{2t} / \partial \omega_{1t-2}$ & $0$ & $0$ & $+$ & $    0.050$ & $    0.048$ & $    0.046$ & $    0.045$ &  & $    0.045$ & $    0.044$ & $    0.043$ & $    0.040$ \\
\bottomrule
\end{tabular}
\caption*{\fsz\emph{Notes:} The table reports the partial effects of wage shocks in the specification with nonlinear wage effects across households observed for $\geq4$ periods. The partial effects, evaluated at the sample average of hours, are calculate along the marginal distribution of the corresponding wage shock. The table reports average effects following a 50\%, 25\%, 10\%, and 1\% reduction in wages ($\omega=\{-0.5,-0.25,-0.1,-0.01\}$), as well as effects following an 1\%, 10\%, 25\%, and 50\% increase ($\omega=\{0.01,0.1,0.25,0.5\}$). The parameter estimates are in appendix table \ref{AppTable::OLS_Results_NonlinearWageEffects}.}
\end{center}
\end{sidewaystable}

We estimate the reduced form specification for hours introducing the squared wage shocks over multiple periods as standalone regressors. We drop the interaction terms to keep things tidy. Table \ref{AppTable::OLS_Results_NonlinearWageEffects} presents the results. Although some squared terms are statistically significant, the linear terms remain mostly unchanged from the baseline. We thus strongly reject full and no commitment. The parameter estimates are hard to interpret, so table \ref{AppTable::PartialEffects_NonlinearWageEffects} reports the partial effects of wage shocks, $\partial \Delta \log h_{jt} / \partial \omega_{kt-\tau}$ for $j,k\in\{1,2\}$ and $\tau\in\{0,1,2\}$, for different magnitudes and signs of the shocks. In most cases, the partial effects have the sign implied by limited commitment. Moreover, negative shocks are often associated with larger shifts in labor supply than positive shocks of the \emph{same} magnitude (inspect e.g. the effects of own and partner shocks at $t-2$), which is a recurrent empirical finding in the literature on the pass-through of income shocks into consumption/labor supply. 

\

\noindent \textbf{Costly renegotiation.} \citet{KatoRiosRull2022MarkovEquilibria} propose a protocol for repeated bargaining in which changing circumstances shift the agents' outside options, but renegotiation costs prevent the spouses from continuously updating their sharing of resources. Their framework is effectively one of no commitment, albeit with renegotiation costs. Intuitively, if renegotiation entails a cost, some minor renegotiations will not take place (e.g. after some minor shock) and the agents will go on with the previous allocation of resources. Behavior under no commitment will thus not respond to all contemporaneous information and renegotiations will be infrequent depending on the size of the cost. This appears to imply that behavior under no commitment may look like behavior under limited commitment, in which renegotiation is occasional following a binding participation constraint. However, the infrequency in \citet{KatoRiosRull2022MarkovEquilibria} is only driven by the size of the current renegotiation cost rather than by alternative histories that drive it in limited commitment.

Our test does \emph{not} rely on the frequency of renegotiation for distinguishing among the commitment alternatives. Instead, it relies on the \emph{sets of variables} that matter for bargaining in each case. While with renegotiation costs some minor shocks do not shift the allocation of resources, (large) current shocks still matter under no commitment while historical shocks do not. \citet{KatoRiosRull2022MarkovEquilibria} admit this themselves by emphasizing that their protocol is attractive for computational reasons in particular, precisely because it does not require ``the carryover of a continuous state variable'', namely of history, while enabling infrequent updating as in limited commitment. By contrast, historical shocks still matter under limited commitment, so history can still be used to distinguish between the two regimes even in the presence of renegotiation costs. In other words, costly renegotiation does not alter the role of contemporaneous vis-\`a-vis historical information for behavior. 

\end{document}